\documentclass[a4paper,12pt,oneside]{article}
\usepackage[utf8]{inputenc}
\usepackage{graphicx}
\usepackage[T1]{fontenc}
\usepackage{amsmath}
\usepackage{amssymb}
\usepackage{float}
\usepackage{hyperref}
\usepackage{cleveref}
\usepackage{lmodern}
\usepackage{empheq}
\usepackage{braket}
\usepackage{xcolor}
\usepackage{cancel}
\usepackage[pagewise]{lineno}
\usepackage{csquotes}
\usepackage{scalerel}
\usepackage{appendix}
\usepackage{afterpage}
\usepackage[mathscr]{eucal}
\usepackage[shortlabels]{enumitem}
\usepackage{import}
\usepackage{tikz}
\usetikzlibrary{shapes.geometric}
\usepackage{booktabs}
\usepackage{mathtools}
\usepackage{titlesec}
\usepackage{caption}
\usepackage{subcaption}
\usepackage{textgreek}
\usepackage{blindtext}
\usepackage{titlesec}
\usepackage{indentfirst}

\titleformat{\section}
{\normalfont\fontsize{12}{12}\selectfont\bfseries} 
{\thesection}
{1em}
{}

\titleformat{\subsection}
{\normalfont\fontsize{12}{12}\selectfont\bfseries} 
{\thesubsection}
{1em}
{}

\newcommand{\orcid}[1]{\href{https://orcid.org/#1}{\includegraphics[width=8pt]{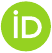}}}

\usepackage[left=3.0cm,right=2.5cm,top=3.0cm,bottom=2.5cm]{geometry}

\hypersetup{
	colorlinks   = true, 
	urlcolor     = blue, 
	linkcolor    = blue, 
	citecolor   = blue 
}

\begin{document}
	
	\numberwithin{equation}{section}
	
	\begin{center}
		\vspace{5mm}
		\large
		\textbf{Extended Scalar Particle Solutions in Black String Spacetimes with Anisotropic Quintessence}\\
		\normalsize
		\vspace{5mm}
		M. L. Deglmann\textsuperscript{*}\footnote{m.l.deglmann@posgrad.ufsc.br}\,\orcid{0000-0002-9737-5997}\,,
		B. V. Simão\footnote{vallin.bruna@posgrad.ufsc.br}\,\orcid{0000-0001-6087-5608}\,,
		C. C. Barros Jr.\footnote{barros.celso@ufsc.br}\,\orcid{0000-0003-2662-1844}.
		\let\thefootnote\relax\footnotetext{*\ Corresponding author: m.l.deglmann@posgrad.ufsc.br.}
		
	\end{center}\begin{center}
		Departamento de Física, Universidade Federal de Santa Catarina (UFSC), Campus Universitário Trindade, Florianópolis, 88035-972, Santa Catarina, Brazil
	\end{center}

	\vspace{5mm}

	\begin{abstract}
		We present novel solutions to the Klein-Gordon equation in a black string spacetime immersed in a Kiselev-like quintessence fluid and surrounded by a cloud of strings (BCK spacetime). 
        The solutions are shown to depend on the quintessence state parameter $\alpha_{\scaleto{Q}{5pt}}$ and extend to a broader radial domain than previously reported. Explicit analytical results are provided for $\alpha_{\scaleto{Q}{5pt}} = 0, 1/2, 1$, thereby encompassing all relevant physical scenarios. 
        These analytical radial solutions are derived using the confluent and biconfluent Heun equations, as well as Bessel equations in specific cases. 
        Constraints on the Heun parameters that produce spectral restrictions are examined, resulting in the $\delta_{\scaleto{N}{5pt}}$-subclass of solutions, which includes the polynomial cases. 
        More importantly, quintessence-induced \enquote{dark phases} are identified for several scenarios; in particular, the regime $\alpha_{\scaleto{Q}{5pt}} = 1$ is emphasized for future comparison with alternative spacetime geometries. These findings contribute to the understanding of scalar particle dynamics and the influence of dark energy on quantum systems in curved spacetime backgrounds.\\
		
	\end{abstract}

	\section{Introduction}	
	
	Over the past decades, multiple independent observations have established strong evidence for the accelerated expansion of the Universe. Astrophysical measurements based on the luminosity distances of Type Ia supernovae \cite{Riess1998, Perlmutter1999}, together with data from the cosmic microwave background (CMB), baryon acoustic oscillations (BAO), and large-scale structure (LSS), consistently indicate the presence of an unknown component, termed dark energy (DE), that drives this late-time accelerated cosmic expansion (see, for instance, \cite{Amendola2010, Steinhardt2003, COPELAND2006}).

	Various hypotheses for dark energy have been proposed \cite{Amendola2010}, including the cosmological constant, which remains the primary candidate in the $\Lambda$CDM cosmological model. Nevertheless, challenges such as the fine-tuning and coincidence problems \cite{Amendola2010} have motivated ongoing research into alternative candidates for dark energy. Among the most interesting ones is the quintessence model, introduced by Caldwell, Dave, and Steinhardt \cite{Caldwell1998}, which has attracted significant attention.

	Subsequently, Kiselev \cite{Kiselev2003} proposed an alternative approach to modeling physical systems involving dark energy by introducing an energy-momentum tensor for a time-independent anisotropic fluid in a spherically symmetric spacetime. In this framework, the pressure components along the radial ($r$), polar ($\theta$), and azimuthal ($\varphi$) directions are all negative. The fluid, initially considered to surround a black hole (BH), is also called quintessence, although it is not directly linked to a scalar-field Lagrangian. Indeed, Visser \cite{Visser2020} has demonstrated that this Kiselev fluid is neither a perfect fluid nor, in the standard cosmological sense, a quintessence field. However, due to its analytical tractability in exploring dark energy-induced implications, this approach has been widely adopted \cite{Ali2020,Saadati2021,AlBadawi2025a,Wang2023,Lungu2025, Dariescu2024}. Thus, we shall consider this type of fluid in our work, and, while we refer to this dark energy candidate simply as \enquote{quintessence}, it must always be understood as a Kiselev-like quintessence fluid. Furthermore, we also add that some studies have incorporated a cloud of strings (CS) \cite{Letelier1979} as an additional matter distribution alongside the Kiselev fluid in the spacetime background (see, e.g., \cite{Chaudhary2024} and \cite{Rayimbaev2024}). This addition also contributes to our spacetime configuration, as will be detailed later.

	The research conducted by Ali et al. \cite{Ali2020} is particularly relevant, as it investigates the properties of a spacetime containing black strings immersed in a quintessential fluid analogous to the Kiselev model. Black strings are solutions to the Einstein field equations (EFEs) that share similarities with Schwarzschild black holes but possess cylindrical symmetry within an anti-de Sitter (AdS) spacetime \cite{Lemos1995, Lemos1996}. Ali and colleagues examined the quintessence equation of state, the thermodynamics of black strings, and the scenario of a rotating black string. To extend this configuration, the present study incorporates a matter component analogous to a cloud of strings\footnote{Originally, a cloud of strings is an effective model describing a disordered distribution of one-dimensional relativistic strings.} (CS) \cite{Letelier1979}, which aligns with the underlying symmetry of the system. This configuration, comprising the black string, the quintessence fluid, and the cloud of strings, is referred to as the BCK spacetime. The selection of these fundamental elements aims to facilitate the investigation of their individual effects on the behavior of quantum systems within this background.

	A previous study \cite{Deglmann2025} examined the properties of the BCK spacetime by identifying admissible ranges for the parameters defining the spacetime geometry, specifically the cloud of strings parameter ($\overline{a}$), the Schwarzschild-like radius ($\rho_{\scaleto{S}{4pt}}$), and the quintessence parameter ($N_{\scaleto{Q}{5pt}}$), additionally analyzing their influence on the metric function $A(\rho)$. In that work, the mapping from spherical to cylindrical coordinates was performed to estimate these parameters, based on typical values from astrophysical objects and the estimated amount of dark energy in the current observable Universe. The investigation subsequently addressed the solution of the Klein-Gordon (KG) equation for a scalar particle near the event horizon, employing the confluent Heun equation\footnote{In general, Heun equations are an excellent tool to handle exact solutions of physical systems. See, for example, the works of \cite{Batic2013, Figueiredo2024, Hortacsu2018, Petroff2007, Schmidt2023, Vieira2016}.} (CHE) \cite{Ronveaux1995}. In addition to the general solution near the event horizon, the study discussed the implications for various predicted scenarios.

	In the specific case where the quintessence state parameter $\alpha_{\scaleto{Q}{5pt}}$ equals $1/2$, a quantum observable termed the \enquote{dark phase} was defined in association with the radial KG solution, which may\footnote{Under certain limits.} be interpreted as the radial wave function of the particle. In principle, this dark phase can also be determined\footnote{The other physical parameters can also change, defining different scenarios.} for any value of $\alpha_{\scaleto{Q}{5pt}}$, offering a novel perspective on the influence of a dark energy candidate on quantum systems. The concept of a dark phase is analogous to the phenomenon observed when scalar particles are situated near rotating objects \cite{Pinho2023}. For each eigenfunction of the wave equation, this phase can be utilized to construct potential observables through wave-function interferometry.

	The present study extends previous analyses by deriving new solutions to the Klein-Gordon equation that are valid over a broader radial domain than the one considered in \cite{Deglmann2025}. These solutions provide insights into particle dynamics across a wider range of scenarios, including cases without event horizons. Given the continuity of the quintessence state parameter ($\alpha_{\scaleto{Q}{5pt}}$), the analysis considers the cases where $\alpha_{\scaleto{Q}{5pt}}$ equals $0$, $1/2$, and $1$. This approach enables the radial wave functions to be expressed as solutions to the confluent and biconfluent Heun equations (CHE and BHE, respectively), with certain special cases represented by Bessel functions. Furthermore, the analysis addresses constraints on quantum energy levels for a specific subclass of solutions.

	In the sequence, we investigate the emergence of dark phases within these extended solutions. Although our analysis is conducted in a cylindrically symmetric spacetime, we still pay particular attention to the limiting case of $\alpha_{\scaleto{Q}{5pt}} = 1$ (which corresponds to $\omega_{\scaleto{Q}{5pt}} = -1$). This specific case is of significant interest as cosmological observations\footnote{Although recent findings from the Dark Energy Spectroscopic Instrument (DESI) favor a dynamical (time-varying) dark energy model \cite{DESI2025}, it is still valuable to consider a specific cosmic epoch where time-dependence may be negligible.} \cite{Amendola2010, PlanckCosmology2014, DESI2025} suggest that the state parameter of present-day dark energy in our Universe is very close to this value.

	There are two main reasons for working with this cylindrical spacetime. First, dark energy-induced quantum effects are expected to be extremely small, making analytical solutions essential for capturing these subtle results. Second, this approach facilitates comparison of the resulting cylindrical dark phases with similar configurations in other spacetime geometries. We also emphasize that, in a recent investigation \cite{Simao2025}, we demonstrated the occurrence of dark phases in the solutions of the Klein-Gordon equation for a scalar particle in the vicinity of the event horizon of a Schwarzschild black hole, surrounded by both a cloud of strings and the Kiselev fluid. For this purpose, we considered first-order approximations to the function $f(r)$, which determines the spacetime metric. This case showed us that, although dark phases occur in this configuration, their \enquote{intensity} is several orders of magnitude smaller than those found in the cylindrical case. In other words, the present work differs from previous ones in that it investigates the emergence of these dark phases in analytical solutions with a broader radial domain, which is determined separately for each physical scenario. Moreover, we also investigate the scenario where the black string is removed, which was not previously possible.

	The structure of this work is as follows: Section \ref{Sec-2-Results-Paper-One} summarizes the key findings from \cite{Deglmann2025}. Sections \cref{Extended_Solutions_Alpha_Q_0,Extended_Solutions_Alpha_Q_One_Half,Extended_Solutions_Alpha_Q_1} investigate the radial Klein-Gordon solution for $\alpha_{\scaleto{Q}{5pt}} = 0, 1/2, 1$, respectively, across an extended domain of the radial coordinate. Section \ref{Dark_Phases_Investigation} analyzes the emergence of dark phases within the context of these extended solutions, and \cref{Conclusion} provides the concluding remarks. Appendix sections \ref{appendix_CHE} and \ref{appendix_BHE} present the Confluent and Biconfluent Heun equations along with their corresponding solutions. Section \ref{appendix_Liouville_Normal_Form} details the Liouville normal forms for both cases.

	\section{Preliminary results}\label{Sec-2-Results-Paper-One}
	
	This section summarizes the main findings of our previous work \cite{Deglmann2025}. The analysis begins with a review of the cylindrically symmetric anti-de Sitter (AdS) spacetime that incorporates a black string (BS), a cloud of strings (CS), and a Kiselev-like quintessence fluid. Collectively, this configuration is referred to as the BCK spacetime. For clarity, we revise the spacetime metric, the criteria for event horizon formation, and the admissible ranges for the physical parameters. With these points established, we then present the general form for the radial Klein-Gordon (KG) equation, including the effective potential $V_{\text{eff}}(\rho)$, which is essential for our following analysis. At last, we discuss the KG solution near the event horizon and highlight the occurrence of quintessence-induced dark phases.
	
	Let us start by considering the interval
	\begin{equation}
		ds^{2} = A(\rho)\,c^{2}dt^{2} - 	\frac{d\rho^{2}}{A(\rho)} - \rho^{2}d\varphi^{2} - \frac{\rho^{2}}{l^{2}}\,dz^{2}\,,\label{metric}
	\end{equation}
	associated with the ansatz of a black string \cite{Ali2020}. Note that $A(\rho) \in \mathbb{R}$, the cylindrical coordinates are the standard $\rho$, $\varphi$, and $z$, and that we have chosen to work in the convention of SI units. In addition, $l$ is the AdS radius, satisfying $l^{2} = -3/\Lambda$, with $\Lambda<0$ being the cosmological constant. The components of the Einstein tensor for this metric take the form of
	\begin{align}
		G_{t}^{\ t} = G_{\rho}^{\ \rho}
		&= \frac{1}{\rho}\,\frac{dA}{d\rho} + \frac{A}{\rho^{2}}\,,\label{Einstein_t_ut}\\[5pt]
		G_{\varphi}^{\ \varphi} = G_{z}^{\ z}
		&= \frac{1}{2}\,\frac{d^{2}A}{d\rho^{2}} + \frac{1}{\rho}\,\frac{dA}{d\rho}\,.\label{Einstein_phi_uphi}
	\end{align}
	Hence, considering the symmetry of the Einstein tensor, we introduce two matter components: a Kiselev-like quintessence fluid and a matter distribution which is similar\footnote{In this work, we consider it simply as a matter distribution.} to a cloud of strings \cite{Letelier1979}, both satisfying the local conservation of energy-momentum. Following \cite{Ali2020}, we describe this Kiselev-like quintessence according to its energy-momentum tensor, whose components are given by
	\begin{equation}
		\begin{aligned}
			T_{t}^{\ t} &= T_{\rho}^{\ \rho} = \rho_{\scaleto{Q}{4pt}}(\rho)\,,\\
			T_{\varphi}^{\ \varphi} &= T_{z}^{\ z} = \alpha_{\scaleto{Q}{5pt}}\,\rho_{\scaleto{Q}{4pt}}(\rho)\,.
		\end{aligned}\label{Quintessence_EM_Tensor}
	\end{equation}
	The auxiliary state parameter $\alpha_{\scaleto{Q}{5pt}}$ is defined as
	\begin{equation}
		\alpha_{\scaleto{Q}{5pt}} = -\frac{1}{2}\left(3\, \omega_{\scaleto{Q}{5pt}} + 1\right)\,, \label{eq:Alpha_Q_Definition}
	\end{equation}
	where $\omega_{\scaleto{Q}{5pt}}$ is the standard state parameter satisfying $-1<\omega_{\scaleto{Q}{5pt}}<-1/3$. Moreover, $\alpha_{\scaleto{Q}{5pt}}$ is a constant (time-independent), and it takes values in the interval $0<\alpha_{\scaleto{Q}{5pt}}<1$. Therefore, we will discuss physical regimes determined solely by the value of $\alpha_{\scaleto{Q}{5pt}}$, without loss of generality. Interestingly, the relation between $\alpha_{\scaleto{Q}{5pt}}$ and $\omega_{\scaleto{Q}{5pt}}$ in this cylindrically symmetric spacetime is identical to the one with spherical symmetry \cite{Simao2025}. In both cases, when $\alpha_{\scaleto{Q}{5pt}} = 0$, $\omega_{\scaleto{Q}{5pt}} = -1/3$.
	
	The non-trivial components of this quintessential fluid are its energy density $\rho_{\scaleto{Q}{4pt}}(\rho)$ and its radial and tangent pressures, identified as $P_{\rho} = - T_{\rho}^{\ \rho}$, $P_{\varphi} = -T_{\varphi}^{\ \varphi}$ and $P_{z} = - T_{z}^{\ z}$, in accordance with our \emph{mostly-minus} sign convention for the metric tensor. Observe that this fluid is anisotropic, unless $\alpha_{\scaleto{Q}{5pt}}=1$ occurs (in the limiting case of $\omega_{\scaleto{Q}{5pt}} = -1$).
	
	Our additional matter component is described by:
	\begin{equation}
		\begin{aligned}
			T_{t}^{\ t} &= T_{\rho}^{\ \rho} = \frac{a}{\rho^{2}}\,,\\
			T_{\varphi}^{\ \varphi} &= T_{z}^{\ z} = 0\,,
		\end{aligned}\label{T-munu-Cloud-of-strings}
	\end{equation}
	where $a$ is a positive real constant associated with the cloud's intensity.
	
	After solving the Einstein equations
    \begin{equation}
       G_{\mu}^{\ \nu} + \Lambda\,\delta_{\mu}^{\ \nu} = \frac{8\pi G}{c^{4}}\,T_{\mu}^{\ \nu}\,,
    \end{equation}
     by combining the components of the Einstein tensor \eqref{Einstein_t_ut} and \eqref{Einstein_phi_uphi} with the energy-momentum tensor from quintessence and the cloud of strings, the solution for $A(\rho)$ reads	
	\begin{equation}
		A(\rho) = \overline{a} + \frac{\rho^{2}}{l^{2}} - \frac{\rho_{\scaleto{S}{4pt}}}{\rho} + N_{\scaleto{Q}{5pt}}\,\rho^{2\alpha_{\scaleto{Q}{3pt}}}\,.\label{Function_A(rho)}
	\end{equation}
	The cloud parameter $\overline{a} = (8\pi G/c^{4})\,a$, the Schwarzschild-like radius $\rho_{\scaleto{S}{4pt}}$, and the quintes\-sence parameter $N_{\scaleto{Q}{5pt}}$ are all positive real constants. In terms of the international system of units (SI), $\overline{a}$ is dimensionless (m$^{0}$), $\rho_{\scaleto{S}{4pt}}$ has units of length (in m), while $N_{\scaleto{Q}{5pt}}$ has units of m$^{-2\alpha_{\scaleto{Q}{3pt}}}$. 
	
\begin{figure}[ht!]
		\begin{center} 
			\includegraphics[width=0.65\textwidth]{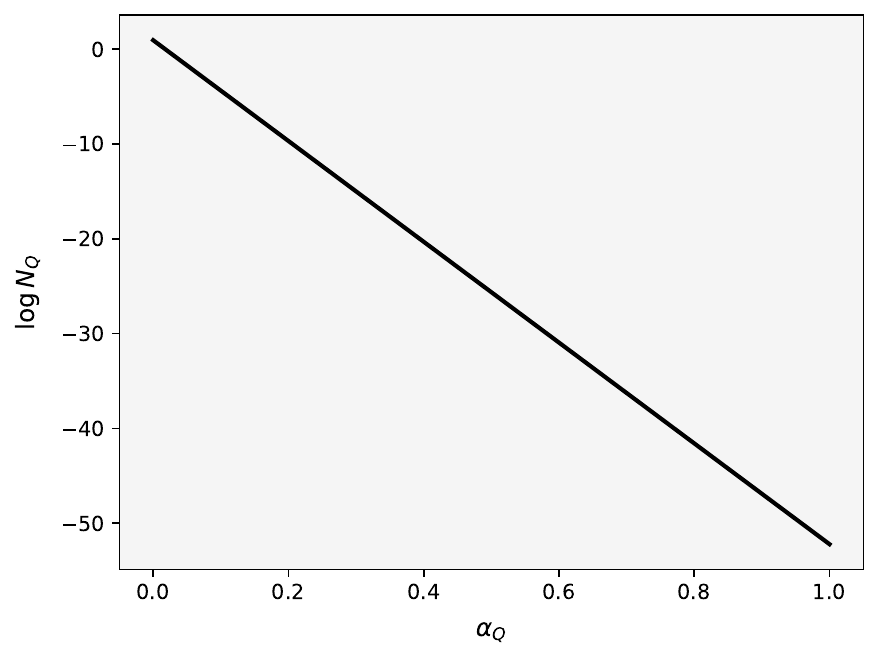}
			\caption{Logarithmic plot of the quintessence parameter $N_{\scaleto{Q}{5pt}}$ as a function of $\alpha_{\scaleto{Q}{5pt}}$. According to \cref{Estimate-N_Q}, there is an exponential decrease in $N_{\scaleto{Q}{5pt}}$ when the state parameter $\alpha_{\scaleto{Q}{5pt}}$ increases. The values on the vertical axis are calculated assuming an observable universe with a radius of $\rho_{\scaleto{obs.}{4pt}} = 3.8 \times 10^{26}$ m, ensuring that this cylindrical model has the same observable volume as our real Universe.}
			\label{N_Q}
		\end{center}
	\end{figure}

	In \cite{Deglmann2025}, we estimated typical values for $\overline{a}$ to be in the range $10^{-6} - 10^{1}$, while $\rho_{\scaleto{S}{4pt}}$ would likely range from $10^{-2}$ m to $10^{15}$ m. Regarding $N_{\scaleto{Q}{5pt}}$, we showed that it would depend on the amount of dark energy in the universe ($E_{\scaleto{DE}{4pt}}$), whose observable radius is $\rho_{\scaleto{obs.}{4pt}}$, and on the state parameter $\alpha_{\scaleto{Q}{5pt}}$. Under these considerations, we showed that $N_{\scaleto{Q}{5pt}}$ decreases exponentially with the increase in $\alpha_{\scaleto{Q}{5pt}}$, so that
	\begin{equation}
		N_{\scaleto{Q}{5pt}} = \frac{2\,G}{c^{4}}\,\frac{E_{\scaleto{DE}{4pt}}}{\rho_{obs.}^{2\alpha_{\scaleto{Q}{5pt}} + 1}}\,.\label{Estimate-N_Q}
	\end{equation}
	This behavior would imply the estimated values of $N_{\scaleto{Q}{5pt}}$ to be those of Figure \ref{N_Q} for a large universe with $\rho_{\scaleto{obs.}{4pt}} = 3.8 \times 10^{26}$ m and $E_{\scaleto{DE}{4pt}} = 2 \times 10^{71}$ kg m$^{2}$ s$^{-2}$. These two values were considered to make the theoretical cylindrical universe have equivalent observable volume and dark energy content as our real spherically symmetric Universe\footnote{For more details, please follow the complete construction in \cite{Deglmann2025}.}. 
    
\begin{figure}[ht!]
		\begin{center} 
			\includegraphics[width=0.65\textwidth, trim={1.5cm 1cm 0cm 1.9cm},clip]{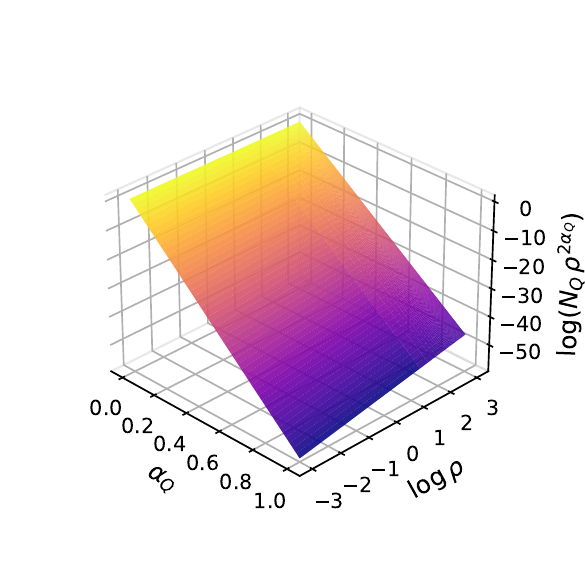}
			\caption{The contribution of the quintessence term $N_{\scaleto{Q}{5pt}}\,{\rho}^{2\alpha_{\scaleto{Q}{5pt}}}$, according to \cref{Estimate-N_Q}, where we use $\rho_{\scaleto{obs.}{4pt}} = l = 3.8 \times 10^{26}$ m. Its significance is more pronounced for small values of the state parameter $\alpha_{\scaleto{Q}{5pt}}$, particularly as $\alpha_{\scaleto{Q}{5pt}}$ approaches zero. On the other hand, as $\alpha_{\scaleto{Q}{5pt}}$ increases toward $\alpha_{\scaleto{Q}{5pt}} = 1$, the term becomes relevant only at very large radial distances, comparable to the radius of the observable universe, $\rho_{\scaleto{obs.}{4pt}}$.}
			\label{N_Q_contribution}
		\end{center}
	\end{figure}

    The quintessence contribution to the metric function $A(\rho)$ is illustrated in Figure \ref{N_Q_contribution}, showing that as $\alpha_{\scaleto{Q}{5pt}}$ increases (towards the upper bound $\alpha_{\scaleto{Q}{5pt}} = 1$), the quintessential term ($N_{\scaleto{Q}{5pt}}\,\rho^{2\alpha_{\scaleto{Q}{3pt}}}$) becomes non-negligible only at extremely large distances.
	For future convenience, we also provide the estimates regarding fixed values of $\alpha_{\scaleto{Q}{5pt}}$, namely $\alpha_{\scaleto{Q}{5pt}} = 0,\,1/2,\,1$, in Table \ref{table_N_Q_values}. These results will improve the understanding of Sections \ref{Extended_Solutions_Alpha_Q_0}, \ref{Extended_Solutions_Alpha_Q_One_Half}, and \ref{Extended_Solutions_Alpha_Q_1}. 
	\begin{table}[ht]
		\begin{center}
			\begin{tabular}{| c | c | c |}
				\hline
				$\alpha_{\scaleto{Q}{5pt}}$ & $\omega_{\scaleto{Q}{5pt}}$ & $N_{\scaleto{Q}{5pt}}$\\
				\hline \hline
				$0$ & $-1/3$ & $8.95$ m$^{0}$\\
				$1/2$ & $-2/3$ & $10^{-26}$ m$^{-1}$\\
				$1$ & $-1$ & $10^{-52}$ m$^{-2}$\\
				\hline
			\end{tabular}
		\end{center}
		\caption{Estimates for $N_{\scaleto{Q}{5pt}}$ according to \cref{Estimate-N_Q}, considering that $\rho_{\scaleto{obs.}{4pt}} = \lvert l \rvert = 3.8 \times 10^{26}$ m and $E_{\scaleto{DE}{4pt}} = 2 \times 10^{71}$ kg m$^{2}$ s$^{-2}$.}
		\label{table_N_Q_values}
	\end{table}
	
	Finally, we recall that the quintessence energy density ($\rho_{\scaleto{Q}{4pt}}$) is determined by 
	\begin{equation*}
		\rho_{\scaleto{Q}{4pt}} = \left(2\, \alpha_{\scaleto{Q}{5pt}} + 1\right)\frac{N_{\scaleto{Q}{5pt}}}{8\pi G/c^{4}}\,\rho^{2\, \alpha_{\scaleto{Q}{3pt}}-2}\,.
	\end{equation*}
	It is relevant to emphasize that, when $\alpha_{\scaleto{Q}{5pt}} = 0$, the quintessence fluid behaves exactly as the cloud of strings \eqref{T-munu-Cloud-of-strings} and this equivalence will be evident in our extended solutions for the regime where $\alpha_{\scaleto{Q}{5pt}}=0$. The associated radial pressure ($P_{\rho}$) is simply $P_{\rho} = -\rho_{\scaleto{Q}{4pt}}$, as determined by \cref{Quintessence_EM_Tensor}. Similarly, the tangent and longitudinal pressures ($P_{\varphi}$ and $P_{z}$, respectively) also follow from \cref{Quintessence_EM_Tensor} and are given by $P_{\varphi} = P_{z} = - \alpha_{\scaleto{Q}{5pt}}\,\rho_{\scaleto{Q}{4pt}}(\rho)$, where $\rho_{\scaleto{Q}{4pt}}$ is given above. Notably, they are indeed negative pressures.

	\subsection{Event horizon}
	
	The BCK spacetime metric, determined by \cref{metric,Function_A(rho)}, admits a nontrivial event horizon $\rho_{\scaleto{+}{4pt}}$ if $\rho_{\scaleto{S}{4pt}} \neq 0$. This holds unless we set $\Lambda = \overline{a} = N_{\scaleto{Q}{5pt}} = 0$. We have obtained the exact analytical solutions to $\rho_{\scaleto{+}{4pt}}$, associated with $\alpha_{\scaleto{Q}{5pt}} = 0,\,1/2,\,1$, in \cite{Deglmann2025}. Moreover, we had also demonstrated that good\footnote{As will be discussed further in Section \ref{Sec-3-Extended-Solutions}, these are good approximations provided $\rho_{\scaleto{+}{4pt}}$ is several orders of magnitude smaller than the AdS radius $l$. Since we do not look for cosmological solutions, this condition is straightforwardly satisfied. For more details, please check Section 5 in \cite{Deglmann2025}.} approximations of these solutions are the ones in Tables \ref{table-EH-cases} (for standard scenarios, where all the parameters are non-trivial) and \ref{table-cloudless-cases} (for cloudless scenarios, where $\overline{a}=0$), which are valid under physically plausible constraints.
	\begin{table}[ht]
		\begin{center}
			\begin{tabular}{ | c | c | c |}
				\hline
				$\alpha_{\scaleto{Q}{5pt}}$ & $\rho_{\scaleto{+}{4pt}}$ & Constraints \\
				\hline\hline
				$0$ & $\rho_{\scaleto{S}{4pt}}/(\overline{a} + N_{\scaleto{Q}{5pt}})$ & $\rho_{\scaleto{+}{4pt}}^{2}/l^{2} \ll 1$\\
				
				$1/2$ & $\rho_{\scaleto{S}{4pt}}/\overline{a}$ & $\rho_{\scaleto{+}{4pt}}^{2}/l^{2} \ll 1$ and  $N_{\scaleto{Q}{5pt}}\rho_{\scaleto{S}{4pt}}/\overline{a}^{2} \ll 1$\\
				
				$1$ & $\rho_{\scaleto{S}{4pt}}/\overline{a}$ &  $N_{\scaleto{QL}{5pt}} \rho_{\scaleto{S}{4pt}}^{2}/\overline{a}^{3} \ll 1$\\
				\hline
			\end{tabular}
		\end{center}
		\caption{Results for $\rho_{\scaleto{+}{4pt}}$ in the standard scenario (where $\overline{a}\neq 0$, $\rho_{\scaleto{S}{4pt}}\neq 0$, and $N_{\scaleto{Q}{5pt}} \neq 0$), according to fixed values of the state parameter $\alpha_{\scaleto{Q}{5pt}}$. In the third row, the dark energy parameter $N_{\scaleto{QL}{5pt}}$ is defined as the sum of $N_{\scaleto{Q}{5pt}}$ and $1/l^{2}$.}
		\label{table-EH-cases}
	\end{table}
	\begin{table}[ht]
		\begin{center}
			\begin{tabular}{ | c | c | c |}
				\hline
				$\alpha_{\scaleto{Q}{5pt}}$ & $\rho_{\scaleto{+}{4pt}}$ & Constraints \\
				\hline\hline
				$0$ & $\rho_{\scaleto{S}{4pt}}/N_{\scaleto{Q}{5pt}}$ & $\rho_{\scaleto{+}{4pt}}^{2}/l^{2} \ll 1$\\
				
				$1/2$ & $\left(\rho_{\scaleto{S}{4pt}}/N_{\scaleto{Q}{5pt}}\right)^{1/2}$ & $\rho_{\scaleto{+}{4pt}}^{2}/l^{2} \ll 1$\\
				
				$1$ & $\left(\rho_{\scaleto{S}{4pt}}/N_{\scaleto{QL}{5pt}}\right)^{1/3}$ &  None\\
				\hline
			\end{tabular}
		\end{center}
		\caption{Determination of $\rho_{\scaleto{+}{4pt}}$ in a scenario without the cloud of strings (\emph{cloudless}), defined by $\overline{a}=0$, with $\rho_{\scaleto{S}{4pt}}\neq 0$ and $N_{\scaleto{Q}{5pt}}\neq 0$, for varying values of $\alpha_{\scaleto{Q}{5pt}}$. Note that the third row result is exact, while $N_{\scaleto{QL}{5pt}}$ is determined by $N_{\scaleto{QL}{5pt}} = N_{\scaleto{Q}{5pt}} + 1/l^{2}$.}
		\label{table-cloudless-cases}
	\end{table}
	
	Our final scenarios are as follows. The first is the absence of quintessence, characterized by $N_{\scaleto{Q}{5pt}}=0$, while $\overline{a}\neq 0$ and $\rho_{\scaleto{S}{4pt}}\neq 0$. The second scenario removes both the quintessential fluid and the cloud of strings, so that $\overline{a} = 0$ and $N_{\scaleto{Q}{5pt}} = 0$, with $\rho_{\scaleto{S}{4pt}}\neq 0$. It should be noted that these two scenarios are special cases of those listed in \cref{table-EH-cases} (first row) and \cref{table-cloudless-cases} (third row), respectively, which we obtain by setting $N_{\scaleto{Q}{5pt}} = 0$.
	
	After reviewing the essential properties of the BCK spacetime, in the next subsection, we address the explicit form of the Klein-Gordon equation in this background.

	\subsection{The Klein-Gordon equation}

    To study scalar bosons within the BCK spacetime, defined by \cref{metric,Function_A(rho)}, we investigate the Klein-Gordon equation, whose general form is
	\begin{equation*}
		\frac{1}{\sqrt{-g}}\,\partial_{\mu}\left(g^{\mu\nu}\sqrt{-g}\,\partial_{\nu}\,\phi\right) + \frac{m_{\phi}^{2}\, c^{2}}{\hbar^{2}}\,\phi = 0\,,\label{Klein-Gordon-equation}
	\end{equation*}
	where $g=\text{det}(g_{\mu\nu})$, while $m_{\phi}$ represents the mass of the spin$-0$ particle. In our specific case, $g_{\mu\nu} = \text{diag}\left(c^{2} A(\rho),\, -A(\rho)^{-1},\, -\rho^{2},\, -\rho^{2}/l^{2}\right)$ and one can show that the associated partial differential equation (PDE) can be reduced to an ordinary differential equation (ODE) by employing the following separation of variables:
	\begin{equation}
		\phi = e^{-i E t/\hbar} e^{i n_{\scaleto{R}{3pt}}\varphi} e^{iz k_{z}/l} \,R(\rho)\,,\label{ansatz-campo-escalar}
	\end{equation}
	where $E$ has the dimension of energy, $n_{\scaleto{R}{4pt}} \in \mathbb{Z}$, while $k_{z} \in \mathbb{R}$ is related to the particle's momentum along the $z-$axis. 
	
	Moreover, it follows from \cref{ansatz-campo-escalar} that the radial equation for $R(\rho)$ is given by
	\begin{equation}
		\partial_{\rho}^{2}\,R(\rho) + \chi(\rho)\,\partial_{\rho}\,R(\rho) + \tau(\rho)\,R(\rho) = 0\,,\label{general-equation-for-R}
	\end{equation}
	with
	\begin{align*}
		\chi(\rho) 
		&\coloneqq \frac{1}{A(\rho)}\,\partial_{\rho}A + \frac{2}{\rho}\,,\\
		\tau(\rho) 
		&\coloneqq \frac{1}{A(\rho)}\left(\frac{\epsilon^{2}}{A(\rho)} - \frac{\kappa^{2}}{\rho^{2}} - \overline{m}_{\phi}^{2}\right)\,.
	\end{align*}
	The parameters $\epsilon$, $\kappa$, and $\overline{m}_{\phi}$ occurring in $\tau(\rho)$ are defined by
	\begin{subequations}
		\begin{align}
			&\epsilon = \frac{E}{\hbar c}\,,\label{def-epsilon}\\
			&\kappa^{2} = n_{\scaleto{R}{4pt}}^{2} + k_{z}^{2}\,,\label{def-kappa}\\
			&\overline{m}_{\phi} = \frac{m_{\phi}\, c}{\hbar}\,.\label{def-m_bar}
		\end{align}
	\end{subequations}
	A good strategy to solve \cref{general-equation-for-R} is to rewrite it in the Liouville normal form \eqref{General_Equation_Liouville_Normal_Form}, where the problem becomes that of determining
	\begin{equation}
		R(\rho) = \frac{R_{0}}{\rho\,\sqrt{A(\rho)}}\,u(\rho)\,, \label{radial-wave-function}
	\end{equation}
	where $R_{0}$ is a normalization constant, while the auxiliary function $u(\rho)$ satisfies 
	\begin{equation}
		u''(\rho) + V_{\text{eff}}(\rho)\,u(\rho) = 0\,.\label{eq-geral-u}
	\end{equation}
	The effective potential $V_{\text{eff}}(\rho)$ is determined by 
	\begin{equation}
		V_{\text{eff}}(\rho) = \frac{A'(\rho)^{2} + 4\epsilon^{2}}{4A(\rho)^{2}} - \frac{1}{A(\rho)}\left[\frac{1}{2}A''(\rho) + \frac{A'(\rho)}{\rho} + \frac{\kappa^{2}}{\rho^{2}} + \overline{m}_{\phi}^{2}\right]\,,\label{Effective-Potential}
	\end{equation}
	where $\epsilon$, $\kappa$, and $\overline{m}_{\phi}$ were conveniently defined by \cref{def-epsilon,def-kappa,def-m_bar}.
		
	When we assume the existence of a nontrivial event horizon, with radius $\rho_{\scaleto{+}{4pt}}$, it is convenient to define a new dimensionless coordinate $x$, so that
	\begin{equation}
		x = \frac{\rho}{\rho_{\scaleto{+}{4pt}}}\,,\label{x-coordinate}
	\end{equation}
	with $x \in \left(0,\,+\infty\right)$. Due to the above definition of $x$, we comprehend that physical solutions containing an event horizon will require $x > 1$ since we want to describe observable particles. In terms of $x$, \cref{eq-geral-u,Effective-Potential} becomes
	\begin{equation}
		u''(x) + \rho_{\scaleto{+}{4pt}}^{2}\,V_{\text{eff}}(x)\,u(x) = 0\,,\label{Eq-para-u}
	\end{equation}
	with
	\begin{equation}
		\rho_{\scaleto{+}{4pt}}^{2}\,V_{\text{eff}}(x) = \frac{A'(x)^{2} + 4\left(\rho_{\scaleto{+}{4pt}}\epsilon\right)^{2}}{4\,A(x)^{2}} 
		- \frac{1}{A(x)}\left[\frac{1}{2}A''(x) + \frac{A'(x)}{x} + \frac{\kappa^{2}}{x^{2}} + \rho_{\scaleto{+}{4pt}}^{2}\,\overline{m}_{\phi}^{2}\right]\,.\label{Potential_in_terms_of_x}
	\end{equation}
	It is worth stressing that the derivatives in \cref{Eq-para-u,Potential_in_terms_of_x} are with respect to $x$, i.e., $d/d x$. 

    The subsequent sections examine the radial Klein-Gordon equation in the BCK spacetime across three distinct regimes of the state parameter $\alpha_{\scaleto{Q}{5pt}}$. 
    In several of these cases, the auxiliary radial equation \eqref{eq-geral-u} (or, equivalently, \eqref{Eq-para-u}) can be reduced to the normal form of either the Confluent or the Biconfluent Heun differential equations (CHE and BHE, respectively). 
    Since Heun equations are not widely known, we have prepared Appendix \ref{appendix} with the main results and properties of these equations and their solutions. Due to their importance, we also anticipate our conventions, where  $\text{HeunC}\left(\alpha,\,\pm\beta,\,\gamma,\,\delta,\,\eta;\,z\right)$ denote the two linearly independent (LI) solutions of the CHE, while $\text{HeunB}\left(\pm\alpha,\,\beta,\,\gamma,\,\delta;\,z\right)$ are LI solutions to the BHE. These solutions are usually called the Confluent and Biconfluent Heun functions, respectively\footnote{Technically, $\text{HeunC}\left(\alpha,\,\beta,\,\gamma,\,\delta,\,\eta;\,z\right)$ should be interpreted as the analytic continuation of the local solution $\text{HeunC}\ell\left(\alpha,\,\beta,\,\gamma,\,\delta,\,\eta;\,z\right)$, as we clarify in \cref{appendix_CHE}. For simplicity, we refer to it as the Confluent Heun function.}, and they depend on a single complex variable, $z$, but are characterized by a set of Heun parameters that can be identified with the physical parameters of the system.
    
    We further note that Appendix \ref{appendix} presents a detailed derivation of the Fuchs-Frobenius solutions for both the CHE and BHE. Since these results are used extensively throughout this work, the reader is strongly encouraged to consult the derivations provided there.
    
	In the next section, we introduce the first Klein-Gordon solution, valid for scalar particles near the event horizon in the BCK spacetime, generally encompassing all values of $\alpha_{\scaleto{Q}{5pt}}$. This result was originally demonstrated in \cite{Deglmann2025}, but we present it here along with new results regarding the regime of $\alpha_{\scaleto{Q}{5pt}}=1$.

	\subsection{Solutions near the event horizon} \label{RWF_NH_All_Alpha_Q_Near}
	
	In this section, we solve the auxiliary equation for $u(x)$, defined by eqs. \eqref{Eq-para-u} and \eqref{Potential_in_terms_of_x}, in the case where the radial coordinate $\rho$ is in the neighborhood of $\rho_{\scaleto{+}{4pt}}$. In terms of the dimensionless coordinate $x=\rho/\rho_{\scaleto{+}{4pt}}$, this condition must be interpreted as the limit where $x\to 1^{+}$. With this result, we determine the radial solution $R(x)$ of the KG equation using the \cref{radial-wave-function}, with which we investigate the effects of the quintessence fluid on scalar particles. Again, it is worth noting that, \textit{a priori}, we could solve the KG equation including the region of $0<x\leq 1$, but, as we mentioned before, this solution would not be related to observable particles. For this reason, it will be disregarded whenever an event horizon exists.
	
	In describing the neighborhood of the event horizon, it is reasonable to approximate $A(\rho)$, given in \eqref{Function_A(rho)}, to the first order in $(\rho-\rho_{\scaleto{+}{4pt}})$ so that
	\begin{equation*}
		A(\rho) = 
		\left.\frac{dA}{d \rho}\right|_{\rho_{\scaleto{+}{4pt}}}\left(\rho - \rho_{\scaleto{+}{4pt}}\right) 
		+ \mathcal{O}\left[(\rho-\rho_{\scaleto{+}{4pt}})^{2}\right]\,.
	\end{equation*}
	This first order approximation leads to
	\begin{equation}
		A(x) = \beta_{\scaleto{+}{4pt}}\left(x - 1\right)\,,\label{A(x)_RWF_NH_All_Alpha_Q}
	\end{equation}
	where
	\begin{equation}
		\beta_{\scaleto{+}{4pt}} = \frac{2 \rho_{\scaleto{+}{4pt}}^{2}}{l^{2}} + \frac{\rho_{\scaleto{S}{4pt}}}{\rho_{\scaleto{+}{4pt}}} + 2\alpha_{\scaleto{Q}{5pt}}\,N_{\scaleto{Q}{5pt}}\,\rho_{\scaleto{+}{4pt}}^{2\alpha_{\scaleto{Q}{5pt}}}\,,\label{def-Beta-Plus}	
	\end{equation}
	implying $\beta_{\scaleto{+}{4pt}}>0$. Observe that we required $\rho_{\scaleto{S}{4pt}} \neq 0$ to promote the necessary condition where the event horizon exists. 
	
	After substituting the expression of \cref{A(x)_RWF_NH_All_Alpha_Q} into \cref{Eq-para-u,Potential_in_terms_of_x} to determine the solution of $u(x)$, we use the relation \eqref{radial-wave-function} to show that
	\begin{equation}
		\begin{aligned}
			R(x) 
			&= x^{(\beta-1)/2}\,\left(x-1\right)^{\gamma/2}\,
			\left[ c_{1}\,
			\text{HeunC}\left(-\alpha,\,\gamma,\,\beta,\,-\delta,\,\delta+\eta;\,1-x\right)\right.\\
			&\left.+\,c_{2} \, \left(x-1\right)^{-\gamma}\,
			\text{HeunC}\left(-\alpha,\,-\gamma,\,\beta,\,-\delta,\,\delta+\eta;\,1-x\right)
			\right]\,,
		\end{aligned}\label{RWF_NH_All_Alpha_Q}
	\end{equation}
	where HeunC$\left(\alpha,\,\beta,\,\gamma,\,\delta,\,\eta;\,x\right)$ must be interpreted as the analytic continuation of the Fuchs-Frobenius solution to the Confluent Heun Equation (CHE), as determined in Appendix \ref{appendix_CHE}. 
    
    The Heun parameters $\alpha$, $\beta$, $\gamma$, $\delta$, and $\eta$ are given by
	\begin{equation}
		\begin{aligned}
			\alpha &= 0\,,\\
			\beta &= \sqrt{1 - \frac{4\,\kappa^{2}}{\beta_{\scaleto{+}{4pt}}}}\,,\\
			\gamma &= i\,\frac{2 \epsilon \rho_{\scaleto{+}{4pt}}}{\beta_{\scaleto{+}{4pt}}}\,,\\
			\delta &= -\frac{\rho_{\scaleto{+}{4pt}}^{2}\,\overline{m}_{\phi}^{2}}{\beta_{\scaleto{+}{4pt}}}\,,\\
			\eta &= -\frac{1}{2} - \frac{\kappa^{2}}{\beta_{\scaleto{+}{4pt}}}\,.
		\end{aligned} \label{Heun_Parameters_RWF_NH_All_Alpha_Q}
	\end{equation}
	
	Now, considering that our approximation of $A(x)$, given by \cref{A(x)_RWF_NH_All_Alpha_Q,def-Beta-Plus}, is valid in the domain of $x \to 1^{+}$, we  follow the general results of Appendix \ref{appendix_CHE} to establish that, in this limit
	$\text{HeunC}\left(-\alpha,\,\pm\gamma,\,\beta,\,-\delta,\,\delta+\eta;\,1-x\to 0^{+}\right) \approx 1$,
	which simplifies \cref{RWF_NH_All_Alpha_Q} considerably. Therefore, by analyzing $\lim_{x\to 1} R(x)$, we conclude that
	\begin{equation}
		R_{\scaleto{NH}{4pt}}(x) = x^{(\beta-1)/2}\left[c_{1} \left(x-1\right)^{\gamma/2}
		+ c_{2} \left(x-1\right)^{-\gamma/2}\right]\,,\label{RWF_NH_Parte_1}
	\end{equation}
	where the constants $c_{1}$ and $c_{2}$ are arbitrary due to the absence of initial or boundary conditions, while the factor $\left(x-1\right)^{\pm\gamma/2}$ acts as a relative phase in $R_{\scaleto{NH}{4pt}}(x)$. Furthermore, the label \textsc{nh} was added to denote that the solution is valid only\footnote{Although this result is only valid within a small range of the coordinate $x=\rho/\rho_{\scaleto{+}{4pt}}$, it is suitable for the radial distance $\rho$. As an example, consider that $\rho_{\scaleto{+}{4pt}} = 10^{10}$ m, while $x$ varies from $1+10^{-3}$ to $1+10^{-2}$. In this case, the near-horizon radial solution is adequate for a distance of up to $9\times 10^{7}$ m.} in the event horizon's vicinity.
	
	We can further rewrite this radial solution in a more convenient way, given by:
	\begin{equation}
		\begin{aligned}
			R_{\scaleto{NH}{4pt}}(x)
			&= c_{1}\,\exp\left[\frac{\beta}{2}\left(x-1\right) + \frac{\gamma}{2} \ln \left(x-1\right)\right]\\ 
			&+ c_{2}\,\exp\left[\frac{\beta}{2}\left(x-1\right) - \frac{\gamma}{2} \ln \left(x-1\right)\right]\,.
		\end{aligned}\label{RWF_NH}
	\end{equation}
	To obtain this result, we considered that $\ln(x) \approx (x-1)$ and $x^{-1/2}\approx 1$ in the region of interest, since the term $x^{-1/2}$ has a negligible overall contribution in $R_{\scaleto{NH}{4pt}}(x)$. Let us also emphasize that, unless $|\beta| \gg |\gamma|$, the dominant contribution in $R_{\scaleto{NH}{4pt}}(x)$ comes from
	\begin{equation}
		R_{\scaleto{NH}{4pt}}(x) \approx c_{1}\,\left(x-1\right)^{\gamma/2} + c_{2}\,\left(x-1\right)^{-\gamma/2}\,.\label{Approx_R_RWF_NH_All_Alpha_Q}
	\end{equation}
	Thus, considering the expression for $\gamma$ shown in \eqref{Heun_Parameters_RWF_NH_All_Alpha_Q}, the real part of $\left(x-1\right)^{\pm \gamma/2}$ is simply $\cos\left[\frac{\epsilon \, \rho_{\scaleto{+}{4pt}}}{\beta_{\scaleto{+}{4pt}}}\ln(x-1)\right]$, showing that the wave function oscillates more rapidly near the event horizon. The same happens to the imaginary part $\sin\left[\frac{\epsilon \, \rho_{\scaleto{+}{4pt}}}{\beta_{\scaleto{+}{4pt}}}\ln(x-1)\right]$, which only differs by a phase of $\pi/2$.
	
	The result of \cref{RWF_NH} is essential for demonstrating that the corresponding radial solutions, for $\alpha_{\scaleto{Q}{5pt}} = 0,\,1/2,\,1$, can be expressed as follows:
	\begin{equation}
		R_{\scaleto{NH}{4pt}}^{\scaleto{\,(\sqcup)}{6pt}}(x) = \Phi_{+}\, \exp\left[-i\,\delta_{+}^{\scaleto{(\sqcup)}{6pt}}\right] + \Phi_{-}\, \exp\left[-i\,\delta_{-}^{\scaleto{(\sqcup)}{6pt}}\right]\, ,\label{RWF_NH_DPs}
	\end{equation}
    and we must remark that the two terms originate from two eigenfunctions of the equation. The functions $\Phi_{\pm}(x)$ are independent of the quintessence parameter, although they may depend on other parameters of the BCK spacetime. In contrast, $\delta_{\pm}^{\scaleto{(\sqcup)}{6pt}}(x)$ are functions that encode the influence of the quintessence fluid through the parameter $N_{\scaleto{Q}{5pt}}$. Since these relative phases are induced by quintessence, we call them \enquote{dark phases}. The symbol \enquote{$\sqcup$} serves as a placeholder to indicate the specific value of $\alpha_{\scaleto{Q}{5pt}}$ associated with the solution, while the corresponding physical scenario is indicated by the subscript.
	
\begin{figure}[ht!]
	\centering
	\begin{minipage}{0.48\linewidth}
		\centering
		\subcaptionbox{}
		{\includegraphics[width=\linewidth, trim={0 {15pt} 0 {15pt}},clip]{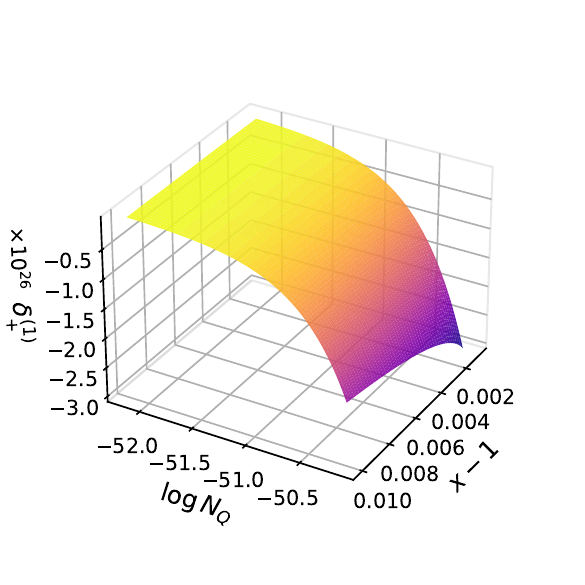}}
	\end{minipage}
	\quad
	\begin{minipage}{0.48\linewidth}
		\centering
		\subcaptionbox{}    
		{\includegraphics[width=\linewidth, trim={0 {15pt} 0 {15pt}},clip]{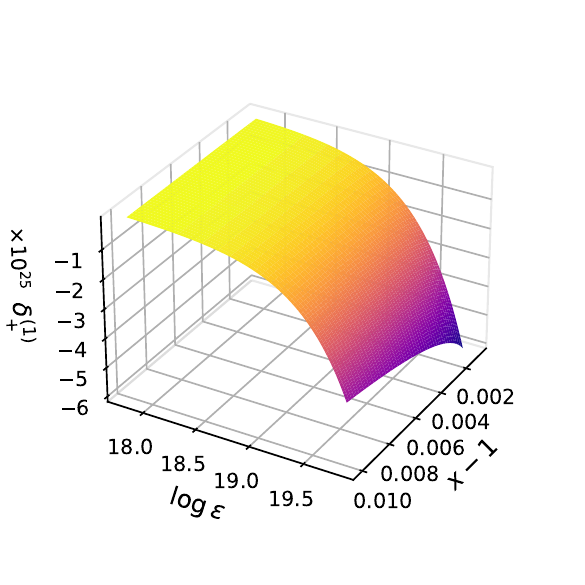}}
	\end{minipage}
	
	\begin{minipage}{0.48\linewidth}
		\centering
		\subcaptionbox{}
		{\includegraphics[width=\linewidth, trim={0 {15pt} 0 {15pt}},clip]{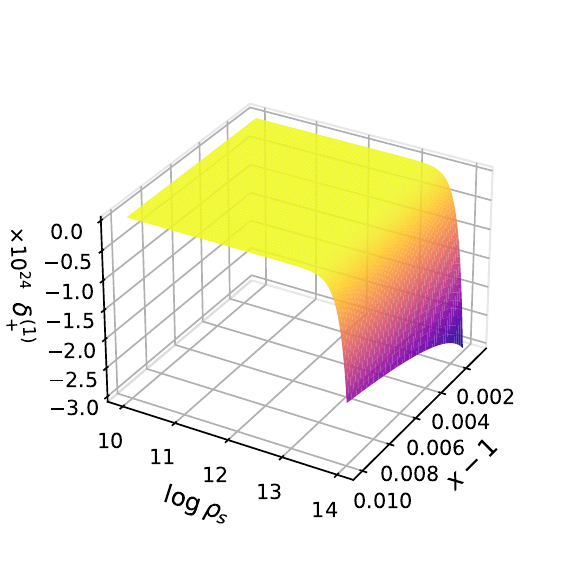}}
	\end{minipage}  
	\quad
	\begin{minipage}{0.48\linewidth}
		\centering
		\subcaptionbox{}
		{\includegraphics[width=\linewidth, trim={0 {15pt} 0 {15pt}},clip]{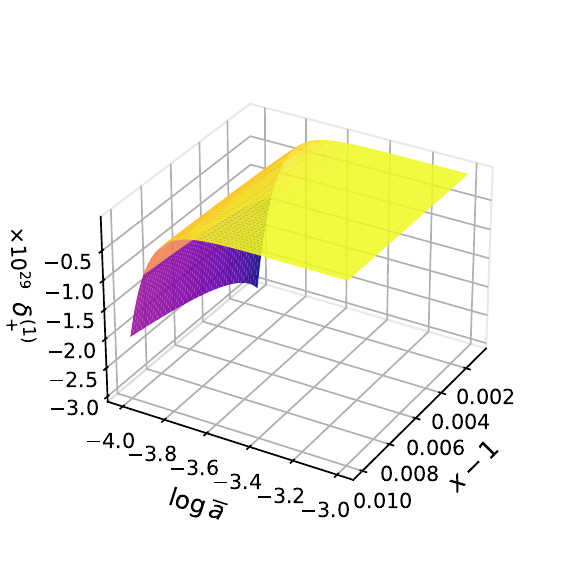}}
	\end{minipage}
	\caption{Plots of the dark phase $\delta_{+}^{\scaleto{(1)}{6pt}}(x)$, according to \cref{eq-DP_1_NH}. All fixed parameters have the following values: $\kappa = 4$, $\overline{m}_{\phi} = 6.37 \times 10^{17}$ m$^{-1}$ (where we use the Higgs boson mass), $\epsilon = 5\,\overline{m}_{\phi}$, $N_{\scaleto{QL}{4pt}} = 6.9 \times 10^{-53}$ m$^{-2}$ (according to \cref{Estimate-N_Q,N_QL_Definition}), $\rho_{\scaleto{S}{4pt}} = 10^{14}$ m, and $\overline{a} = 10^{-3}$.}
	\label{DP_1_NH}
\end{figure}

	As an example, we provide the explicit case of a standard scenario with $\alpha_{\scaleto{Q}{5pt}}=1$. In this case, the radial solution near the event horizon becomes
	\begin{equation*}
		R_{\scaleto{ST+NH}{4pt}}^{\scaleto{\,(1)}{6pt}}(x) = \Phi_{+}\, \exp\left[-i\,\delta_{+}^{\scaleto{(1)}{6pt}}\right] + \Phi_{-}\, \exp\left[-i\,\delta_{-}^{\scaleto{(1)}{6pt}}\right]\,,\label{RWF_1_NH}
	\end{equation*}
	where $\Phi_{\pm}(x)$ is given by
	\begin{equation*}
		\Phi_{\pm}(x) = c_{\pm}\,\exp\left[\pm\,i\,\frac{\epsilon\,\rho_{\scaleto{S}{4pt}}}{\overline{a}^{2}}\,\ln \left(x-1\right) + i\,\sqrt{\frac{\kappa^{2}}{\overline{a}} - \frac{1}{4}} \,\left(x-1\right)\,\right]\,,\label{Phi_pm}
	\end{equation*}
	while
	\begin{equation}
		\delta_{\pm}^{\scaleto{(1)}{6pt}}(x) = \left[\pm \frac{2\,\epsilon \rho_{\scaleto{S}{4pt}}}{\overline{a}^{2}}\, \ln \left(x-1\right) + \frac{\kappa^{2}/\overline{a}}{\sqrt{\kappa^{2}/\overline{a} - 1/4}}\, \left(x-1\right)\right]\frac{N_{\scaleto{Q}{5pt}}\,\rho_{\scaleto{S}{4pt}}^{2}}{\overline{a}^{3}}
		\label{eq-DP_1_NH}
	\end{equation}
	are the associated dark phases. This result is valid in the physical regime where $N_{\scaleto{QL}{5pt}} \rho_{\scaleto{S}{4pt}}^{2}/\overline{a}^{3} \ll 1$, according to Table \ref{table-EH-cases} (third row). To illustrate the behavior of these dark phases, we provide a numerical example of $\delta_{+}^{\scaleto{(1)}{6pt}}(x)$ in Figure \ref{DP_1_NH}.
	
	It is important to emphasize that, although the general solution of \cref{RWF_NH} is significant, its validity is restricted to the vicinity of the event horizon ($|x-1| \ll 1$)\footnote{The neighborhood is small with respect to the variable $x$, since we formally assume that $x\to 1^{+}$. Yet, the transformation $\rho = x\,\rho_{\scaleto{+}{4pt}}$ implies that, in terms of the radial coordinate $\rho$, these distances are considerably larger.}. While the occurrence of dark phases has been demonstrated for this class of solutions, it remains uncertain whether this phenomenon persists in solutions defined over a broader radial domain. The following sections address this question by extending the radial domain to values up to several times the radius of the event horizon. Additionally, since the auxiliary state parameter $\alpha_{\scaleto{Q}{5pt}}$ is continuous and the extended solutions depend on it, the analysis will focus on the radial KG solutions for three specific values: $\alpha_{\scaleto{Q}{5pt}} = 0,\,1/2,\,1$.

    \pagebreak

	\section{Extended solutions}\label{Sec-3-Extended-Solutions}
	
	In this section, we investigate radial solutions of the Klein-Gordon equation for an extended radial domain, in which the radial distance $\rho$ can be up to tens of times greater than the radius of the event horizon. These extended solutions need to be solved on a case-by-case basis, thus resulting in a large number of solutions. For each value of $\alpha_{\scaleto{Q}{5pt}}$, we must solve the respective radial KG equation in three possible scenarios, namely: the standard scenario, the cloudless scenario, and the one where the black string is removed (referred to as the horizonless scenario). Therefore, before starting the analysis, it is necessary to clarify the assumptions that allow us to obtain these solutions.
	
	For the cases where $\alpha_{\scaleto{Q}{5pt}} = 0$ or $1/2$, we consider a single assumption: that both $\rho$ and $\rho_{\scaleto{+}{4pt}}$ are significantly smaller than the AdS radius $l$, with $\rho > \rho_{\scaleto{+}{4pt}}$. Thus, the condition $\rho_{\scaleto{+}{4pt}}^{2}/l^{2} < \rho^{2}/l^{2} \ll 1$ is satisfied, allowing us to neglect the term $\rho^{2}/l^{2}$ in $A(\rho)$. This assumption means that we are considering particles in a large extended domain, but not on a cosmological scale. Under this condition, the function $A(\rho)$, originally given by \cref{Function_A(rho)}, becomes:
	\begin{equation}
		A(\rho) = \overline{a} - \frac{\rho_{\scaleto{S}{4pt}}}{\rho} + N_{\scaleto{Q}{5pt}}\,\rho^{2\alpha_{\scaleto{Q}{5pt}}}\,.\label{A_Assumption}
	\end{equation}
    Note that these approximations improve as the value of $\rho^{2}/l^{2}$ decreases, and since we are not investigating cosmological solutions, this condition is easily met over a large radial domain, where $\rho^{2}/l^{2}\ll 1$. For example, if we consider $l\sim 10^{26}$ m, following our previous considerations from \cite{Deglmann2025}, $\rho$ could reach values up to $10^{20}$ m (or even greater) without approaching a regime where the quotient $\rho^{2}/l^{2}$ is not negligible. However, it is important to recall that the quintessence parameter $N_{\scaleto{Q}{5pt}}$ decreases exponentially as $\alpha_{\scaleto{Q}{5pt}}$ increases \cite{Deglmann2025}. Consequently, in the limiting case of $\alpha_{\scaleto{Q}{5pt}}=1$, the above consideration does not hold, and we must therefore take all the terms in $A(\rho)$ into account.
    
    Therefore, when $\alpha_{\scaleto{Q}{5pt}}=1$ we investigate the combined contribution of $N_{\scaleto{QL}{5pt}} = N_{\scaleto{Q}{5pt}} + 1/l^{2}$, which does not constrain the radial domain. These solutions will be detailed in \cref{Extended_Solutions_Alpha_Q_1}.
	
	For clarity, we shall label these solutions with a descriptive subscript from Table \ref{naming-solutions} so that, given the label, we can easily identify the conditions leading to it.
	\begin{table}[ht]
		\begin{center}
			\begin{tabular}{ | c | c | c |}
				\hline
				Label & Scenario & Conditions on $\overline{a}$, $\rho_{\scaleto{S}{4pt}}$, and $N_{\scaleto{Q}{5pt}}$\\
				\hline\hline
				\textsc{st} & Standard & All parameters are non-zero\\
				
				\textsc{cl} & Cloudless & $\overline{a} = 0$, $\rho_{\scaleto{S}{4pt}}\neq 0$\\
				
				\textsc{hl} & Horizonless & $\rho_{\scaleto{S}{4pt}} = 0$\\
				
				\textsc{qf} & Free from quintessence & $N_{\scaleto{Q}{5pt}} = 0$\\
				\hline
			\end{tabular}
		\end{center}
		\caption{Labels for the extended solutions. In the fourth row, \textsc{qf} was chosen to denote the \enquote{quintessence-free} scenario.}
		\label{naming-solutions}
	\end{table}
	To clarify the value of the state parameter $\alpha_{\scaleto{Q}{5pt}}$ for each solution, we use a superscript notation. For example, $R_{\scaleto{ST}{4pt}}^{\scaleto{(1/2)}{6pt}}(x)$ represents the radial solution for the standard case (where $\overline{a}\neq 0$, $\rho_{\scaleto{S}{4pt}}\neq 0$, and $N_{\scaleto{Q}{5pt}}\neq 0$) for $\alpha_{\scaleto{Q}{5pt}} = 1/2$. We also observe that, in principle, it is possible to have the combination of certain scenarios, such as the horizonless case together with the one which is free from quintessence (\textsc{hl+qf}).

    As the reader can see, the necessity to solve each case of $R(\rho)$ separately, based on both the scenarios (\textsc{st, cl, hl}) and the regimes (determined by the values of $\alpha_{\scaleto{Q}{5pt}}$), produces a substantial number of extended solutions. To organize them systematically, we present a tree diagram in Figure \ref{tree-diagram} that categorizes all solutions and highlights additional constrained cases (ACCs).

    \tikzstyle{level 1}=[level distance=30mm, sibling distance=38mm]
    \tikzstyle{level 2}=[level distance=50mm, sibling distance=13mm]
    \tikzstyle{level 3}=[level distance=38mm]

\begin{figure}[ht]
\begin{tikzpicture}[grow=right,->]
	\begin{scope}[yshift=0]
	   \node {$R(\rho)$}
	child {node[align=center] {$\alpha_{\scaleto{Q}{5pt}} = 1$\\(Limiting case)}
		child {node {$R_{\scaleto{HL}{4pt}}^{\scaleto{(1)}{6pt}}(\rho)$}
			child[align=left,-] {node {One ACC}}
		}
		child {node {$R_{\scaleto{CL}{4pt}}^{\scaleto{(1)}{6pt}}(x)$}
			child[align=left,-] {node {Two ACCs}}
		}
		child {node {$R_{\scaleto{ST}{4pt}}^{\scaleto{(1)}{6pt}}(x)$}
			child[align=left,-] {node {One ACC}}
		}
	}
	child {node {$\alpha_{\scaleto{Q}{5pt}} = 1/2$}
		child {node {$R_{\scaleto{HL}{4pt}}^{\scaleto{(1/2)}{6pt}}(\rho)$}
			child[align=left, -] {node {Extra: $R_{\scaleto{HL+QF}{4pt}}^{\scaleto{(1/2)}{6pt}}(\rho)$}}
		}
		child {node {$R_{\scaleto{CL}{4pt}}^{\scaleto{(1/2)}{6pt}}(x)$}
			child[align=left, -] {node {Two ACCs}}
		}
		child {node {$R_{\scaleto{ST}{4pt}}^{\scaleto{(1/2)}{6pt}}(x)$}
		}
	}
	child {node[align=center] {$\alpha_{\scaleto{Q}{5pt}} = 0$\\(Limiting case)}
		child {node {$R_{\scaleto{HL}{4pt}}^{\scaleto{(0)}{6pt}}(\rho)$}
		}
		child {node {$R_{\scaleto{CL}{4pt}}^{\scaleto{(0)}{6pt}}(x)$}
		}
		child {node {$R_{\scaleto{ST}{4pt}}^{\scaleto{(0)}{6pt}}(x)$}
			child[align=left, -] {node {Extra: $R_{\scaleto{QF}{4pt}}(x)$}}
		}
	};
	\end{scope}
\end{tikzpicture}
\caption{Tree diagram illustrating the complete set of extended radial solutions. We emphasize that each radial domain depends on the value of $\alpha_{\scaleto{Q}{5pt}}$, but, in some cases, there might be further specific considerations. The detailed derivation for each case is given in the respective subsection. Finally, the acronym ACC denotes Additional Constrained Case, referring to complementary solutions subject to constraints on physical parameters.}
\label{tree-diagram}
\end{figure}
       
	The following sections examine these extended solutions, beginning with the case $\alpha_{\scaleto{Q}{5pt}} = 0$, which represents the lower bound of $\alpha_{\scaleto{Q}{5pt}}$. This limiting case illustrates significant characteristics that are common to the entire set of solutions.

	\subsection{Lower bound for the state parameter} \label{Extended_Solutions_Alpha_Q_0}

	To begin our investigation of extended solutions, we first consider the case where $\alpha_{\scaleto{Q}{5pt}} = 0$, corresponding to the limiting case of $\omega_{\scaleto{Q}{5pt}} = -1/3$. This limit represents the maximum contribution of quintessence to $A(\rho)$ when its effect is indistinguishable from that of the cloud of strings. In order to determine these solutions, we initially investigate the expression for $A(\rho)$, which we subsequently use to determine the effective potential $\rho_{\scaleto{+}{4pt}}^{2}\,V_{\text{eff}}(\rho)$. Then, solving the ODE for the auxiliary function $u(x)$, given by \cref{Eq-para-u}, we finally obtain the desired solution using the relation between $u(x)$ and $R(x)$, in accordance with \cref{radial-wave-function}. 
	
	Considering the step-by-step approach presented above, we first consider the standard scenario (in which $\overline{a}\neq 0$, $N_{\scaleto{Q}{5pt}} \neq 0$, and $\rho_{\scaleto{S}{4pt}} \neq 0$), where the result of \cref{A_Assumption} implies that:
	\begin{equation*}
		A(\rho) = \left(\overline{a} + N_{\scaleto{Q}{5pt}}\right) - \frac{\rho_{\scaleto{S}{4pt}}}{\rho}\,.
	\end{equation*}
	Recalling that, under the general assumption $\rho_{\scaleto{+}{4pt}}^{2}/l^{2} < \rho^{2}/l^{2} \ll 1$, the event horizon $\rho_{\scaleto{+}{4pt}}$ is determined by
	\begin{equation*}
		\rho_{\scaleto{+}{4pt}} = \frac{\rho_{\scaleto{S}{4pt}}}{\overline{a} + N_{\scaleto{Q}{5pt}}}\,,
	\end{equation*}
	we conclude that
	\begin{equation}
		A(x) = \left(\overline{a} + N_{\scaleto{Q}{5pt}}\right)\left(1 - \frac{1}{x}\right)\,,\label{A(x)_Alpha_Q_0}
	\end{equation}
	where, again, $x=\rho/\rho_{\scaleto{+}{4pt}}$ is the dimensionless radial variable.
	
	In the sequence, we substitute the result from \cref{A(x)_Alpha_Q_0} into the effective potential \eqref{Potential_in_terms_of_x}, which yields
	\begin{align}
		\rho_{\scaleto{+}{4pt}}^{2}\,V_{\text{eff}}(x) =  \frac{1/4}{x^{2}\left(x-1\right)^{2}} 
		+ \left(\frac{\epsilon\,\rho_{\scaleto{+}{4pt}}^{2}}{\rho_{\scaleto{S}{4pt}}}\right)^{2}\frac{x^{2}}{\left(x-1\right)^{2}} -\kappa^{2}\frac{\rho_{\scaleto{+}{4pt}}/\rho_{\scaleto{S}{4pt}}}{x\left(x-1\right)} -\frac{\rho_{\scaleto{+}{4pt}}^{3}\,\overline{m}_{\phi}^{2}}{\rho_{\scaleto{S}{4pt}}}\,\frac{x}{x-1}\,.\label{Potential_V_eff_Alpha_Q_0}
	\end{align}
	To rewrite $\rho_{\scaleto{+}{4pt}}^{2}\,V_{\text{eff}}(x)$ in a more convenient way, we introduce the following definitions
	\begin{equation}
		\begin{aligned}
			\epsilon_{0} &= \frac{\epsilon\rho_{\scaleto{+}{4pt}}^{2}}{\rho_{\scaleto{S}{4pt}}} = \frac{\epsilon\, \rho_{\scaleto{S}{4pt}}}{\left(\overline{a} + N_{\scaleto{Q}{5pt}}\right)^{2}}\,,\\
			\kappa_{0}^{2} &= \frac{\kappa^{2}\rho_{\scaleto{+}{4pt}}}{\rho_{\scaleto{S}{4pt}}} = \frac{\kappa^{2}}{\left(\overline{a} + N_{\scaleto{Q}{5pt}}\right)}\,,\\
			\quad m_{0}^{2} &= \frac{\overline{m}_{\phi}^{2}\,\rho_{\scaleto{+}{4pt}}^{3}}{\rho_{\scaleto{S}{4pt}}} = \frac{\overline{m}_{\phi}^{2}\, \rho_{\scaleto{S}{4pt}}^{2}}{\left(\overline{a} + N_{\scaleto{Q}{5pt}}\right)^{3}}\,,
		\end{aligned}\label{Redef_Parameters_Case_0}
	\end{equation}
	and then, performing a partial fraction decomposition of  \cref{Potential_V_eff_Alpha_Q_0}, we can express the resulting effective potential as
	\begin{equation}
		\rho_{\scaleto{+}{4pt}}^{2}\,V_{\text{eff}}(x) = \frac{\kappa_{0}^{2} + 1/2}{x} +  \frac{2\epsilon_{0}^{2} - \kappa_{0}^{2} - m_{0}^{2} - 1/2}{x-1} + \frac{1/4}{x^{2}} + \frac{\epsilon_{0}^{2} + 1/4}{\left(x-1\right)^{2}} + \left(\epsilon_{0}^{2} - m_{0}^{2}\right)\,.\label{Effective_Potential_Alpha_Q_0_ST}
	\end{equation}
	
	Comparing \cref{Eq-para-u,Effective_Potential_Alpha_Q_0_ST} with the normal form of the CHE, as determined by \cref{Sol-Geral-Normal-ECH_z-0,Parameters_A_to_E_CHE}, we find that
	\begin{equation}
		\begin{aligned}
			u_{\scaleto{ST}{4pt}}^{\scaleto{(0)}{6pt}}(x) 
			&= e^{\alpha x/2}\,x^{1/2}\,\left(x-1\right)^{(1 + \gamma)/2}
			\left[c_{1}\,
			\text{HeunC}\left(-\alpha,\,\gamma,\,\beta,\,-\delta,\,\delta+\eta;\,1-x\right) \right.\\
			&\left.+\, c_{2}\,\left(x-1\right)^{-\gamma}
			\text{HeunC}\left(-\alpha,\,-\gamma,\,\beta,\,-\delta,\,\delta+\eta;\,1-x\right)\right] \,, \label{u(x)_Alpha_Q_0_ST}
		\end{aligned}
	\end{equation}
	where the subscript \enquote{\textsc{st}} denotes the standard scenario, as defined in \cref{naming-solutions}, while the superscript \enquote{$(0)$} indicates that $\alpha_{\scaleto{Q}{5pt}} = 0$. The associated Heun parameters are
	\begin{equation}
		\begin{aligned}
			\alpha &= 2\sqrt{m_{0}^{2} - \epsilon_{0}^{2}}\,,\\
			\beta &= 0\,,\\
			\gamma &= 2 i \,\epsilon_{0}\,,\\
			\delta &= 2\epsilon_{0}^{2} - m_{0}^{2}\,,\\
			\eta &= - \kappa_{0}^{2}\,,
		\end{aligned} \label{Heun_Parameters_Alpha_Q_0_ST}
	\end{equation}
	where $\epsilon_{0}$, $m_{0}$ e $\kappa_{0}$ were defined by \cref{Redef_Parameters_Case_0}.

	Hence, by combining \cref{radial-wave-function,A(x)_Alpha_Q_0,u(x)_Alpha_Q_0_ST} we obtain that the formal solution $R_{\scaleto{ST}{4pt}}^{\scaleto{(0)}{6pt}}(x)$ is given by
	\begin{equation}
		\begin{aligned}
			R_{\scaleto{ST}{4pt}}^{\scaleto{(0)}{6pt}}(x) 
			&= e^{\alpha x/2}\,
			\left(x-1\right)^{\gamma/2} 
			\left[ c_{1}
			\text{HeunC}\left(-\alpha,\,\gamma,\,\beta,\,-\delta,\,\delta+\eta;\,1-x\right) \right.\\
			&\left.+\, c_{2}\,\left(x-1\right)^{-\gamma}
			\text{HeunC}\left(-\alpha,\,-\gamma,\,\beta,\,-\delta,\,\delta+\eta;\,1-x\right)\right]\,.
		\end{aligned} \label{Radial_WF_Alpha_Q_0_ST}
	\end{equation}
	This solution, corresponding to $\alpha_{\scaleto{Q}{5pt}} = 0$, is valid for $x= \rho/\rho_{\scaleto{+}{4pt}}$ provided that $\rho_{\scaleto{+}{4pt}}<\rho \ll |l|$. For example, with $|l| = 3.8 \times 10^{26}$ m, $N_{\scaleto{Q}{5pt}} = 8.95$, $\overline{a} = 10^{-4}$, and $\rho_{\scaleto{S}{4pt}} = 10^{10}$ m, we obtain that $\rho_{\scaleto{+}{4pt}} \approx 1.12 \times 10^{9}$ m. Consequently, the approximation for $A(x)$ remains valid up to $10^{14}$ times the event horizon radius. This contrasts with the solution described by \eqref{Heun_Parameters_RWF_NH_All_Alpha_Q}, which is valid only for $|x-1| \approx 0$.
	
	Finally, we 	examine the solutions of the $\delta_{N}$ subclass, which are discussed in \cref{sec-vínculos-espectrais}, and, by definition, satisfy the constraint of \cref{Delta_Condition}, that is,
	\begin{equation}
		N + 1 + i\epsilon_{0}  = \frac{m_{0}^{2} - 2\epsilon_{0}^{2}}{2\sqrt{m_{0}^{2} - \epsilon_{0}^{2}}}\,,\label{energy-levels-st-0}
	\end{equation}
	with $N \in \mathbb{Z}$. It is important to note that this subclass includes the polynomial solutions of the CHE, which are regular at the two finite singularities of the differential equation. Therefore, the above result does not provide all the energy eigenstates associated with $R_{\scaleto{ST}{4pt}}^{\scaleto{(0)}{6pt}}(x)$, only those belonging to $\delta_{N}$.
	
	In the next subsection, we will demonstrate that the solution \eqref{Radial_WF_Alpha_Q_0_ST} and the constraint \eqref{energy-levels-st-0} exhibit only minor variations when the cloud of strings is absent.

	\subsubsection{Special case without the cloud of strings}

	The radial solution for the cloudless scenario, $R_{\scaleto{CL}{4pt}}^{\scaleto{(0)}{6pt}}(x)$, is simply obtained from \cref{Heun_Parameters_Alpha_Q_0_ST,Radial_WF_Alpha_Q_0_ST}, by setting $\overline{a} = 0$. The same procedure applies to the particular case of \cref{energy-levels-st-0}. 
	
	In this scenario, $R_{\scaleto{CL}{4pt}}^{\scaleto{(0)}{6pt}}(x)$ depends solely on the quintessence parameter $N_{\scaleto{Q}{5pt}}$ and the Schwarzschild-like radius $\rho_{\scaleto{S}{4pt}}$. Consequently, the convenient parameters in \cref{Redef_Parameters_Case_0}, appearing in $R_{\scaleto{ST}{4pt}}^{\scaleto{(0)}{6pt}}(x)$, reduce to the following expressions:
	\begin{align*}
		\epsilon_{\scaleto{\text{CL}}{4pt}}^{\scaleto{(0)}{6pt}} = \frac{\epsilon \rho_{\scaleto{S}{4pt}}}{N_{\scaleto{Q}{5pt}}^{2}}\,,
		\quad
		\kappa_{\scaleto{\text{CL}}{4pt}}^{\scaleto{(0)}{6pt}} = \sqrt{\frac{\kappa^{2}}{N_{\scaleto{Q}{5pt}}}}\,,
		\quad m_{\scaleto{\text{CL}}{4pt}}^{\scaleto{(0)}{6pt}} = \sqrt{\frac{\overline{m}_{\phi}^{2}\,\rho_{\scaleto{S}{4pt}}^{2}}{N_{\scaleto{Q}{5pt}}^{3}}}\,.
	\end{align*}
	where the subscript \scaleto{CL}{6pt} denotes the cloudless scenario, as defined in Table \ref{naming-solutions}.
	
	Notably, this cloudless case closely resembles the standard general case of $\alpha_{\scaleto{Q}{5pt}} = 0$ because, as aforementioned, both the cloud of strings and the quintessence parameters exhibit identical behavior in this limiting case.
	
	In the next subsection, we show the consequences of removing the black string and how this first horizonless scenario behaves.

	\subsubsection{Removing the black string at the lower bound} \label{HL-0}
	
	The third possible solution when $\alpha_{\scaleto{Q}{5pt}}=0$ is the one where $\rho_{\scaleto{S}{4pt}} = 0$, while $\overline{a}$ and/or $N_{\scaleto{Q}{5pt}}$ are nontrivial. This condition implies that $A(\rho)$, given by \cref{A_Assumption}, becomes simply
	\begin{equation}
		A(\rho) = \overline{a} + N_{\scaleto{Q}{5pt}}\,, \label{Alpha_Q_0_R_s_0}
	\end{equation}
	provided that $\rho_{\scaleto{+}{4pt}}<\rho \ll |l|$, allowing us to neglect the $\rho^{2}/l^{2}$ term in \eqref{Function_A(rho)}. Additionally, since $\rho_{\scaleto{S}{4pt}}=0$ eliminates the event horizon, we need to express the metric function $A(\rho)$ in terms of $\rho$ rather than $x = \rho/\rho_{\scaleto{+}{4pt}}$. Hence, substituting \cref{Alpha_Q_0_R_s_0} into \cref{Effective-Potential} yields the effective potential:
	\begin{equation}
		V_{\text{eff}}(\rho) = \frac{\epsilon^{2}}{\left(\overline{a} + N_{\scaleto{Q}{5pt}}\right)^{2}} - \frac{\overline{m}_{\phi}^{2}}{\left(\overline{a} + N_{\scaleto{Q}{5pt}}\right)} - \frac{\kappa^{2}/\left(\overline{a} + N_{\scaleto{Q}{5pt}}\right)}{\rho^{2}}\,.\label{Alpha_Q_0_R_s_0-Eff-Potential}
	\end{equation}
	
	Considering the differential equation for $u(\rho)$, defined by \cref{Eq-para-u}, the effective potential \eqref{Alpha_Q_0_R_s_0-Eff-Potential} above, and the general result of \cref{Eq_u_Normal_Form}, we obtain that the solution for $u_{\scaleto{\text{HL}}{4pt}}^{\scaleto{(0)}{6pt}}(\rho)$ is given by
	\begin{equation}
		\begin{aligned}
			u_{\scaleto{\text{HL}}{4pt}}^{\scaleto{(0)}{6pt}}(\rho) 
			&= e^{\alpha \rho/2}\,\rho^{(1+\beta)/2}\,\left(\rho - 1\right)^{(1 + \gamma)/2}
			\left[ c_{1}\,
			\text{HeunC}\left(\alpha,\,\beta,\,\gamma,\,\delta,\,\eta;\,\rho\right) \right.\\
			&\left.+\, c_{2}\, \rho^{-\beta}\,
			\text{HeunC}\left(\alpha,\,-\beta,\,\gamma,\,\delta,\,\eta;\,\rho\right)\right]\,,
		\end{aligned}\label{u-HL_0}
	\end{equation}
	where the subindex \enquote{\textsc{hl}} stands for horizonless, in accordance with \cref{naming-solutions}. The associated Heun parameters are:
	\begin{equation}
		\begin{aligned}
			\alpha &= \frac{2 i}{\overline{a} + N_{\scaleto{Q}{5pt}}}\,\sqrt{\epsilon^{2} - \left(\overline{a}+N_{\scaleto{Q}{5pt}}\right)\overline{m}_{\phi}^{2}}\,,\\
			\beta &= 2\,\sqrt{\frac{\kappa^{2}}{\overline{a} + N_{\scaleto{Q}{5pt}}} + \frac{1}{4}}\,,\\
			\gamma &= 1\,,\\
			\delta &= 0\,,\\
			\eta &= \frac{1}{2}\,.
		\end{aligned} \label{Heun_Parameters_HL_0}
	\end{equation}
	Thus, the complete solution describing the radial wave function $R_{\scaleto{\text{HL}}{4pt}}^{\scaleto{(0)}{6pt}}(\rho)$ is
	\begin{equation}
		\begin{aligned}
			R_{\scaleto{\text{HL}}{4pt}}^{\scaleto{(0)}{6pt}}(\rho) 
			&= e^{\alpha \rho/2}\,\rho^{(\beta - 1)/2}\,\left(\rho - 1\right)^{(1 + \gamma)/2}
			\left[ c_{1}\,
			\text{HeunC}\left(\alpha,\,\beta,\,\gamma,\,\delta,\,\eta;\,\rho\right) \right.\\
			&\left.+\, c_{2}\, \rho^{-\beta}\,
			\text{HeunC}\left(\alpha,\,-\beta,\,\gamma,\,\delta,\,\eta;\,\rho\right)\right]\,,\label{RWF_HL_0_Full}
		\end{aligned}
	\end{equation}
	where $c_{1}$ and $c_{2}$ are constants.
	
	Within the set of solutions associated with the Heun parameters \eqref{Heun_Parameters_HL_0}, those belonging to the subclass $\delta_{N}$, that is, those obeying the constraint \eqref{Delta_Condition}, satisfy one of the following constraints:
	\begin{equation}
		\epsilon^{2} = \overline{m}_{\phi}^{2}\left(\overline{a} + N_{\scaleto{Q}{5pt}}\right)\,,\label{Vínculo_Alpha_0}
	\end{equation}
	or
	\begin{equation*}
		N + 1 + \frac{1}{2} + \sqrt{ \frac{\kappa^{2}}{\overline{a} + N_{\scaleto{Q}{5pt}}} + \frac{1}{4}} = 0\,.\label{Vínculo_Alpha_neq_0}
	\end{equation*}
	This happens because \cref{Heun_Parameters_HL_0} establishes that $\delta = 0$ and, consequently, there are two ways to satisfy condition \eqref{Delta_Condition}. The first requirement is equivalent to fixing the parameter $\alpha$ as $\alpha=0$, from which the result \eqref{Vínculo_Alpha_0} follows. On the other hand, a closer look into the second condition (which follows from $N+1+(\beta+\gamma)/2=0$) shows us that this requirement cannot be satisfied since $N$ must be a positive integer.
	
	As an alternative description, we can solve \cref{eq-geral-u} with the effective potential \eqref{Alpha_Q_0_R_s_0-Eff-Potential} as a Bessel equation \cite{Olver2010,Butkov1968}. Considering these results, the ODE for the auxiliary function $u_{\scaleto{\text{HL}}{4pt}}^{\scaleto{(0)}{6pt}}(\rho)$ becomes:
	\begin{equation*}
		\frac{d^{2}}{d\,\rho^{2}}\,u_{\scaleto{\text{HL}}{4pt}}^{\scaleto{(0)}{6pt}}(\rho) + \left[ \frac{\epsilon^{2}}{\left(\overline{a} + N_{\scaleto{Q}{5pt}}\right)^{2}} - \frac{\overline{m}_{\phi}^{2}}{\left(\overline{a} + N_{\scaleto{Q}{5pt}}\right)} - \frac{\kappa^{2}/\left(\overline{a} + N_{\scaleto{Q}{5pt}}\right)}{\rho^{2}}\right]
		u_{\scaleto{\text{HL}}{4pt}}^{\scaleto{(0)}{6pt}}(\rho) = 0\,.
	\end{equation*}
	This leads to the solution:
	\begin{equation}
		u_{\scaleto{\text{HL}}{4pt}}^{\scaleto{(0)}{6pt}}(\rho) = \rho^{1/2} \left[c_{1}\,\text{J}_{m}\left(k \rho\right) + c_{2}\,\text{Y}_{m}\left(k \rho\right)\right]\,,\label{u_HL_0_Bessel}
	\end{equation}
	where J$_{m}(x)$ and Y$_{m}(x)$ are the $m$-th order Bessel functions of the first and second kinds, respectively, while $k$ and $m$ are determined by
	\begin{align}
		k &= \frac{\epsilon^{2}}{\left(\overline{a} + N_{\scaleto{Q}{5pt}}\right)^{2}} - \frac{\overline{m}_{\phi}^{2}}{\left(\overline{a} + N_{\scaleto{Q}{5pt}}\right)}\,,\label{def-k}\\
		m &= \sqrt{\frac{1}{4} + \frac{\kappa^{2}}{\overline{a}+N_{\scaleto{Q}{5pt}}}}\,.\label{def-m}
	\end{align}
	As usual, $c_{1}$ and $c_{2}$ are arbitrary constants. We observe that, unlike the solutions of the CHE, the properties of Bessel functions are widely known and are detailed in the references \cite{Olver2010,Butkov1968,Abramowitz1968}.
	
	Subsequently, using the relation \eqref{radial-wave-function} along with \cref{Alpha_Q_0_R_s_0,u_HL_0_Bessel}, the radial solution $R_{\scaleto{\text{HL}}{4pt}}^{\scaleto{(0)}{6pt}}(\rho)$ becomes
	\begin{equation}
		R_{\scaleto{\text{HL}}{4pt}}^{\scaleto{(0)}{6pt}}(\rho) = \rho^{-1/2} \left[c_{1}\,\text{J}_{m}\left(k \rho\right) + c_{2}\,\text{Y}_{m}\left(k \rho\right)\right]\,,\label{RWF_HL_Alpha_Q_0_Bessel}
	\end{equation}
	where $c_{1}$ and $c_{2}$ are still arbitrary and distinct from other past occurrences (such as in \cref{u_HL_0_Bessel}).
	Furthermore, since $\kappa^{2} = n_{\scaleto{R}{4pt}}^{2} + k_{z}^{2}$, we know that $\kappa = 0$ (equivalent to $m=1/2$) occurs only for particles that do not change their position in the spatial coordinates $\varphi$ and $z$, moving only in the radial direction. Otherwise, $\kappa \neq 0$, implying in $m>1/2$.
	
	Regarding the regularity of the solution \eqref{RWF_HL_Alpha_Q_0_Bessel} at the origin, we set $c_{2}=0$ to remove the divergent behavior of $\text{Y}_{m}\left(k \rho\right)$. We emphasize that, in the absence of the \emph{black string}, the term \enquote{origin} refers to the point $\rho = 0$, degenerate to $\varphi$, where the center of the removed \emph{black string} would be. The spatial coordinate $z$, in turn, is arbitrary, since the \emph{black string} is infinitely long and aligned with the $z$ axis. It is important to remember that the energy densities of the quintessence and the cloud of \emph{strings} are also relative to this origin at $\rho = 0$. Therefore, even with $\rho_{\scaleto{S}{4pt}} = 0$, we have a privileged spatial position, around which the cloud of strings and the quintessence fluid are distributed.
	
	In the sequence, we investigate a scenario without quintessence, denoted by \textsc{qf}, in accordance with Table \ref{naming-solutions}. Formally, we could allocate this case separately from all the others since the absence of quintessence makes it independent of the state parameter $\alpha_{\scaleto{Q}{5pt}}$. However, in practice, this solution was developed after the previous scenarios (where $\alpha_{\scaleto{Q}{5pt}} = 0$), to facilitate comparisons.

	\subsubsection{In the absence of quintessence}\label{sec-QL}
	
	To analyze the scenario without quintessence, we consider that $N_{\scaleto{Q}{5pt}} = 0$, so that the system is independent of the state parameter $\alpha_{\scaleto{Q}{5pt}}$, but we keep $\overline{a}$ and $\rho_{\scaleto{S}{4pt}}$ non-zero. Furthermore, we maintain the assumption that $\rho_{\scaleto{+}{4pt}}< \rho \ll |l|$, thus neglecting the terms associated with the cosmological constant in \cref{Function_A(rho)}. We also highlight that the solution for the quintessence-free scenario is a reasonable approximation of the cases where $\alpha_{\scaleto{Q}{5pt}} > 1/2$, when the terms $\rho^{2}/l^{2} + N_{\scaleto{Q}{5pt}}\rho^{2\alpha_{\scaleto{Q}{3pt}}}$ are infinitely small compared to $\overline{a} - \rho_{\scaleto{S}{4pt}}/\rho$.

	Under these assumptions, it follows that \cref{A_Assumption} for $A(\rho)$ can be written in the form:
	\begin{equation*}
		A(x) = \overline{a} - \frac{\rho_{\scaleto{S}{4pt}}}{\rho_{\scaleto{+}{4pt}}}\,\frac{1}{x}\,.
	\end{equation*}
	Since
	\begin{equation*}
		\rho_{\scaleto{+}{4pt}} = \frac{\rho_{\scaleto{S}{4pt}}}{\overline{a}}\,,
	\end{equation*}
	we obtain that
	\begin{equation}
		A(x) = \overline{a}\left(1 - \frac{1}{x}\right)\,.\label{A(x)_QL}
	\end{equation}
	As expected, the result \eqref{A(x)_QL} has the same form as \cref{A(x)_Alpha_Q_0}, under the substitution of $N_{\scaleto{Q}{5pt}} + \overline{a}$ to simply $\overline{a}$. In other words, it is equivalent to fixing $N_{\scaleto{Q}{5pt}} = 0$. This is possible because $\alpha_{\scaleto{Q}{5pt}} = 0$ is the only regime where the parameters $\overline{a}$ and $N_{\scaleto{Q}{5pt}}$ play the same role in the metric function. 
	
	Consequently, the effective potential \eqref{Effective_Potential_Alpha_Q_0_ST} becomes
	\begin{equation}
		\rho_{\scaleto{+}{4pt}}^{2}\,V_{\text{eff}}(x) 
		= \frac{\kappa_{\scaleto{QF}{4pt}}^{2} + 1/2}{x} 
		+ \frac{2\epsilon_{\scaleto{QF}{4pt}}^{2} - \kappa_{\scaleto{QF}{4pt}}^{2} - m_{\scaleto{QF}{4pt}}^{2} - 1/2}{x-1} 
		+ \frac{1/4}{x^{2}} 
		+ \frac{\epsilon_{\scaleto{QF}{4pt}}^{2} + 1/4}{\left(x-1\right)^{2}} 
		+ \epsilon_{\scaleto{QF}{4pt}}^{2} - m_{\scaleto{QF}{4pt}}^{2}\,,\label{V_eff_QL}
	\end{equation}
	where the subscript \enquote{\textsc{qf}} indicates the absence of quintessence. The parameters $\epsilon_{\scaleto{QF}{4pt}}$, $\kappa_{\scaleto{QF}{4pt}}$, and $m_{\scaleto{QF}{4pt}}$ follow from \cref{Redef_Parameters_Case_0}:
	\begin{equation}
		\epsilon_{\scaleto{QF}{4pt}} 
		= \frac{\epsilon\,\rho_{\scaleto{S}{5pt}}}{\overline{a}^{2}}\,,\qquad
		\kappa_{\scaleto{QF}{4pt}} 
		= \frac{\kappa}{\overline{a}^{1/2}}\,,\qquad 
		m_{\scaleto{QF}{4pt}} 
		= \frac{\rho_{\scaleto{S}{5pt}} \,\overline{m}_{\phi}}{\overline{a}^{3/2}}\label{Parameters_Absence_Quintessence}\,.
	\end{equation}
	Therefore, we only need to fix $N_{\scaleto{Q}{5pt}}=0$ in the radial solution \cref{Radial_WF_Alpha_Q_0_ST} to obtain $R_{\scaleto{QF}{4pt}}(x)$, which reads
	\begin{equation}
		\begin{aligned}
			R_{\scaleto{QF}{4pt}}(x)
			&= e^{\alpha x/2}\,
			\left(x-1\right)^{\gamma/2} 
			\left[ c_{1}
			\text{HeunC}\left(-\alpha,\,\gamma,\,\beta,\,-\delta,\,\delta+\eta;\,1-x\right) \right.\\
			&\left.+\, c_{2}\,\left(x-1\right)^{-\gamma}
			\text{HeunC}\left(-\alpha,\,-\gamma,\,\beta,\,-\delta,\,\delta+\eta;\,1-x\right)\right]\,.\label{RWF_QL}
		\end{aligned}
	\end{equation}
	where $c_{1}$ e $c_{2}$ are constants. Analogously to the case of \cref{Heun_Parameters_Alpha_Q_0_ST}, the Heun parameters for the quintessence-free scenario are:
	\begin{equation}
		\begin{aligned}
			\alpha &= 2 i\,\frac{\rho_{\scaleto{S}{4pt}}}{\overline{a}^{2}} \sqrt{\epsilon^{2} - \overline{a} \,\overline{m}_{\phi}^{2}}\,,\\
			\beta &= 0\,,\\
			\gamma &= 2i\,\frac{\epsilon \rho_{\scaleto{S}{4pt}}}{\overline{a}^{2}}\,,\\
			\delta &= \frac{2 \rho_{\scaleto{S}{4pt}}^{2}}{\overline{a}^{4}}\left(\epsilon^{2} - \frac{1}{2}\,\overline{a}\,\overline{m}_{\phi}^{2}\right)\,,\\
			\eta &= - \frac{\kappa^{2}}{\overline{a}}\,.
		\end{aligned} \label{Heun_Parameters_QL}
	\end{equation}
	
	Although this solution is valid in an extended domain, we can investigate its behavior when $x \to 1^{+}$,  where we obtain that
	\begin{equation}
		R_{\scaleto{QF}{4pt}}(x)
		= c_{1}
		\left(x-1\right)^{\gamma/2}
		+ c_{2}\,\left(x-1\right)^{-\gamma/2}\,,\label{Radial_WF_Free_from_Quintessence}
	\end{equation}
	as long as the constant term $e^{\alpha/2}$ is absorbed by $c_{1}$ and $c_{2}$. Notably, this result is equivalent to our previous solution near the event horizon. This can be understood by means of \cref{Heun_Parameters_RWF_NH_All_Alpha_Q,RWF_NH_Parte_1}.
	
	Having examined all cases for $\alpha_{\scaleto{Q}{5pt}} = 0$, we now turn to the physically significant case of $\alpha_{\scaleto{Q}{5pt}} = 1/2$. This midpoint value corresponds to a scenario where the quintessence's influence on the metric is only significant at large distances.
	
	\subsection{Solutions for the intermediate state parameter} \label{Extended_Solutions_Alpha_Q_One_Half}

	To examine the implications of a scalar particle within the BCK spacetime, with $\alpha_{\scaleto{Q}{5pt}} = 1/2$ ($\omega_{\scaleto{Q}{5pt}} = -2/3$), let us consider the result of \cref{A_Assumption} and its general assumption $\rho_{\scaleto{+}{4pt}}<\rho \ll |l|$. In this case, the event horizon $\rho_{\scaleto{+}{4pt}}$ is determined by
	\begin{equation}
		\overline{a} - \frac{\rho_{\scaleto{S}{4pt}}}{\rho_{\scaleto{+}{4pt}}} + N_{\scaleto{Q}{5pt}}\,\rho_{\scaleto{+}{4pt}} = 0\,,\label{Constraint_EH_Alpha_Meio}
	\end{equation}
	Then, for any nontrivial\footnote{Since we have just examined the quintessence-free scenario in the last subsection.} value of $N_{\scaleto{Q}{5pt}}$, $A(x)$ can be expressed as
	\begin{equation*}
		A(x) = \overline{a} - \frac{\rho_{\scaleto{S}{4pt}}}{\rho_{\scaleto{+}{4pt}}}\,\frac{1}{x} + N_{\scaleto{Q}{5pt}}\,\rho_{\scaleto{+}{4pt}}\,x\,,
	\end{equation*}
	provided $\rho_{\scaleto{+}{4pt}} \neq 0$, which requires $\rho_{\scaleto{S}{4pt}}\neq 0$. Using the constraint from \cref{Constraint_EH_Alpha_Meio}, we can rewrite the above expression for $A(x)$ as
	\begin{equation}
		A(x) = \overline{a}\left(1 - \frac{1}{x}\right)\left[1 + \frac{N_{\scaleto{Q}{5pt}} \,\rho_{\scaleto{+}{4pt}}}{\overline{a}}
		\left(1 + x\right)\right]\,.\label{A(x)-Alpha_Q-Meio}
	\end{equation}
	Moreover, it follows directly from \cref{Constraint_EH_Alpha_Meio} that
	\begin{equation*}
		\frac{N_{\scaleto{Q}{5pt}}\,\rho_{\scaleto{+}{4pt}}}{\overline{a}} = \frac{1}{2}\left(\sqrt{1 + \frac{4\,N_{\scaleto{Q}{5pt}} \rho_{\scaleto{S}{4pt}}}{\overline{a}^{2}}} - 1\right)\,,
	\end{equation*}
	and, given the ranges of $\overline{a}$, $\rho_{\scaleto{S}{4pt}}$, and $N_{\scaleto{Q}{5pt}}$ presented in \cref{Sec-2-Results-Paper-One}, it is reasonable to consider solutions satisfying
	\begin{equation}
		\frac{N_{\scaleto{Q}{5pt}} \,\rho_{\scaleto{S}{4pt}} }{\overline{a}^{2}} \ll 1\,,\label{Small_Parameter_Condition}
	\end{equation}
	so that
	\begin{equation*}
		\frac{N_{\scaleto{Q}{5pt}} \,\rho_{\scaleto{+}{4pt}}}{\overline{a}} = \frac{N_{\scaleto{Q}{5pt}} \,\rho_{\scaleto{S}{4pt}}}{\overline{a}^{2}} + \mathcal{O}\left(\frac{N_{\scaleto{Q}{5pt}}\,\rho_{\scaleto{S}{4pt}} }{\overline{a}^{2}}\right)^{2}\,.
	\end{equation*}
	\vspace{1mm}
	
	\noindent In this regime, \cref{A(x)-Alpha_Q-Meio} becomes
	\begin{equation}
		A(x) = \overline{a}\left(1 - \frac{1}{x}\right)\left[1 + \frac{N_{\scaleto{Q}{5pt}} \,\rho_{\scaleto{S}{4pt}}}{\overline{a}^{2}}
		\left(1 + x\right)\right]\,, \label{A(x)_Metade_Intervalo}
	\end{equation}
	which represents a first-order correction to the case where quintessence is absent, given by \cref{A(x)_QL}. Notably, this correction increases with $x$, indicating that dark energy's influence becomes significant at large distances.
	
	From \cref{A(x)_Metade_Intervalo}, the derivatives $A'(x)$ and $A''(x)$ are:
	\begin{align}
		A'(x) 
		&= \frac{\overline{a}}{x^{2}}
		\left[1 + \frac{N_{\scaleto{Q}{5pt}}\,\rho_{\scaleto{S}{4pt}}}{\overline{a}^{2}}
		\left(1 + x^{2}\right)\right]\,,\label{A_First_Deriv_Alpha_Meio}\\
		A''(x)
		&= -\frac{2 \overline{a}}{x^{3}}\left(1 + \frac{N_{\scaleto{Q}{5pt}}\,\rho_{\scaleto{S}{4pt}}}{\overline{a}^{2}}\right)\,.\label{A_Second_Deriv_Alpha_Meio}
	\end{align}
	Subsequently, we use \cref{A(x)_Metade_Intervalo,A_First_Deriv_Alpha_Meio,A_Second_Deriv_Alpha_Meio} to determine the effective potential $\rho_{\scaleto{+}{4pt}}^{2}\,V_{\text{eff}}(x)$, given by \cref{Potential_in_terms_of_x} in the regime of \cref{Small_Parameter_Condition}. For convenience, we give the explicit results written in terms of their partial fraction decompositions:
	\begin{subequations}
		\begin{align}
			\frac{A'(x)^{2}}{4\,A(x)^{2}} 
			&=
			\frac{\left(1- N_{\scaleto{Q}{5pt}}\,\rho_{\scaleto{S}{4pt}}/\overline{a}^{2}\right)/2}{x} 
			+ \frac{1/4}{x^{2}} - \frac{\left(1- N_{\scaleto{Q}{5pt}}\,\rho_{\scaleto{S}{4pt}}/\overline{a}^{2}\right)/2}{x-1} 
			+ \frac{1/4}{\left(x-1\right)^{2}}\,,\label{PartFrac_1_meio_Intervalo}\\[7pt]
			\frac{\rho_{\scaleto{+}{4pt}}^{2}\epsilon^{2}}{A(x)^{2}} 
			&= \left(\frac{\epsilon \rho_{\scaleto{S}{4pt}}}{\overline{a}^{2}}\right)^{2}\left[
			\frac{2 - 10\,N_{\scaleto{Q}{5pt}}\,\rho_{\scaleto{S}{4pt}}/\overline{a}^{2}}{x-1} 
			+ \frac{1 - 4 \,N_{\scaleto{Q}{5pt}}\,\rho_{\scaleto{S}{4pt}}/\overline{a}^{2}}{\left(x-1\right)^{2}} + \left(1 - \frac{6\,N_{\scaleto{Q}{5pt}}\,\rho_{\scaleto{S}{4pt}}}{\overline{a}^{2}}\right)\right.\notag\\
			&- \left. \frac{2\,N_{\scaleto{Q}{5pt}}\,\rho_{\scaleto{S}{4pt}}}{\overline{a}^{2}}\,x\right]\,,\label{PartFrac_2_meio_Intervalo} 
		\end{align}
		as well as
		\begin{align}
			-\frac{1}{A(x)}\left[\frac{1}{2}A''(x) + \frac{A'(x)}{x}\right] &= -\frac{N_{\scaleto{Q}{5pt}}\,\rho_{\scaleto{S}{4pt}}/\overline{a}^{2}}{x-1}\,, \label{PartFrac_3_meio_Intervalo} 
		\end{align}
		and
		\begin{align}
			-\frac{\kappa^{2}}{x^{2}\,A(x)} 
			&= \frac{\kappa^{2}/\overline{a}\left(1 - N_{\scaleto{Q}          {4pt}}\,\rho_{\scaleto{S}{4pt}}/\overline{a}^{2}\right)}{x} - \frac{\kappa^{2}/\overline{a}\left(1 - 2\,N_{\scaleto{Q}{5pt}}\,\rho_{\scaleto{S}{4pt}}/\overline{a}^{2}\right)}{x-1}\,,\label{PartFrac_4_meio_Intervalo}\\[7pt]
			-\frac{\rho_{\scaleto{+}{4pt}}^{2}\,\overline{m}_{\phi}^{2}}{A(x)} 
			&= -\left(\frac{\rho_{\scaleto{S}{4pt}}\,\overline{m}_{\phi}}{\overline{a}^{3/2}}\right)^{2}\left[
			\frac{1 - 2 N_{\scaleto{Q}{5pt}}\,\rho_{\scaleto{S}{4pt}}/\overline{a}^{2}}{x-1} 
			+ \left(1 - \frac{2 N_{\scaleto{Q}{5pt}}\,\rho_{\scaleto{S}{4pt}}}{\overline{a}^{2}}\right) - \frac{N_{\scaleto{Q}{5pt}}\,\rho_{\scaleto{S}{4pt}}}{\overline{a}^{2}}\,x\right]\,.\label{PartFrac_5_meio_Intervalo}
		\end{align}
	\end{subequations}
	
	After adding \cref{PartFrac_1_meio_Intervalo,PartFrac_2_meio_Intervalo,PartFrac_3_meio_Intervalo,PartFrac_4_meio_Intervalo,PartFrac_5_meio_Intervalo}, we conclude that the effective potential \eqref{Potential_in_terms_of_x} can be written as
	\begin{equation}
		\rho_{\scaleto{+}{4pt}}^{2}\,V_{\text{eff}}(x) = \frac{A}{x} + \frac{B}{x-1} + \frac{C}{x^{2}} + \frac{D}{\left(x-1\right)^{2}} + E + F\,x\,,\label{V_eff_Meio_Estendido}
	\end{equation}
	while the coefficients $A$--$F$ are given by
	\begin{subequations}
		\begin{align}
			A &= \left(\frac{1}{2} + \frac{\kappa^{2}}{\overline{a}}\right)\left(1 - \frac{N_{\scaleto{Q}{5pt}}\,\rho_{\scaleto{S}{4pt}}}{\overline{a}^{2}}\right)\,,\label{Coeff_A_Meio_Intervalo}\\
			B &= -\frac{1}{2}\left(1+\frac{N_{\scaleto{Q}{5pt}}\,\rho_{\scaleto{S}{4pt}}}{\overline{a}^{2}}\right) - \frac{1}{\overline{a}}\left(\kappa^{2} + \frac{\rho_{\scaleto{S}{4pt}}^{2}\,\overline{m}_{\phi}^{2}}{\overline{a}^{2}}\right)\left(1 - \frac{2\,N_{\scaleto{Q}{5pt}}\,\rho_{\scaleto{S}{4pt}}}{\overline{a}^{2}}\right)\notag\\
			&+ 2\left(\frac{\epsilon\,\rho_{\scaleto{S}{4pt}}}{\overline{a}^{2}}\right)^{2}\left(1 - \frac{5\,N_{\scaleto{Q}{5pt}}\,\rho_{\scaleto{S}{4pt}}}{\overline{a}^{2}}\right)\,,\label{Coeff_B_Meio_Intervalo}\\
			C &= \frac{1}{4}\,,\label{Coeff_C_Meio_Intervalo}\\
			D &= \frac{1}{4} + \left(\frac{\epsilon\,\rho_{\scaleto{S}{4pt}}}{\overline{a}^{2}}\right)^{2}
			\left(1 - \frac{4\,N_{\scaleto{Q}{5pt}}\,\rho_{\scaleto{S}{4pt}}}{\overline{a}^{2}}\right)\,\label{Coeff_D_Meio_Intervalo}\\
			E &= \left(\frac{\epsilon\,\rho_{\scaleto{S}{4pt}}}{\overline{a}^{2}}\right)^{2}\left(1 - \frac{6\,N_{\scaleto{Q}{5pt}}\,\rho_{\scaleto{S}{4pt}}}{\overline{a}^{2}}\right)- \frac{1}{\overline{a}}\,\frac{\rho_{\scaleto{S}{4pt}}^{2}\,\overline{m}_{\phi}^{2}}{\overline{a}^{2}}\left(1 - \frac{2\,N_{\scaleto{Q}{5pt}}\,\rho_{\scaleto{S}{4pt}}}{\overline{a}^{2}}\right)\,\label{Coeff_E_Meio_Intervalo}\\
			F &= \frac{N_{\scaleto{Q}{5pt}}\,\rho_{\scaleto{S}{4pt}}^{3}}{\overline{a}^{5}} \left(\overline{m}_{\phi}^{2} - \frac{2\epsilon^{2}}{\overline{a}}\right)\label{Coeff_F_Meio_Intervalo}\,.
		\end{align}
	\end{subequations}
	
	The comparison between \cref{Coeff_A_Meio_Intervalo,Coeff_B_Meio_Intervalo,Coeff_C_Meio_Intervalo,Coeff_D_Meio_Intervalo,Coeff_E_Meio_Intervalo,Coeff_F_Meio_Intervalo} with  \cref{V_eff_QL,Parameters_Absence_Quintessence} reveals that the effective potential at the midpoint of the interval represents a first-order correction to the scenario where quintessence is absent. Furthermore, in the physically plausible limit where $N_{\scaleto{Q}{5pt}}\,\rho_{\scaleto{S}{4pt}}/\overline{a}^{2} \to 0$, it follows that \cref{Coeff_A_Meio_Intervalo,Coeff_B_Meio_Intervalo,Coeff_C_Meio_Intervalo,Coeff_D_Meio_Intervalo,Coeff_E_Meio_Intervalo,Coeff_F_Meio_Intervalo} reduce to the case of \cref{V_eff_QL}. 
	
	Although the term $F$, defined in \eqref{Coeff_F_Meio_Intervalo}, presents a new difficulty to the determination of an analytical solution, we can consider the special case where $F=0$, leading to the constraint
	\begin{equation}
		\epsilon^{2} = \frac{1}{2}\,\overline{a}\,\overline{m}_{\phi}^{2}\,.\label{Vínculo_Energia_Comum}
	\end{equation}
    With this procedure, we can obtain an analytical solution to this particular case\footnote{We emphasize that this constrained case does not encompass the most general solution for the standard scenario of $\alpha_{\scaleto{Q}{5pt}}=1/2$, whose analytical solution lies beyond the purpose of this study.}, where the above constraint simplifies the coefficients \eqref{Coeff_B_Meio_Intervalo} and \eqref{Coeff_E_Meio_Intervalo} to
	\begin{subequations}
		\begin{align}
			\overline{B} &= -\frac{1}{2}\left(1+\frac{N_{\scaleto{Q}{5pt}}\,\rho_{\scaleto{S}{4pt}}}{\overline{a}^{2}}\right) - \frac{\kappa^{2}}{\overline{a}}\left(1 - \frac{2\,N_{\scaleto{Q}{5pt}}\,\rho_{\scaleto{S}{4pt}}}{\overline{a}^{2}}\right)-6\left(\frac{\epsilon\,\rho_{\scaleto{S}{4pt}}}{\overline{a}^{2}}\right)^{2}  \frac{N_{\scaleto{Q}{5pt}}\,\rho_{\scaleto{S}{4pt}}}{\overline{a}^{2}}\,,\label{Coeff_Bbar_Meio_Intervalo}\\
			\overline{E} &= -\left(\frac{\epsilon\,\rho_{\scaleto{S}{4pt}}}{\overline{a}^{2}}\right)^{2}\left(1 + \frac{2\,N_{\scaleto{Q}{5pt}}\,\rho_{\scaleto{S}{4pt}}}{\overline{a}^{2}}\right)\,\label{Coeff_Ebar_Meio_Intervalo}\,,
		\end{align}
	\end{subequations}
	while \eqref{Coeff_A_Meio_Intervalo}, \eqref{Coeff_C_Meio_Intervalo}, and \eqref{Coeff_D_Meio_Intervalo} remain unchanged. 
	
	The next step consists of comparing \cref{Eq-para-u} and the constrained\footnote{The constrained version is the one where $F=0$ so that \cref{Vínculo_Energia_Comum} holds.} version of \cref{V_eff_Meio_Estendido}, which define the ODE for the auxiliary function $u(x)$, with the results \eqref{Eq_u_Normal_Form}, \eqref{Parameters_A_to_E_CHE}, and \eqref{Sol-Geral-Normal-ECH_z-1} regarding the confluent Heun equation. By doing so, the solution for $\alpha_{\scaleto{Q}{5pt}} = 1/2$ in the standard scenario can be expressed as:
	\begin{equation}
		\begin{aligned}
			u_{\scaleto{ST}{4pt}}^{\scaleto{(1/2)}{6pt}}(x) 
			&= e^{\alpha x/2}\,x^{(1+\beta)/2}\,\left(x - 1\right)^{(1 + \gamma)/2}
			\left[c_{1}\,
			\text{HeunC}\left(-\alpha,\,\gamma,\,\beta,\,-\delta,\,\delta+\eta;\,1-x\right) \right.\\
			&\left.+\, c_{2}\,\left(x-1\right)^{-\gamma}
			\text{HeunC}\left(-\alpha,\,-\gamma,\,\beta,\,-\delta,\,\delta+\eta;\,1-x\right)\right] \,.\label{u_ST_Meio}
		\end{aligned}
	\end{equation}
	The associated Heun parameters are:
	\begin{equation}
		\begin{aligned}
			\alpha &= 2 \,\frac{\epsilon \rho_{\scaleto{S}{4pt}}}{\overline{a}^{2}} \,\sqrt{1 + \frac{2\,N_{\scaleto{Q}{5pt}}\,\rho_{\scaleto{S}{4pt}}}{\overline{a}^{2}}}\,,\\
			\beta &= 0\,,\\
			\gamma &= 2i\,\frac{\epsilon \rho_{\scaleto{S}{4pt}}}{\overline{a}^{2}} \,\sqrt{1 - \frac{4\,N_{\scaleto{Q}{5pt}}\,\rho_{\scaleto{S}{4pt}}}{\overline{a}^{2}}}\,,\\
			\delta &= -\frac{N_{\scaleto{Q}{5pt}}\,\rho_{\scaleto{S}{4pt}}}{\overline{a}^{2}} \left(1 - \frac{\kappa^{2}}{\overline{a}} + 6\,\frac{\epsilon^{2} \rho_{\scaleto{S}{4pt}}^{2}}{\overline{a}^{4}}\right)\,,\\
			\eta &= \frac{N_{\scaleto{Q}{5pt}}\,\rho_{\scaleto{S}{4pt}}}{\overline{a}^{2}}\left(\frac{1}{2} + \frac{\kappa^{2}}{\overline{a}}\right) -\frac{\kappa^{2}}{\overline{a}}\,.
		\end{aligned} \label{Heun_Parameters_Meio_Intervalo}
	\end{equation}
	\vspace{1mm}
	
	Then, using the relation between $u_{\scaleto{ST}{4pt}}^{\scaleto{(1/2)}{6pt}}(x)$ and $R_{\scaleto{ST}{4pt}}^{\scaleto{(1/2)}{6pt}}(x)$, determined by \cref{radial-wave-function}, we conclude that:
	\begin{equation}
		\begin{aligned}
			R_{\scaleto{ST}{4pt}}^{\scaleto{(1/2)}{6pt}}(x) 
			&= \left[1 - \frac{1}{2}\frac{N_{\scaleto{Q}{5pt}}\,\rho_{\scaleto{S}{4pt}}}{\overline{a}^{2}}(1+x)\right] e^{\alpha x/2}\left(x - 1\right)^{\gamma/2}
			\left[c_{1}\,
			\text{HeunC}\left(-\alpha,\,\gamma,\,\beta,\,-\delta,\,\delta+\eta;\,1-x\right) \right.\\
			&\left.+\, c_{2}\,\left(x-1\right)^{-\gamma}
			\text{HeunC}\left(-\alpha,\,-\gamma,\,\beta,\,-\delta,\,\delta+\eta;\,1-x\right)\right]\,,
		\end{aligned}
		\label{Radial_WF_ST_Middle}
	\end{equation}\\
	
	\noindent where $A(x)^{-1/2}$, contained in the relation \eqref{radial-wave-function}, was approximated to the first order in $N_{\scaleto{Q}{5pt}}\,\rho_{\scaleto{S}{4pt}}/\overline{a}^{2}$. We emphasize that there is no loss of generality in this procedure, since \cref{A(x)_Metade_Intervalo} was derived precisely for the regime \eqref{Small_Parameter_Condition}. Thus, considering only linear terms in $N_{\scaleto{Q}{5pt}}\,\rho_{\scaleto{S}{4pt}}/\overline{a}^{2}$ in the expression of $A(x)^{-1/2}$ is consistent with our results.
	
	Notably, by imposing the energy constraint of \cref{Vínculo_Energia_Comum} to the solution described by \cref{Heun_Parameters_QL}, we obtain the result of \cref{Heun_Parameters_Meio_Intervalo} in the limit $N_{\scaleto{Q}{5pt}}\,\rho_{\scaleto{S}{4pt}}/\overline{a}^{2} \to 0$. This equivalence arises because, under these constraints, both solutions describe the same physical background.
	
	Finally, it should be clear that for the radial wave function determined by \cref{Radial_WF_ST_Middle}, we shall not impose the constraint for $\delta_{N}$ solutions since we had already used the energy condition \eqref{Vínculo_Energia_Comum} to obtain an analytical solution in terms of the CHE. However, note that the situation will be different in our next case, where we consider the cloudless scenario of $\alpha_{\scaleto{Q}{5pt}} = 1/2$.

	\subsubsection{Middle of the interval: cloudless regime}
	
	To investigate the cloudless scenario associated with $\alpha_{\scaleto{Q}{5pt}}=1/2$ (corresponding to $\omega_{\scaleto{Q}{5pt}} = -2/3$), we must obtain a convenient expression for $A(x)$. To do so, we primarily recall that the event horizon radius in this situation follows from \cref{Constraint_EH_Alpha_Meio}, yielding:
	\begin{equation*}
		\rho_{\scaleto{+}{4pt}} = \left(\frac{\rho_{\scaleto{S}{4pt}}}{N_{\scaleto{Q}{5pt}}}\right)^{1/2}\,.
	\end{equation*}
	Considering the admissible values for $\rho_{\scaleto{S}{4pt}}$ and $N_{\scaleto{Q}{5pt}}$ established in \cref{Sec-2-Results-Paper-One}, the condition $\rho_{\scaleto{+}{4pt}} \ll |l|$ remains valid, because we considered a large universe with $\rho_{\scaleto{obs.}{4pt}} \sim |l| \sim 10^{26}$ m. In this situation, the metric function $A(x)$ simplifies to
	\begin{equation}
		A(x) = \left(N_{\scaleto{Q}{5pt}}\,\rho_{\scaleto{S}{4pt}}\right)^{1/2}\left(x - \frac{1}{x}\right)\,. \label{A(x)_CL_Meio}
	\end{equation}
	Keeping in mind that $N_{\scaleto{Q}{5pt}}$ is on the order of $10^{-26}$ m$^{-1}$ (when $\alpha_{\scaleto{Q}{5pt}} = 1/2$) and that the Schwarzschild radius, $\rho_{\scaleto{S}{4pt}}$, typically varies between $10^{-2}$ m -- $10^{15}$ m, we see that the coefficient $\left(N_{\scaleto{Q}{5pt}}\,\rho_{\scaleto{S}{4pt}}\right)^{1/2}$ should be located between $10^{-14}$ and $10^{-5}$, approximately. The fact that this multiplicative factor is small, yet global, implies that we can explore regions of the extended domain by considering only which power of $x$ is dominant in $\left(x - \tfrac{1}{x}\right)$, for a given region.
	
	The above expression for $A(x)$ indicates that, unless $x$ is small, a good approximation for \cref{A(x)_CL_Meio} is
	\begin{equation}
		A(x) = \left(N_{\scaleto{Q}{5pt}}\,\rho_{\scaleto{S}{4pt}}\right)^{1/2} \, x\,. \label{metric-cloudless-meio-intervalo-approx}
	\end{equation}
	And since we have already obtained the solution for the radial wave function when $x\to 1^{+}$, which is determined by \cref{RWF_NH,Heun_Parameters_RWF_NH_All_Alpha_Q} with $\beta_{\scaleto{+}{4pt}} = 2\left(N_{\scaleto{Q}{5pt}}\,\rho_{\scaleto{S}{4pt}}\right)^{1/2}$ and $\rho_{\scaleto{+}{4pt}} = \left(\rho_{\scaleto{S}{4pt}}/N_{\scaleto{Q}{5pt}}\right)^{1/2}$, we now extend our analysis to the region where \cref{metric-cloudless-meio-intervalo-approx} is valid. We also emphasize that, as $x$ increases, this approximation becomes increasingly better. As an example, we could say that $x>5$ would constitute a reasonable domain to which \cref{metric-cloudless-meio-intervalo-approx} is valid.
	
	Then, by substituting the metric function \eqref{metric-cloudless-meio-intervalo-approx} into the general expression of $\rho_{\scaleto{+}{4pt}}^{2}\, V_{\text{eff}}(x)$, given by \cref{Potential_in_terms_of_x}, we obtain an effective potential of the form
	\begin{equation}
		\rho_{\scaleto{+}{4pt}}^{2}\,V_{\text{eff}}(x) = -\frac{\overline{m}_{\phi}^{2}\,\rho_{\scaleto{S}{4pt}}^{1/2}}{N_{\scaleto{Q}{5pt}}^{3/2}}\,\frac{1}{x} 
		+ \left(\frac{\epsilon^{2}}{N_{\scaleto{Q}{5pt}}^{2}} - \frac{3}{4}\right)\frac{1}{x^{2}} 
		- \frac{\kappa^{2}}{\left(N_{\scaleto{Q}{5pt}}\,\rho_{\scaleto{S}{4pt}}\right)^{1/2}}\,\frac{1}{x^{3}}\,.\label{Potential-Cloudless-Middle}
	\end{equation}
	
	Observe that each term in \cref{Potential-Cloudless-Middle} is proportional to a power of $N_{\scaleto{Q}{5pt}}^{-1}$, which is on the order of $10^{26}$ m. While the coefficient of the second term is certainly of greater magnitude (since it depends on $N_{\scaleto{Q}{5pt}}^{-2}$),  we should investigate the relative magnitudes of the coefficients accompanying $x^{-3}$ and $x^{-1}$. To do so, we calculate the ratio between them:
	\begin{equation}
		\frac{\kappa^{2}/\left(N_{\scaleto{Q}{5pt}}\,\rho_{\scaleto{S}{4pt}}\right)^{1/2}}{\overline{m}_{\phi}^{2}\,\rho_{\scaleto{S}{4pt}}^{1/2}/N_{\scaleto{Q}{5pt}}^{3/2}} = \frac{N_{\scaleto{Q}{5pt}}}{\rho_{\scaleto{S}{4pt}}}\,\left(\frac{\kappa^{2}}{\overline{m}_{\phi}^{2}}\right)\,,\label{CL-potential-ratio}
	\end{equation}
	where $N_{\scaleto{Q}{5pt}}/\rho_{\scaleto{S}{4pt}}$ typically ranges from $10^{-24}$ m$^{-2}$ to $10^{-41}$ m$^{-2}$. Since $\kappa^{2} = n_{\scaleto{R}{4pt}}^{2} + k_{z}^{2}$, with $n_{\scaleto{R}{4pt}} \in \mathbb{Z}$ and $k_{z}^{2} \propto \overline{m}_{\phi}^{2}$, the contribution of the term $x^{-3}$ in \eqref{Potential-Cloudless-Middle} is insignificant when compared to the term of $x^{-1}$. Consequently, we can approximate the equation of $u_{\scaleto{CL}{4pt}}^{\scaleto{(1/2)}{6pt}}(x)$, considering only the first two terms of the potential, so that:
	\begin{equation*}
		\frac{d^{2}}{d\,x^{2}}\,u_{\scaleto{CL}{4pt}}^{\scaleto{(1/2)}{6pt}}(x) + \left[-\frac{\overline{m}_{\phi}^{2}\,\rho_{\scaleto{S}{4pt}}^{1/2}}{N_{\scaleto{Q}{5pt}}^{3/2}}\,\frac{1}{x} 
		+ \left(\frac{\epsilon^{2}}{N_{\scaleto{Q}{5pt}}^{2}} - \frac{3}{4}\right)\frac{1}{x^{2}}\right]u_{\scaleto{CL}{4pt}}^{\scaleto{(1/2)}{6pt}}(x) = 0\,.
	\end{equation*}
	
	Although we can determine the solution in terms of the CHE, in this particular case it is more convenient to describe $u_{\scaleto{CL}{4pt}}^{\scaleto{(1/2)}{6pt}}(x)$ in terms of Bessel functions \cite{Olver2010,Butkov1968,Abramowitz1968}. By doing so, the solution takes the form:
	\begin{equation}
		u_{\scaleto{CL}{4pt}}^{\scaleto{(1/2)}{6pt}}(x) = x^{1/2}\left[c_{1}\,\text{I}_{s}\left(2\sqrt{b\,x}\right) + c_{2}\,\text{K}_{s}\left(2\sqrt{b\,x}\right)\right]\,,
	\end{equation}
	where $\text{I}_{s}(x)$ and $\text{K}_{s}(x)$ are, respectively, the modified Bessel functions of the first and second kinds, of order $s$. Furthermore, the scaling factor $b$ and the order $s$ are determined by:
	\begin{align}
		b 
		&= \frac{\overline{m}_{\phi}^{2}\,\rho_{\scaleto{S}{4pt}}^{1/2}}{N_{\scaleto{Q}{5pt}}^{3/2}}\,,\label{def-b}\\
		s
		&= 2i\,\sqrt{\frac{\epsilon^{2}}{N_{\scaleto{Q}{5pt}}^{2}} - 1}\,.\label{def-s}
	\end{align}
	
	Note that, for $N_{\scaleto{Q}{5pt}}\neq 0$, the parameter $b$ is a positive real number, while $s$ is purely imaginary, except when $\epsilon = 0$. Therefore, according to the relation \eqref{radial-wave-function}, the radial solution of the Klein-Gordon equation in the cloudless scenario (\textsc{cl}) with $\alpha_{\scaleto{Q}{5pt}}=1/2$ takes the form of
	\begin{equation}
		R_{\scaleto{CL}{4pt}}^{\scaleto{(1/2)}{6pt}}(x) = x^{-1}\left[c_{1}\,\text{I}_{s}\left(2\sqrt{b\,x}\right) + c_{2}\,\text{K}_{s}\left(2\sqrt{b\,x}\right)\right]\,.\label{RWF_Middle_Bessel}
	\end{equation}
	Since the term associated with $\text{I}_{s}\left(2\sqrt{b\,x}\right)$ diverges as $x\to\infty$, it is possible to set $c_{1}=0$ to select a well-behaved solution. However, this condition is not strictly necessary, as the domain of the solution in Eq. \eqref{RWF_Middle_Bessel} is bounded above by the requirement $x^{2}\,\rho_{\scaleto{+}{4pt}}^{2}\ll l^{2}$.
	
	Subsequently, we will examine additional cases under specific constraints, considering the effective potential \eqref{Potential-Cloudless-Middle}.

	\subsubsection*{Additional Constrained Cases:}

	The first particular case is the one of a massless scalar field ($\overline{m}_{\phi} = 0$), where the dominant term in the effective potential \eqref{Potential-Cloudless-Middle} is given by:
	\begin{equation*}
		\rho_{\scaleto{+}{4pt}}^{2}\,V_{\text{eff}}(x) = \left(\frac{\epsilon^{2}}{N_{\scaleto{Q}{5pt}}^{2}} - \frac{3}{4}\right)\frac{1}{x^{2}}\,.
	\end{equation*}
	Once again, we could write the solution in terms of the CHE, but it is simpler to solve the associated differential equation as a Cauchy-Euler case, from which it follows that
	\begin{equation}
		u_{\scaleto{CL}{4pt}}^{\scaleto{(1/2,\,\triangle)}{6pt}}(x) = x^{1/2} \left[c_{1}\,x^{s/2} + c_{2}\, x^{-s/2}\right]\,,\label{U_WF_CL_Middle_Extra_1}
	\end{equation}
	where $s$ is defined by \cref{def-s}. Notably, since $s$ is purely imaginary, the real part of this solution behaves as $\text{Re}\left[u_{\scaleto{CL}{4pt}}^{\scaleto{(1/2,\,\triangle)}{6pt}}(x)\right] \propto x^{1/2}\,\cos \left(\tfrac{s}{2}\,\ln x\right)$ (if setting $c_{1}$ or $c_{2}$ to zero, for simplicity).
	
	In our second particular case, where $\epsilon^{2} = 3\,N_{\scaleto{Q}{5pt}}^{2}/4$, with $\overline{m}_{\phi}\neq 0$, the effective potential \eqref{Potential-Cloudless-Middle} reduces to:
	\begin{equation*}
		\rho_{\scaleto{+}{4pt}}^{2}\,V_{\text{eff}}(x) = -\frac{\overline{m}_{\phi}^{2}\,\rho_{\scaleto{S}{4pt}}^{1/2}}{N_{\scaleto{Q}{5pt}}^{3/2}}\,\frac{1}{x}\,, \label{Cloudless_V_eff_Restricted}
	\end{equation*}
	implying that the solution to \cref{Eq-para-u} is:
	\begin{equation}
		u_{\scaleto{CL}{4pt}}^{\scaleto{(1/2,\,\square)}{6pt}}(x) =  \sqrt{b\, x}\left[c_{1}\, \text{I}_{1} \left(2\,\sqrt{b\, x}\right) + c_{2}\, \text{K}_{1} \left(2\,\sqrt{b\, x}\right)\right]\,,\label{U_WF_CL_Middle_Extra_2}
	\end{equation}
	where the factor $b$ is defined by \cref{def-b}. 
	
	The corresponding radial solutions for these two special cases, namely  \cref{U_WF_CL_Middle_Extra_1,U_WF_CL_Middle_Extra_2}, are obtained via:
	\begin{equation}
		R_{\scaleto{CL}{4pt}}^{\scaleto{(1/2,\,\triangle/\square)}{6pt}}(x) = x^{-1}\,u_{\scaleto{CL}{4pt}}^{\scaleto{(1/2,\,\triangle/\square)}{6pt}}(x)\,.\label{Radial_WF_CL_Middle_Extras}
	\end{equation}
	We emphasize that the constants $c_{1}$ and $c_{2}$ absorbed any normalization constant $R_{0}$ in \cref{Radial_WF_CL_Middle_Extras}.
	
	The last remaining scenario, with $\alpha_{\scaleto{Q}{5pt}} = 1/2$, involves removing the \emph{black string}, that is, a scenario where $\rho_{\scaleto{S}{4pt}} = 0$. This will be explored in the next subsection.

	\subsubsection{Intermediate Case: Absence of the black string} \label{HL-1/2}
	
	We now examine the particle’s radial solution in the absence of a black string, that is, with $\rho_{\scaleto{S}{4pt}} = 0$. We keep the state parameter fixed at $\alpha_{\scaleto{Q}{5pt}} = 1/2$ and the domain condition $\rho \ll |l|$. In practice, this system models a central mass distribution (due to the cloud of \emph{strings}) in the presence of an anisotropic quintessence fluid, but without an event horizon. And since $\rho_{\scaleto{S}{4pt}}=0$, we use the radial coordinate $\rho$ instead of $x=\rho/\rho_{\scaleto{+}{4pt}}$ in this scenario.
	
	Under these assumptions, the metric function $A(\rho)$ becomes:
	\begin{equation}
		A(\rho) = \overline{a} \left(1 + \frac{N_{\scaleto{Q}{5pt}}}{\overline{a}}\,\rho\right)\,.\label{A_HL_Meio_Intervalo}
	\end{equation}
	Recall that, in this regime, the physical values of $N_{\scaleto{Q}{5pt}}/\overline{a}$ typically range from $10^{-20}$ m$^{-1}$ to $10^{-27}$ m$^{-1}$. As a consequence, we only retain first-order approximations in $N_{\scaleto{Q}{5pt}}/\overline{a}$ when deriving the explicit effective potential of \cref{Effective-Potential}. This approach mirrors the procedure in \cref{Extended_Solutions_Alpha_Q_One_Half}, although with a simplified form in the present case. The resulting effective potential is given by
	\begin{equation}
		V_{\text{eff}}(\rho) = \frac{\epsilon^{2}}{\overline{a}^{2}}\left(1 - \frac{2\,N_{\scaleto{Q}{5pt}}}{\overline{a}}\,\rho\right) - \frac{N_{\scaleto{Q}{5pt}}/\overline{a}}{\rho} - \left(\frac{\kappa^{2}/\overline{a}}{\rho^{2}} + \frac{\overline{m}_{\phi}^{2}}{\overline{a}}\right)\left(1 - \frac{N_{\scaleto{Q}{5pt}}}{\overline{a}}\,\rho\right)\,.\label{Potential_Horizonless_Meio_Intervalo}
	\end{equation}
	
    The metric function $A(\rho)$, satisfying \cref{A_HL_Meio_Intervalo}, represents a cloud of strings--dominated regime, with a minor perturbation introduced by the presence of quin\-tes\-sence. 
    This interpretation extends to the effective potential \eqref{Potential_Horizonless_Meio_Intervalo}, which decomposes into terms associated with the cloud of strings and a small quintessence-\-induced correction. 
    
    By rearranging the terms in \cref{Potential_Horizonless_Meio_Intervalo} according to powers of $\rho$, we obtain:
	\begin{equation}
		V_{\text{eff}}(\rho) = -\frac{\kappa^{2}/\overline{a}}{\rho^{2}} + \left(\frac{\kappa^{2}}{\overline{a}} - 1\right) \frac{N_{\scaleto{Q}{5pt}}/\overline{a}}{\rho} + \frac{\epsilon^{2} - \overline{a}\,\overline{m}_{\phi}^{2}}{\overline{a}^{2}} - \left(\epsilon^{2} - \frac{1}{2}\, \overline{a}\, \overline{m}_{\phi}^{2} \right) \frac{2 N_{\scaleto{Q}{5pt}}}{\overline{a}^{3}}\, \rho\,.\label{V_eff_HL_Meio}
	\end{equation}

	To solve the differential equation \eqref{eq-geral-u} for $u_{\scaleto{HL}{4pt}}^{\scaleto{(1/2)}{6pt}}(\rho)$ using the effective potential above, we apply the results from \cref{Eq_u_Normal_Form,Parameters_A_to_E_CHE}. This leads to a result expressed in terms of the solutions of the CHE, provided that 
	\begin{equation*}
		\epsilon^{2} = \frac{1}{2}\,\overline{a}\, \overline{m}_{\phi}^{2}\,,
	\end{equation*}
    which is identical to the constraint\footnote{Relative to the standard scenario of $\alpha_{\scaleto{Q}{5pt}} = 1/2$.} of \cref{Vínculo_Energia_Comum}. Under this condition, the solution for $u_{\scaleto{HL}{4pt}}^{\scaleto{(1/2)}{6pt}}(\rho)$ takes the form of
	\begin{equation*}
		\begin{aligned}
			u_{\scaleto{HL}{4pt}}^{\scaleto{(1/2)}{6pt}}(\rho) &= e^{\alpha \rho/2}\,\rho^{(1+\beta)/2}\,\left(\rho - 1\right)^{(1 + \gamma)/2}
			\left[ c_{1}
			\text{HeunC}\left(\alpha,\,\beta,\,\gamma,\,\delta,\,\eta;\,\rho\right) \right.\\
			&\left.+\, c_{2} \, \rho^{-\beta}\,
			\text{HeunC}\left(\alpha,\,-\beta,\,\gamma,\,\delta,\,\eta;\,\rho\right)\right] \,,
		\end{aligned}
	\end{equation*}
    while the Heun parameters are:
	\begin{equation}
		\begin{aligned}
			\alpha &= \sqrt{\frac{2\,\overline{m}_{\phi}^{2}}{\overline{a}}}\,,\\
			\beta &= 2\,\sqrt{\frac{\kappa^{2}}{\overline{a}} + \frac{1}{4}}\,,\\
			\gamma &= 1\,,\\
			\delta &= \frac{N_{\scaleto{Q}{5pt}}}{\overline{a}} \left(\frac{\kappa^{2}}{\overline{a}} - 1\right)\,,\\
			\eta &= \frac{1}{2} - \frac{N_{\scaleto{Q}{5pt}}}{\overline{a}} \left(\frac{\kappa^{2}}{\overline{a}} - 1\right)\,.
		\end{aligned} \label{Heun_Parameters_Middle_Alpha_Q}
	\end{equation}
	
    The radial solution $R_{\scaleto{HL}{4pt}}^{\scaleto{(1/2)}{6pt}}(\rho)$ follows from the relation \eqref{radial-wave-function}, considering the result for $u_{\scaleto{HL}{4pt}}^{\scaleto{(1/2)}{6pt}}(\rho)$ and the metric function of \cref{A_HL_Meio_Intervalo}. Therefore,
	\begin{equation*}
		\begin{aligned}
			R_{\scaleto{HL}{4pt}}^{\scaleto{(1/2)}{6pt}}(\rho) &= \left(1 - \frac{N_{\scaleto{Q}{5pt}}}{2 \overline{a}}\,\rho\right) e^{\alpha \rho/2}\,\rho^{(\beta - 1)/2}\,\left(\rho - 1\right)^{(1 + \gamma)/2}
			\left[ c_{1}
			\text{HeunC}\left(\alpha,\,\beta,\,\gamma,\,\delta,\,\eta;\,\rho\right) \right.\\
			&\left.+\, c_{2} \, \rho^{-\beta}\,
			\text{HeunC}\left(\alpha,\,-\beta,\,\gamma,\,\delta,\,\eta;\,\rho\right)\right] \,.
		\end{aligned}
	\end{equation*}
    To guarantee that $R_{\scaleto{HL}{4pt}}^{\scaleto{(1/2)}{6pt}}(\rho)$ is regular at the origin, we fix $c_{2} = 0$. Then, 
    \begin{equation}
			R_{\scaleto{HL}{4pt}}^{\scaleto{(1/2)}{6pt}}(\rho) = \overline{R_{0}}\,\rho^{\beta/2}\,\rho^{-1/2}\,e^{\alpha\rho/2}\left(1 - \frac{N_{\scaleto{Q}{5pt}}}{2 \overline{a}}\,\rho\right)\left(\rho - 1\right) \text{HeunC}\left(\alpha,\,\beta,\,\gamma,\,\delta,\,\eta;\,\rho\right) \,,\label{Radial_WF_Intermediate_HL}
	\end{equation}
    where, for simplification, we had used that $\gamma=1$, according to \cref{Heun_Parameters_Middle_Alpha_Q}, and that $\overline{R_{0}}$ is the overall normalization constant which had absorbed $c_{1}$. Since $\beta/2 \geq 1/2$, $R_{\scaleto{HL}{4pt}}^{\scaleto{(1/2)}{6pt}}(0) = 0$ when $\beta/2 \neq 1/2$. Conversely, if $\beta/2 = 1/2$ (that is, $\kappa = 0$), then $R_{\scaleto{HL}{4pt}}^{\scaleto{(1/2)}{6pt}}(0) = -\overline{R_{0}}$.
	
	On the other hand, when there is no quintessence ($N_{\scaleto{Q}{5pt}} = 0$), the effective potential \eqref{Potential_Horizonless_Meio_Intervalo} reduces to
	\begin{equation}
		V_{\text{eff}}(\rho) = \frac{\epsilon^{2} - \overline{a}\,\overline{m}_{\phi}^{2}}{\overline{a}^{2}} - \frac{\kappa^{2}/\overline{a}}{\rho^{2}}\,,\label{Potential_String_Clouds_Alone}
	\end{equation}
	regardless the value of $\epsilon^{2}$. Accordingly, if the only matter component in the Universe is the cloud of strings, we solve \cref{eq-geral-u} for the auxiliary function $u_{\scaleto{HL+QF}{4pt}}(\rho)$ under the effective potential of \cref{Potential_String_Clouds_Alone}, and, after employing the relation \eqref{radial-wave-function}, we obtain that the radial solution $R_{\scaleto{HL+QF}{4pt}}(\rho)$ for the spin$-0$ particle is given by
	\begin{equation}
		R_{\scaleto{HL+QF}{4pt}}(\rho) = \rho^{-1/2}\left[c_{1}\,\text{J}_{n}\left(\sqrt{\frac{\epsilon^{2} - \overline{a}\,\overline{m}_{\phi}^{2}}{\overline{a}^{2}}}\,\rho\right) + c_{2}\,\text{Y}_{n}\left(\sqrt{\frac{\epsilon^{2} - \overline{a}\,\overline{m}_{\phi}^{2}}{\overline{a}^{2}}}\,\rho\right)\right]\,,\label{Radial_WF_HL_F}
	\end{equation}
	with
	\begin{equation}
		n = \sqrt{\frac{1}{4} + \frac{\kappa^{2}}{\overline{a}}}\,,\label{order-n}
	\end{equation}
	where $\text{J}_{n}(\rho)$ and $\text{Y}_{n}(\rho)$ are the Bessel functions of the first and second kinds, respectively. Here, the subscript \enquote{\textsc{hl+qf}} labels the scenario with neither the black string nor the quintessence fluid. Additionally, for solutions near the origin ($\rho\sim 0$), we set $c_{2}=0$ to ensure regularity. At last, note that the result for $R_{\scaleto{HL+QF}{4pt}}(\rho)$ is independent of $\alpha_{\scaleto{Q}{5pt}}$, as expected, given the absence of quintessence.
	
	In the last particular scenario, defined by $\overline{a}\neq 0$, $\rho_{\scaleto{S}{4pt}} = 0$, $N_{\scaleto{Q}{5pt}} = 0$, and $\epsilon^{2} = \overline{a}\,\overline{m}_{\phi}^{2}$, the effective potential of \cref{Potential_String_Clouds_Alone} simplifies to a single term proportional to  $\rho^{-2}$. Under these conditions, \cref{Eq-para-u} reduces to a Cauchy-Euler equation. By solving this ODE for $u_{\scaleto{HL+QF}{4pt}}^{\,\diamond}(\rho)$ and employing the relation \eqref{radial-wave-function}, we obtain the radial solution:
	\begin{equation}
		R_{\scaleto{HL+QF}{4pt}}^{\,\diamond}(\rho) = c_{1}\, \rho^{-1/2 + n} + c_{2}\, \rho^{-1/2 - n}\,, \label{Radial_WF_HL_F_Losango}
	\end{equation}	
	where $n \in \mathbb{R}$ is defined by \cref{order-n} and satisfies $n \geq 1/2$. This ensures the radial wave function satisfies $R_{\scaleto{HL+QF}{4pt}}^{\,\diamond}(0) = 0$ only  if $c_{2} = 0$. Therefore, to describe particles near the origin\footnote{As we have already discussed, the origin is located at $\rho = 0$, where the center of the \emph{black string} would be if it were not removed. Around this center, the cloud of \emph{strings} and the anisotropic fluid of quintessence are distributed.}, we set $c_{2} = 0$. Moreover, due to the domain restriction where $\rho \ll |l|$, we need not set $c_{1}=0$ for particles which are distant from the origin. Finally, we observe that the symbol $\diamond$ is used to distinguish this particular case of \enquote{\textsc{hl+qf}} characterized by the defined spectrum $\epsilon^{2} = \overline{a}\,\overline{m}_{\phi}^{2}$. 
	
	After deriving the radial solutions \cref{Radial_WF_ST_Middle,Radial_WF_CL_Middle_Extras,Radial_WF_Intermediate_HL,Radial_WF_HL_F,Radial_WF_HL_F_Losango}, for the specific case of $\alpha_{\scaleto{Q}{5pt}} = 1/2$, and established their validity over a broad range of the radial coordinates, we now proceed to investigate the upper bound of $\alpha_{\scaleto{Q}{5pt}} = 1$ in the following section. The comprehensive set of wave functions obtained in these analysis will provide valuable insights into the behavior of solutions for other values of $\alpha_{\scaleto{Q}{5pt}}$.

	\subsection{Solutions for the upper bound} \label{Extended_Solutions_Alpha_Q_1}
	
	This section explores solutions to the Klein-Gordon radial equation, defined by \cref{general-equation-for-R}, for the limiting case of $\alpha_{\scaleto{Q}{5pt}}=1$ (i.e., $\omega_{\scaleto{Q}{5pt}} = -1$). We adopt here the same methodology used in the previous analyses, which consists of three sequential steps: first, we determine the appropriate form for $A(\rho)$ (or $A(x)$) and derive the effective potential for each of the scenarios; then, we solve the ODE for the auxiliary function $u(\rho)$ (or $u(x)$); finally, we use the relation \eqref{radial-wave-function}, from which we obtain the associated radial solution.
	
	Again, we investigate the behavior of these solutions in three distinct scenarios: the standard case (\textsc{st}), the cloudless case (\textsc{cl}), and the horizonless case (\textsc{hl}). For each scenario, we obtain the radial solution, which can be interpreted as the scalar particle's wave function, using either Bessel or Heun equations, providing results on the dynamics of spin--$0$ particles in this spacetime.
	
	To obtain the radial solution, firstly in the standard scenario, we rewrite the exact expression of $A(\rho)$, given in \cref{Function_A(rho)}, in terms of the dimensionless variable $x$, to show that:
	\begin{equation}
		A(\rho) = \overline{a} - \frac{\rho_{\scaleto{S}{4pt}}}{\rho_{\scaleto{+}{4pt}}}\frac{1}{x} + N_{\scaleto{QL}{5pt}}\,\rho_{\scaleto{+}{4pt}}^{2} x^{2}\,,\label{A(x)_Upper_Bound}
	\end{equation}
	where
	\begin{equation}
		N_{\scaleto{QL}{5pt}} = N_{\scaleto{Q}{5pt}} + 1/l^{2}\,.\label{N_QL_Definition}
	\end{equation}
	It is important to highlight that, in this limiting case of $\alpha_{\scaleto{Q}{5pt}} = 1$, the contributions from quintessence and the cosmological constant are combined in $A(\rho)$ via the definition of $N_{\scaleto{QL}{5pt}}$. This happens because when $\alpha_{\scaleto{Q}{5pt}}=1$ both the cosmological constant and the quintessence terms depend on $\rho^{2}$, implying it is not necessary to constrain the radial domain to handle the term of $\rho^{2}/l^{2}$ in \cref{Function_A(rho)}. Consequently, the solutions associated with $\alpha_{\scaleto{Q}{5pt}}=1$ will be, \textit{a priori}, valid throughout the whole coordinate domain, that is, $\rho \in [0,\,\infty)$.
	
	As shown in our previous work \cite{Deglmann2025}, when $\alpha_{\scaleto{Q}{5pt}}=1$, with $\overline{a}\neq 0$, $\rho_{\scaleto{S}{4pt}}\neq 0$, and $N_{\scaleto{QL}{5pt}} \neq 0$, the exact solution to the event horizon radius ($\rho_{\scaleto{+}{4pt}}$) is given by:
	\begin{equation}
		\rho_{\scaleto{+}{4pt}} = \Delta^{1/3} - \frac{\overline{a}}{3N_{\scaleto{QL}{5pt}}}\,\Delta^{-1/3}\,,\label{EH_Alpha_Q_1_Complete}
	\end{equation}
	where
	\begin{equation*}
		\Delta = \left(\frac{\overline{a}}{3\,N_{\scaleto{QL}{5pt}}}\right)^{3/2}
		\left[\left(1 + \frac{27}{4}\,\frac{N_{\scaleto{QL}{5pt}}\,\rho_{\scaleto{S}{4pt}}^{2}}{\overline{a}^{3}}\right)^{1/2} + \left(\frac{27}{4}\,\frac{N_{\scaleto{QL}{5pt}}\,\rho_{\scaleto{S}{4pt}}^{2}}{\overline{a}^{3}}\right)^{1/2}\right]\,. \label{Delta}
	\end{equation*}
	Hence, considering the plausible regime where
	\begin{equation}
		\frac{N_{\scaleto{QL}{5pt}}\,\rho_{\scaleto{S}{4pt}}^{2}}{\overline{a}^{3}} \ll 1\,,\label{plausible-regime-upper-bound}
	\end{equation}
	which follows from the estimates for the physical parameters $N_{\scaleto{QL}{5pt}}$, $\rho_{\scaleto{S}{4pt}}$, and $\overline{a}$, the event horizon radius given by \cref{EH_Alpha_Q_1_Complete} reduces\footnote{In accordance with the result of Table \ref{table-EH-cases}.} to
	\begin{equation*}
		\rho_{\scaleto{+}{4pt}} = \frac{\rho_{\scaleto{S}{4pt}}}{\overline{a}}\,,\label{EH_Alpha_Q_1}
	\end{equation*}
	implying that
	\begin{equation}
		N_{\scaleto{QL}{5pt}}\,\rho_{\scaleto{+}{4pt}}^{2} = \frac{N_{\scaleto{QL}{5pt}}\rho_{\scaleto{S}{4pt}}^{2}}{\overline{a}^{2}}\,.\label{NQL_R_2_a_3}
	\end{equation}

	Substituting the above results into \cref{A(x)_Upper_Bound}, we obtain that \cref{A(x)_Upper_Bound} can be written as
	\begin{equation}
		A(x) = \overline{a} \left(1 - \frac{1}{x} + \frac{N_{\scaleto{QL}{5pt}}\rho_{\scaleto{S}{4pt}}^{2}}{\overline{a}^{3}}\,x^{2}\right)\,.\label{A(x)_Upper_Bound_Complete}
	\end{equation}
	The result of \cref{A(x)_Upper_Bound_Complete} can be seen as a first-order correction to the case where quintessence is absent, given by \cref{A(x)_QL}. For the sake of clarity, we recall that $\overline{a}$ would typically range from $10^{-6}$ to $10^{1}$, with $\rho_{\scaleto{S}{4pt}}$ taking values from $10^{-2}$ to $10^{15}$ m, while $N_{\scaleto{QL}{5pt}}$ would be on the order of $10^{-52}$ m$^{-2}$, so that $N_{\scaleto{QL}{5pt}}\,\rho_{\scaleto{S}{4pt}}^{2}/\overline{a}^{3}$ would be in the range $(10^{-60},\,10^{-4})$, satisfying the constraint of \cref{NQL_R_2_a_3}.
	
	In principle, we could compute the effective potential $\rho_{\scaleto{+}{4pt}}^{2}\,V_{\text{eff}}(x)$ and search analytical solutions of $R_{\scaleto{ST}{4pt}}^{\scaleto{\,(1)}{6pt}}(x)$, with $A(x)$ in the form of \cref{A(x)_Upper_Bound_Complete}. However, the resulting ODE for the auxiliary function $u_{\scaleto{ST}{4pt}}^{\scaleto{(1)}{6pt}}(x)$ does not present known analytical solutions, even if we explore additional restrictions on $\epsilon$ and $\overline{m}_{\phi}$. Therefore, we solve the problem by dividing it into two disjoint regions: the first one being near the event horizon ($x\to 1^{+}$) and the second satisfying $x\gg \overline{a}/\left(N_{\scaleto{QL}{5pt}}\,\rho_{\scaleto{S}{4pt}}^{2}\right)^{1/3}$ so that the term $1/x$ contained in \cref{A(x)_Upper_Bound_Complete} can be neglected in comparison to the quadratic term.
	
	Moreover, since we had already obtained the solution near the event horizon, in eqs. \eqref{RWF_1_NH} and \eqref{eq-DP_1_NH}, we will now investigate the case when the scalar particle is farther away from $x=1$.

	\subsubsection{Far from the event horizon}

	For larger values of $x$, away from the event horizon (located at $x=1$), we can simplify the expression of $A(x)$, given by \cref{A(x)_Upper_Bound_Complete}, to determine the radial solution in this regime. As $x$ increases, the term $1/x$ decreases, implying we can approximate $A(x)$ as
	\begin{equation}
		A(x) = \overline{a} \left(1 + \frac{N_{\scaleto{QL}{5pt}}\rho_{\scaleto{S}{4pt}}^{2}}{\overline{a}^{3}}\,x^{2}\right)\,,\label{A(x)_Upper_Bound_Far}
	\end{equation}
	provided that 
	\begin{equation}
		x \gg \frac{\overline{a}}{\left(N_{\scaleto{QL}{5pt}}\,\rho_{\scaleto{S}{4pt}}^{2}\right)^{1/3}}\,.\label{Região_ST_1_Far_from_EH}
	\end{equation}
	
	For example, let us consider that: $N_{\scaleto{QL}{5pt}} = 6.9 \times 10^{-53}$ m$^{-2}$, $\rho_{\scaleto{S}{4pt}} = 10^{15}$ m and $\overline{a} = 10^{-6}$. With this choice of parameters, $\overline{a}/\left(N_{\scaleto{QL}{5pt}}\,\rho_{\scaleto{S}{4pt}}^{2}\right)^{1/3} \approx 24.4$, implying the approximation \eqref{A(x)_Upper_Bound_Far} would be excellent for $x\gg 24.4$. Notably, with this same configuration, $\rho_{\scaleto{+}{4pt}} \sim 10^{21}$ m, making this solution suitable for particles which are extremely distant from the center of the \emph{black string} (positioned at the origin of spatial coordinates), since the radius of the event horizon itself is large. In any case, this estimate does not diminish the importance of the analytical solution.

\begin{figure}[ht]
		\begin{center}
			\includegraphics[width=0.65\textwidth]{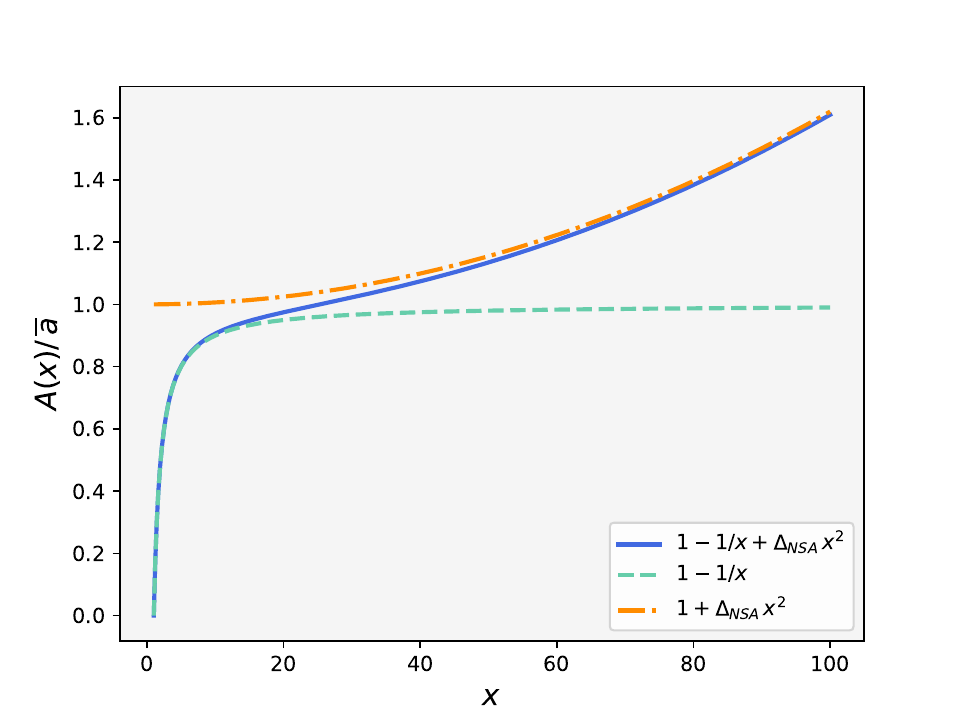}
			\caption{Comparison of $A(x)/\overline{a}$ for $x \in [1, 100]$, with a fixed value of $\Delta_{\scaleto{NSA}{4pt}} = N_{\scaleto{QL}{5pt}}\,\rho_{\scaleto{S}{4pt}}^{2}/\overline{a}^{3} = 6.2 \times 10^{-5}$. The solid line shows the exact case of \cref{A(x)_Upper_Bound_Complete}, the dashed line represents the case without quintessence, and the dash-dotted line neglects the $1/x$ term. This comparison highlights the impact of quintessence and the $1/x$ term on the behavior of $A(x)/\overline{a}$. We stress that, as the value of $N_{\scaleto{QL}{5pt}}\,\rho_{\scaleto{S}{4pt}}^{2}/\overline{a}^{3}$ decreases, it becomes more difficult to distinguish these lines.}
			\label{Plot_comparative_metric}
		\end{center}
	\end{figure}

	In Fig. \ref{Plot_comparative_metric}, we show a comparison between the exact expression of \cref{A(x)_Upper_Bound_Complete}, the approximation of \cref{A(x)_Upper_Bound_Far}, and the quintessence-free case given by \cref{A(x)_QL}. For this figure, we choose $N_{\scaleto{QL}{5pt}}\,\rho_{\scaleto{S}{4pt}}^{2}/\overline{a}^{3} = 6.9 \times 10^{-5}$, since we used the same values of $N_{\scaleto{QL}{5pt}}$, $\rho_{\scaleto{S}{4pt}}$, and $\overline{a}$ from the previous estimate (referring to \cref{Região_ST_1_Far_from_EH}). In the comparative plot, we show the variable $x$ taking values in the interval $[1,\,100]$ and highlight that, as $N_{\scaleto{QL}{5pt}}\,\rho_{\scaleto{S}{4pt}}^{2}/\overline{a}^{3}$ decreases, larger values of $x$ are required to observe a noticeable difference between the approximation of \cref{A(x)_Upper_Bound_Far} and the exact solution of \cref{A(x)_Upper_Bound_Complete}.
	Regarding Fig. \ref{Plot_comparative_metric}, it is evident that, for smaller values of $x$, such as $1<x<15$, it is reasonable to use the radial solution \eqref{RWF_QL}, which considers the term $1/x$.
	
	The effective potential $\rho_{\scaleto{+}{4pt}}^{2}\,V_{\text{eff}}(x)$, defined by \cref{Potential_in_terms_of_x}, following from the expression \eqref{A(x)_Upper_Bound_Far}, takes the form of
	\begin{equation*}
		\begin{aligned}
			\rho_{\scaleto{+}{4pt}}^{2}\,V_{\text{eff}}(x) &= \left(\frac{\epsilon\rho_{\scaleto{S}{4pt}}}{\overline{a}^{2}}\right)^{2} \left(1 - \frac{2 N_{\scaleto{QL}{5pt}}\rho_{\scaleto{S}{4pt}}^{2}}{\overline{a}^{3}}\,x^{2}\right)\\
			&- \frac{3 N_{\scaleto{QL}{5pt}}\rho_{\scaleto{S}{4pt}}^{2}}{\overline{a}^{3}} - \left(\frac{\kappa^{2}/\overline{a}}{x^{2}} + \frac{\rho_{\scaleto{S}{4pt}}^{2} \,\overline{m}_{\phi}^{2}}{\overline{a}^{3}}\right) \left(1 - \frac{N_{\scaleto{QL}{5pt}}\rho_{\scaleto{S}{4pt}}^{2}}{\overline{a}^{3}}\,x^{2}\right)\,.
		\end{aligned}
	\end{equation*}
	We can rearrange the above expression in terms of powers of $x$ to see that
	\begin{equation}
		\begin{aligned}
			\rho_{\scaleto{+}{4pt}}^{2}\,V_{\text{eff}}(x) 
			&= - \frac{\kappa^{2}/\overline{a}}{x^{2}} + \frac{\rho_{\scaleto{S}{4pt}}^{2}}{\overline{a}^{4}}\left[\left(\epsilon^{2} - \overline{a}\,\overline{m}_{\phi}^{2}\right) - N_{\scaleto{QL}{5pt}}  \left(3 \overline{a} - \kappa^{2}\right)\right]\\
			&- \left(2\epsilon^{2} - \overline{a}\, \overline{m}_{\phi}^{2}\right) \frac{N_{\scaleto{QL}{5pt}}\rho_{\scaleto{S}{4pt}}^{4}}{\overline{a}^{7}}\,x^{2}\,.\label{Potential_Upper_Bound_Far}
		\end{aligned}
	\end{equation}
	
	The solution to \cref{Eq-para-u} with a potential in the form of \eqref{Potential_Upper_Bound_Far} can be determined in comparison with the Biconfluent Heun equation (BHE), defined by \cref{Biconfluent_Normal_Form,Biconfluent_Parameters}. To do so, we first use the change of variables
	\begin{equation}
		z = \left(2\,Q\right)^{1/4}\,x\,,\label{z_Def_1}
	\end{equation}
	with
	\begin{equation}
		Q = \left(2\epsilon^{2} - \overline{a}\, \overline{m}_{\phi}^{2}\right) \frac{N_{\scaleto{QL}{5pt}}\,\rho_{\scaleto{S}{4pt}}^{4}}{\overline{a}^{7}}\,,\label{Q_Def_1}
	\end{equation}
	to put the respective equation for $u_{\scaleto{ST}{4pt}}^{\scaleto{\,(1)}{6pt}}(x)$ into the form of \cref{Biconfluent_Normal_Form}.	It follows from the definition \eqref{z_Def_1} that the derivatives in $x$ and $z$ are related by
	\begin{equation*}
		\frac{d^{2}}{d x^{2}} = \left(2\,Q\right)^{1/2}\, \frac{d^{2}}{d z^{2}}\,.
	\end{equation*} 
	Hence, \cref{Eq-para-u} for $u_{\scaleto{ST}{4pt}}^{\scaleto{(1)}{6pt}}(z)$ becomes:
	\begin{equation}
		\frac{d^{2}}{d z^{2}}\, u_{\scaleto{ST}{4pt}}^{\scaleto{(1)}{6pt}} + \frac{\rho_{\scaleto{+}{4pt}}^{2} \,V_{\text{eff}}(z)}{(2Q)^{1/2}}\,u_{\scaleto{ST}{4pt}}^{\scaleto{(1)}{6pt}}(z) = 0\,,\label{Eq_u_of_z}
	\end{equation}
	where $\rho_{\scaleto{+}{4pt}}^{2} \,V_{\text{eff}}(z)$ follows from \cref{z_Def_1,Potential_Upper_Bound_Far} so that
	\begin{equation*}
		\frac{\rho_{\scaleto{+}{4pt}}^{2}\, V_{\text{eff}}(z)}{(2Q)^{1/2}} = -\frac{1}{2}\,z^{2} 
		+ \frac{1}{\left(2\, \overline{a} N_{\scaleto{QL}{5pt}}\right)^{1/2}}\,
		\frac{\left(\epsilon^{2} - \overline{a}\, \overline{m}_{\phi}^{2}\right) - \overline{a}N_{\scaleto{QL}{5pt}} \left(3 - \kappa^{2}/\overline{a}\right)}{\left(2\,\epsilon^{2} - \overline{a}\, \overline{m}_{\phi}^{2}\right)^{1/2}}
		- \frac{\kappa^{2}/\overline{a}}{z^{2}}\,,
	\end{equation*}
	provided $\left(2\,\epsilon^{2} - \overline{a}\, \overline{m}_{\phi}^{2}\right) \neq 0$, which ensures that $Q$ is non-zero, allowing for the \-change of variables. 
    
    Therefore, we use the results from the normal form of the BHE, given by \cref{Biconfluent_Normal_Form,Biconfluent_Parameters,EBH_Normal_Form_Solution}, to determine the solution for the auxiliary function $u_{\scaleto{ST}{4pt}}^{\scaleto{(1)}{6pt}}(z)$:\\
	\begin{equation}
		\begin{aligned}
			u_{\scaleto{ST}{4pt}}^{\scaleto{(1)}{6pt}}(z) 
			&= z^{(1+\alpha)/2}\, e^{-\beta z/2}\, e^{-z^{2}/2}
			\left[c_{1}\,\text{HeunB}\left(\alpha,\,\beta,\,\gamma,\,\delta;\,z\right)\right.\\ 
			&\left.+\, c_{2} \,z^{-\alpha}\, \text{HeunB}\left(-\alpha,\,\beta,\,\gamma,\,\delta;\,z\right)\right]\,,
		\end{aligned} \label{u_far_st_1}
	\end{equation}
	where $c_{1}$ and $c_{2}$ are arbitrary constants. The corresponding Heun parameters $\alpha$, $\beta$, $\gamma$, and $\delta$ are:
	\begin{equation}
		\begin{aligned}
			\alpha 
			&= \sqrt{1 + \frac{4\kappa^{2}}{\overline{a}}}\,,\\
			\beta 
			&= 0\,,\\
			\gamma 
			&= \frac{1}{\left(2\, \overline{a} N_{\scaleto{QL}{5pt}}\right)^{1/2}}\,
			\frac{\left(\epsilon^{2} - \overline{a}\, \overline{m}_{\phi}^{2}\right) - \overline{a}N_{\scaleto{QL}{5pt}} \left(3 - \kappa^{2}/\overline{a}\right)}{\left(2\,\epsilon^{2} - \overline{a}\, \overline{m}_{\phi}^{2}\right)^{1/2}}\,,\\
			\delta 
			&= 0\,.
		\end{aligned} \label{Parameters_HeunB_Upper_Bound_Far}	
	\end{equation}

	Given the above, we use the relation \eqref{radial-wave-function} to obtain the radial wave function $R_{\scaleto{ST}{4pt}}^{\scaleto{(1)}{6pt}}(z)$, which has the form:
	\begin{equation}
		\begin{aligned}
			R_{\scaleto{ST}{4pt}}^{\scaleto{(1)}{6pt}}(z) 
			&= \left(1-\frac{N_{\scaleto{QL}{5pt}}\,\rho_{\scaleto{S}{4pt}}^{2}}{2\,\overline{a}^{3}}\frac{z^{2}}{\sqrt{2Q}}\right)\,z^{(-1+\alpha)/2}\, e^{-z^{2}/2}
			\left[c_{1}\,\text{HeunB}\left(\alpha,\,\beta,\,\gamma,\,\delta;\,z\right)\right.\\ 
			&\left.+\, c_{2} \,z^{-\alpha}\, \text{HeunB}\left(-\alpha,\,\beta,\,\gamma,\,\delta;\,z\right)\right]\,,
		\end{aligned}\label{RWF_1_Far_EH_z}
	\end{equation}
	with $c_{1}$ and $c_{2}$ still being arbitrary constants, distinct from those in \cref{u_far_st_1}. Moreover, it is important to emphasize that this solution is valid for scalar particles which are far from the event horizon and whose spectrum satisfies $2\,\epsilon^{2} - \overline{a}\, \overline{m}_{\phi}^{2} \neq 0$. Here, $z$ and $Q$ are defined by \cref{z_Def_1,Q_Def_1}, respectively. We also recall that, to obtain the above result via the relation \cref{radial-wave-function}, we used a first-order approximation of $A(x)^{-1/2}$ with respect to $N_{\scaleto{QL}{5pt}}\,\rho_{\scaleto{S}{4pt}}^{2}/\overline{a}^{3}$, in accordance with \cref{A(x)_Upper_Bound_Far}.
	
	For clarity, we can rewrite \cref{RWF_1_Far_EH_z} in terms of the original variable $x$. By doing so, we observe that
	\begin{equation}
		\begin{aligned}
			R_{\scaleto{ST}{4pt}}^{\scaleto{(1)}{6pt}}(x) 
			&= \left(1 - \frac{N_{\scaleto{QL}{5pt}}\,\rho_{\scaleto{S}{4pt}}^{2}}{2\,\overline{a}^{3}}\, x^{2}\right) x^{(\alpha-1)/2}\, e^{-x^{2}/\lambda_{c}^{2}}		
			\left[\overline{c_{1}}\,\text{HeunB}\left(\alpha,\,\beta,\,\gamma,\,\delta;\,(2Q)^{-1/4}\,x\right)\right.\\ 
			&\left.+\, \overline{c_{2}} \,x^{-\alpha}\, \text{HeunB}\left(-\alpha,\,\beta,\,\gamma,\,\delta;\,(2Q)^{-1/4}\,x\right)\right]\,,
		\end{aligned}\label{RWF_1_Far_EH_x}
	\end{equation}
	where the constants $\overline{c}_{1}$ and $\overline{c}_{2}$ had absorbed any multiplicative factors resulting from the change of variables from $z$ back to $x$. In addition, the parameter $\lambda_{c}$ is dimensionless and acts as a characteristic \enquote{length}, defined by:
	\begin{equation*}
		\lambda_{c}^{-2} = \left(\epsilon^{2} - \frac{1}{2}\,\overline{a}\,\overline{m}_{\phi}^{2}\right)^{1/2} \left(\frac{N_{\scaleto{QL}{5pt}}\,\rho_{\scaleto{S}{4pt}}^{4}}{\overline{a}^{7}}\right)^{1/4}\,.
	\end{equation*}
	
	An interesting property of the biconfluent Heun equation, defined by \cref{Biconfluent_Heun_Equation}, is that, unlike the confluent case, there is not a singularity located at $z=1$ (assuming that $z$ is the independent variable of the ODE). On the other hand, the BHE has singularities at $z=0$ (regular) and as $z\to\infty$ (irregular).  These characteristics ensure the convergence of the Fuchs-Frobenius (local) solution for all $|z|<\infty$.
	
	Within the set of possible values for the parameters \eqref{Parameters_HeunB_Upper_Bound_Far}, it is always interesting to investigate the solutions of the $\delta_{N}$ subclass, which satisfy \cref{Delta_Condition_EBH} by definition. This $\delta_{N}$--condition implies that:
	\begin{equation}
		\begin{aligned}
			\frac{1}{\left(2\, \overline{a} N_{\scaleto{QL}{5pt}}\right)^{1/2}}\,
			\frac{\left(\epsilon^{2} - \overline{a}\, \overline{m}_{\phi}^{2}\right) - \overline{a}N_{\scaleto{QL}{5pt}} \left(3 - \kappa^{2}/\overline{a}\right)}{\left(2\,\epsilon^{2} - \overline{a}\, \overline{m}_{\phi}^{2}\right)^{1/2}} \mp \left(1 + \frac{4\kappa^{2}}{\overline{a}}\right)^{1/2} = 2N + 2\,.
		\end{aligned}\label{Regime_Físico_1_Far}
	\end{equation}
	As discussed in the appendix, this condition is necessary but not sufficient for obtaining the polynomials related to the BHE. However, we are not interested in the explicit form of these polynomials, but rather in the constraints relating all $\delta_{N}$ solutions.
	
	As an example, let us consider the particular case of $\epsilon^{2} = \overline{a}\,\overline{m}_{\phi}^{2}$, so that the \cref{Regime_Físico_1_Far} simplifies to
	\begin{equation}
		\left(\frac{N_{\scaleto{QL}{5pt}}}{8\,\overline{m}_{\phi}^{2}}\right)^{1/2} \left(\frac{\kappa^{2}}{\overline{a}} - 3\right) \mp \left(\frac{1}{4} + \frac{\kappa^{2}}{\overline{a}}\right)^{1/2} = N + 1\,,
	\end{equation}
	implying in restrictions for $\kappa^{2} = n_{\scaleto{R}{4pt}}^{2} + k_{z}^{2}$. Certainly, we could examine other relationships between $\epsilon^{2}$ and $\overline{a}\,\overline{m}_{\phi}^{2}$, as long as $2\,\epsilon^{2} \neq \overline{a}\, \overline{m}_{\phi}^{2}$ for the change of variables \eqref{z_Def_1} to remain valid.
	
	On the other hand, if $\epsilon^{2} = \overline{a}\, \overline{m}_{\phi}^{2}/2$ occurs, we need to return to the beginning of the problem and impose this constraint directly on the effective potential \eqref{Potential_Upper_Bound_Far}. With this restriction, we have that
	\begin{equation}
		\begin{aligned}
			\rho_{\scaleto{+}{4pt}}^{2}\,V_{\text{eff}}(x) 
			&= - \frac{\kappa^{2}/\overline{a}}{x^{2}} - V_{0}\,,\label{Potential_Upper_Bound_Simpler}
		\end{aligned}
	\end{equation}
	where $V_{0}$ is defined as
	\begin{equation}
		V_{0} = \frac{\rho_{\scaleto{S}{4pt}}^{2}}{2\,\overline{a}^{3}}\left[\overline{m}_{\phi}^{2} - 2 N_{\scaleto{QL}{5pt}}\left(\kappa^{2}/\overline{a} - 3\right)\right]\,.\label{def-V_0}
	\end{equation}
	
	Subsequently, we solve the auxiliary equation \eqref{Eq-para-u} for $u_{\scaleto{ST}{4pt}}^{\scaleto{(1,\, \oslash)}{6pt}}(x)$ using the potential \eqref{Potential_Upper_Bound_Simpler} and write the linearly independent solutions in terms of the modified Bessel functions, $\text{I}_{n}(x)$ and $\text{K}_{n}(x)$, of order $n$. This procedure yields
	\begin{equation*}
		u_{\scaleto{ST}{4pt}}^{\scaleto{(1,\, \oslash)}{6pt}}(x) = \sqrt{x} \left[c_{1}\,\text{I}_{n}\left(\sqrt{V_{0}} \,x\right) + c_{2}\,\text{K}_{n}\left(\sqrt{V_{0}} \,x\right)\right]\,,
	\end{equation*}
	where $n$ is real and given by \cref{order-n}, satisfying $n \geq 1/2$. Note that the symbol \enquote{$\oslash$} was included as a label to distinguish this particular case, where $\epsilon^{2} = \overline{a}\,\overline{m}_{\phi}^{2}/2$, from the general solution given by \cref{RWF_1_Far_EH_x}.
	
	From the result above for $u_{\scaleto{ST}{4pt}}^{\scaleto{(1,\, \oslash)}{6pt}}(x)$, we can determine the associated radial wave function using the relation \eqref{radial-wave-function}. Therefore, we obtain:
	\begin{equation}
		R_{\scaleto{ST}{4pt}}^{\scaleto{(1,\, \oslash)}{6pt}}(x) = x^{-1/2}\left(1 - \frac{N_{\scaleto{QL}{5pt}}\rho_{\scaleto{S}{4pt}}^{2}}{2\,\overline{a}^{3}}\,x^{2}\right) 
		\left[c_{1}\,\text{I}_{n}\left(\sqrt{V_{0}} \,x\right) + c_{2}\,\text{K}_{n}\left(\sqrt{V_{0}} \,x\right)\right]\,,\label{RWF_ST_1_Constrained}
	\end{equation}
	where, again, we use an approximation for $A(x)^{-1/2}$ up to first order in $N_{\scaleto{QL}{5pt}}\,\rho_{\scaleto{S}{4pt}}^{2}/\overline{a}^{3}$. Furthermore, considering that the domain of the solution \eqref{RWF_ST_1_Constrained} is unbounded, we set $c_{1} = 0$ to ensure proper behavior\footnote{After all, only $\text{K}_{n}(x)$ is a decreasing function of the argument.} when $x\to \infty$. Finally, we recall that the divergence of $\text{K}_{n}(x)$ when $x\to 0$ is not a problem, since $x = 0$ does not belong to the domain of the solution. Thus, we conclude that:
	\begin{equation}
		R_{\scaleto{ST}{4pt}}^{\scaleto{(1,\, \oslash)}{6pt}}(x) = \frac{R_{0}}{\sqrt{x}}\left(1 - \frac{N_{\scaleto{QL}{5pt}}\rho_{\scaleto{S}{4pt}}^{2}}{2\,\overline{a}^{2}}\,x^{2}\right) \text{K}_{n}\left(\sqrt{V_{0}} \,x\right)\,,\label{RWF_1_Constrained_Bessel}
	\end{equation}
	where $R_{0}$ is a normalization constant, while $n$ and $V_{0}$ are defined by \cref{order-n,def-V_0}, respectively. We also recall that $\sqrt{V_{0}}$ is a real number, provided that $\overline{m}_{\phi}^{2} \geq 2 N_{\scaleto{QL}{5pt}} \left(\kappa^{2}/\overline{a} - 3\right)$, which is physically reasonable since $\overline{m}_{\phi}$ is on the order of $6.37 \times 10^{17}$ m$^{-1}$, while $N_{\scaleto{QL}{5pt}}$ should be near $6.9 \times 10^{-53}$ m$^{-2}$. In other words, since the admissible values for $\overline{a}$ would be in the interval $[10^{-6},\,10^{0}]$, it would not be possible for the term of $\kappa^{2}$ to be so large as to make $V_{0}$ negative.
	
	The main conclusion we reach in both solutions of this standard scenario is that a particle further away from the event horizon (with $x \gg 1/x$) does not perceive the existence of this horizon, which is quite plausible. Furthermore, the effect of the quintessence fluid is quite subtle and almost entirely encoded in the term $\left(1 - \frac{N_{\scaleto{QL}{5pt}}\rho_{\scaleto{S}{4pt}}^{2}}{2\,\overline{a}^{3}}\,x^{2}\right)$. Thus, the influence of dark energy grows with $x^{2}$, being greater the further away the scalar particle is.
	
	Finally, now that we have examined the standard scenario, we proceed with the investigation of the case where the cloud of \emph{strings} is removed from the system.

	\subsubsection{Cloudless at the upper bound}
	
	In the preceding analysis, we solved the limiting case of $\alpha_{\scaleto{Q}{5pt}}=1$ in the standard scenario. However, solutions \eqref{RWF_1_Far_EH_x} and \eqref{RWF_ST_1_Constrained} do not hold if $\overline{a} = 0$. Thus, to investigate the cloudless scenario, let us consider the general result for $A(\rho)$, given by \cref{Function_A(rho)}, as we set $\overline{a} = 0$ and use the definition of $N_{\scaleto{QL}{5pt}}$ from \cref{N_QL_Definition} to obtain that
	\begin{equation*}
		A(\rho) = N_{\scaleto{QL}{5pt}} \,\rho^{2} - \frac{\rho_{\scaleto{S}{4pt}}}{\rho}\,.
	\end{equation*}
	
	After requiring that $A(\rho_{\scaleto{+}{4pt}}) = 0$ in the above equation, we conclude that $\rho_{\scaleto{+}{4pt}} = (\rho_{\scaleto{S}{4pt}}/N_{\scaleto{QL}{5pt}})^{1/3}$, which is an exact result, in accordance with the results of table \ref{table-cloudless-cases}. This result implies we can write the above equation for $A(\rho)$ in terms of the dimensionless variable $x$, resulting in
	\begin{equation}
		A(x) = \left(N_{\scaleto{QL}{5pt}}\,\rho_{\scaleto{S}{4pt}}^{2}\right)^{1/3}\left(x^{2} - \frac{1}{x}\right)\,.\label{Metric_Cloudless_1}
	\end{equation}
	Notably, this expression for $A(x)$ is valid for all $x>1$, that is, the whole radial domain.
	
	Similarly to the standard scenario, it is not possible to obtain a valid analytical solution for the entire domain of $A(x)$. Therefore, we must solve the problem in two disjoint regions of the domain. The first region is the one close to the event horizon, while the second is defined by the condition $x^{2}\gg 1/x$, which has no upper bound. Recalling that, as $x \to 1^{+}$, we recover the conditions of \eqref{RWF_NH_All_Alpha_Q}, describing the solution near the horizon (\textsc{nh}), with $\beta_{\scaleto{+}{4pt}} = 3\,(N_{\scaleto{QL}{5pt}}\,\rho_{\scaleto{S}{4pt}}^{2})^{1/3}$, the only remaining task is to explore the case of $x^{3}\gg 1$. In this case, it is a good approximation of \cref{Metric_Cloudless_1} to consider 
	\begin{equation}
		A(x) = \left(N_{\scaleto{QL}{5pt}}\,\rho_{\scaleto{S}{4pt}}^{2}\right)^{1/3}\,x^{2}\,.\label{A(x)_Cloudless_Upper_Bound_Far}
	\end{equation}
	This approximation is good for values of $x$ away from the event horizon and improves as $x$ increases. As an example, it is reasonable to take $x>5$, for instance, so that the relative error in $A(x)$ is less than $1\%$.
	
	In the sequence, we consider the general expression for the effective potential, defined by \eqref{Potential_in_terms_of_x}, with $A(x)$ approximated by \eqref{A(x)_Cloudless_Upper_Bound_Far}, to show that:
	\begin{equation}
		\rho_{\scaleto{+}{4pt}}^{2}\,V_{\text{eff}}(x) = 
		-\left(2 + \frac{\overline{m}_{\phi}^{2}}{N_{\scaleto{QL}{5pt}}}\right)\frac{1}{x^{2}} 
		+ \left[\frac{\epsilon^{2}}{\left(N_{\scaleto{QL}{5pt}}^{2}\,\rho_{\scaleto{S}{4pt}}\right)^{2/3}} 
		- \frac{\kappa^{2}}{\left(N_{\scaleto{QL}{5pt}}\,\rho_{\scaleto{S}{4pt}}^{2}\right)^{1/3}}\right]\frac{1}{x^{4}}\,.\label{Potential_CL_1_Far}
	\end{equation}
	Then, after investigating the ODE for $u_{\scaleto{CL}{4pt}}^{\scaleto{(1)}{6pt}}(x)$, given by \cref{Eq-para-u}, with the potential of \cref{Potential_CL_1_Far}, we obtain that:
	\begin{equation}
		\begin{aligned}
			u_{\scaleto{CL}{4pt}}^{\scaleto{(1)}{6pt}}(x) = x^{1/2} \left[c_{1}\,\text{J}_{\ell}\left(\frac{\sqrt{m}}{x}\right) + c_{2}\, \text{Y}_{\ell}\left(\frac{\sqrt{m}}{x}\right)\right]\,,
		\end{aligned}\label{u_Cloudless_1}
	\end{equation}
	where, again, $\text{J}_{\ell}\left(x\right)$ and $\text{Y}_{\ell}\left(x\right)$ are the $\ell$-th order Bessel functions of the first and second kinds, respectively, while $\ell$ and $m$ are determined by
	\begin{equation}
		\begin{aligned}
			\ell
			&= \sqrt{\frac{9}{4} + \frac{\overline{m}_{\phi}^{2}}{N_{\scaleto{QL}{5pt}}}\,}\,,\\
			m 
			&= \frac{\epsilon^{2}}{\left(N_{\scaleto{QL}{5pt}}^{2}\,\rho_{\scaleto{S}{4pt}}\right)^{2/3}} 
			- \frac{\kappa^{2}}{\left(N_{\scaleto{QL}{5pt}}\,\rho_{\scaleto{S}{4pt}}^{2}\right)^{1/3}}\,.
		\end{aligned}\label{ell-m}
	\end{equation}
	By using the results of \cref{u_Cloudless_1,A(x)_Cloudless_Upper_Bound_Far} into the relation \cref{radial-wave-function}, we show that $R_{\scaleto{CL}{4pt}}^{\scaleto{(1)}{6pt}}(x)$ has the form of
	\begin{equation}
		R_{\scaleto{CL}{4pt}}^{\scaleto{(1)}{6pt}}(x) = x^{-3/2}
		\left[c_{1}\,\text{J}_{\ell}\left(\frac{\sqrt{m}}{x}\right) + c_{2}\, \text{Y}_{\ell}\left(\frac{\sqrt{m}}{x}\right)\right]\,,\label{RWF_CL_Far_General}
	\end{equation}
	where $\ell$ and $m$ are defined by \eqref{ell-m}, while the constants $c_{1}$ and $c_{2}$ absorbed the remaining multiplicative factors arising from the relation \eqref{radial-wave-function}. Also note that, if we are interested in the region of large $x$, i.e., $x\gg 1$, we must set $c_{2} = 0$ to prevent the radial wave function from diverging.
	
    Conversely, if $m$, defined by \cref{ell-m},  is set to $m = 0$, implying the restricted case where
	\begin{equation}
		\epsilon^{2} = \kappa^{2}\,N_{\scaleto{QL}{5pt}}\,,\label{Epsilon_Constraint_CL_1_Far}
	\end{equation}
	the effective potential \eqref{Potential_CL_1_Far}, valid for $x>1$, 
	simplifies to
	\begin{equation*}
		\rho_{\scaleto{+}{4pt}}^{2}\,V_{\text{eff}}(x) = 
		-\left(2 + \frac{\overline{m}_{\phi}^{2}}{N_{\scaleto{QL}{5pt}}}\right)\frac{1}{x^{2}}\,,
	\end{equation*}
	implying that \cref{Eq-para-u}, for $u_{\scaleto{CL}{4pt}}^{\scaleto{(1,\,\boxtimes)}{6pt}}(x)$, is a Cauchy-Euler equation whose solution is given by
	\begin{equation*}
		u_{\scaleto{CL}{4pt}}^{\scaleto{(1,\,\boxtimes)}{6pt}}(x) = c_{1}\,x^{1/2+\ell} + c_{2}\, x^{1/2-\ell}\,,
	\end{equation*}
	where $\ell$ is given by \cref{ell-m} and the symbol \enquote{$\boxtimes$} is used only to distinguish this particular case, satisfying the constraint of \cref{Epsilon_Constraint_CL_1_Far}, from the general solution of \cref{RWF_CL_Far_General}. 
	
	By employing \cref{radial-wave-function}, we obtain that  $R_{\scaleto{CL}{4pt}}^{\scaleto{(1,\,\boxtimes)}{6pt}}(x)$ is given by
	\begin{equation*}
		R_{\scaleto{CL}{4pt}}^{\scaleto{(1,\,\boxtimes)}{6pt}}(x) = c_{1}\,x^{-3/2+\ell} + c_{2}\, x^{-3/2-\ell}\,,
	\end{equation*}
	where, once more, $c_{1}$ and $c_{2}$ absorbed the multiplicative constants arising from the relation \ref{radial-wave-function}. Considering that the domain is unbounded ($x>1$), we require that $c_{1} = 0$ since $\ell > 3/2$, implying
	\begin{equation}
		R_{\scaleto{CL}{4pt}}^{\scaleto{(1,\,\boxtimes)}{6pt}}(x) = R_{0}\, x^{-3/2-\ell}\,,
		\label{RWF_CL_1_Constrained}
	\end{equation}
	where we renamed $c_{2}$ to $R_{0}$ as the remaining normalization constant.

	\subsubsection{Horizonless at the upper bound} \label{HL-1}

	In this final case, the black string contribution is removed by setting $\rho_{\scaleto{S}{4pt}} = 0$, while retaining the cloud of strings ($\overline{a}\neq 0$) and the quintessence fluid (with $\alpha_{\scaleto{Q}{5pt}} = 1$). Since $N_{\scaleto{QL}{5pt}}$, as defined by \cref{N_QL_Definition}, incorporates both the effect of $N_{\scaleto{Q}{5pt}}$ and $1/l^{2}$, it is unnecessary to restrict the radial domain, which must\footnote{This follows from the fact that $\rho_{\scaleto{S}{4pt}}=0$, implying that $x$ cannot be defined.} be described in terms of $\rho$ rather than $x$.
	
	This choice of scenario simplifies the general expression of \cref{Function_A(rho)}, for $A(\rho)$, which is reduced to
	\begin{equation}
		A(\rho) = \overline{a}\left(1 + \frac{N_{\scaleto{QL}{5pt}}}{\overline{a}}\,\rho^{2}\right)\,,\label{HL_Alpha_1}
	\end{equation}
	where $N_{\scaleto{QL}{5pt}}/\overline{a}$ would typically range from $10^{-52}$ m$^{-2}$ to $10^{-46}$ m$^{-2}$, implying its contribution is negligible unless $\rho\gg 1$, specifically of the order of $\rho_{\scaleto{obs.}{4pt}} \sim |l|$. We stress that \cref{HL_Alpha_1} is quite similar to \cref{A(x)_Upper_Bound_Far}, but with a different domain since in the horizonless scenario $\rho$ starts at $\rho = 0$ and cannot be written in terms of the dimensionless variable $x$. The result from \cref{HL_Alpha_1} can also be interpreted as a tiny correction to the case without quintessence, where only $\overline{a}$ is nonzero.
	
	To obtain the effective potential $V_{\text{eff}}(\rho)$, we substitute \cref{HL_Alpha_1} into the general expression of \cref{Effective-Potential} to show that
	\begin{equation}
		V_{\text{eff}}(\rho) = \frac{\epsilon^{2}}{\overline{a}^{2}}\left(1 - \frac{2\,N_{\scaleto{QL}{5pt}}}{\overline{a}}\,\rho^{2}\right) - \frac{3\,N_{\scaleto{QL}{5pt}}}{\overline{a}} - \left(\frac{\kappa^{2}/\overline{a}}{\rho^{2}} + \frac{\overline{m}_{\phi}^{2}}{\overline{a}}\right)\left(1 - \frac{N_{\scaleto{QL}{5pt}}}{\overline{a}}\,\rho^{2}\right)\,,\label{Potential_HL_Alpha_1}
	\end{equation}
	where we have neglected quadratic and higher order terms in $N_{\scaleto{QL}{5pt}}$ since it is of the order $10^{-52}$ m$^{-2}$. Note that this result is very similar to the case of \cref{Potential_Horizonless_Meio_Intervalo}, where $\alpha_{\scaleto{Q}{5pt}}$ was in the middle of its interval. This similarity follows from the fact that, in both cases, quintessence is just a tiny correction to the respective $A(\rho)$. 
	
	To investigate the solution associated with \cref{Potential_HL_Alpha_1}, we rewrite it more conveniently:
	\begin{equation}
		V_{\text{eff}}(\rho) = -\left(\frac{2\epsilon^{2} - \overline{a}\, \overline{m}_{\phi}^{2}}{\overline{a}^{2}}\right) \frac{N_{\scaleto{QL}{5pt}}}{\overline{a}}\,\rho^{2} + \left(\frac{\epsilon^{2} - \overline{a}\, \overline{m}_{\phi}^{2}}{\overline{a}^{2}}\right) - \frac{\kappa^{2}/\overline{a}}{\rho^{2}}\,.\label{Potential_HL_Alpha_1_Shorter}
	\end{equation}
	We recall that the terms $3N_{\scaleto{QL}{5pt}}/\overline{a}$ and $\kappa^{2}\,N_{\scaleto{QL}{5pt}}/\overline{a}^{2}$ were excluded due to their negligible contribution to the effective potential.

	After considering \cref{eq-geral-u} for $u_{\scaleto{HL}{4pt}}^{\scaleto{\,(1)}{6pt}}(\rho)$, with the effective potential \eqref{Potential_HL_Alpha_1_Shorter}, we see that our problem is almost in the form of a BHE such as \cref{Biconfluent_Normal_Form}. Thus, we can change variables from $\rho$ to $z = \left(2\,P\right)^{1/4} \rho$, analogously to \cref{z_Def_1}, where now
	\begin{equation}
		P = \left(\frac{2\epsilon^{2} - \overline{a}\, \overline{m}_{\phi}^{2}}{\overline{a}^{3}}\right) N_{\scaleto{QL}{5pt}}\,,\label{def-P}
	\end{equation}
	provided $2\epsilon^{2} - \overline{a}\, \overline{m}_{\phi}^{2} \neq 0$. Under this change of variables, we use the result of \cref{Eq_u_of_z}, simply replacing \enquote{$Q$} for \enquote{$P$}, to determine the term corresponding to $V_{\text{eff}}(z)$: 
	\begin{equation}
		\frac{V_{\text{eff}}(z)}{(2P)^{1/2}} = -\frac{1}{2}\,z^{2} + \frac{1}{\sqrt{2} \overline{a}}\,\frac{\epsilon^{2} - \overline{a}\, \overline{m}_{\phi}^{2}}{\left(2\epsilon^{2} - \overline{a}\, \overline{m}_{\phi}^{2}\right)^{1/2}} - \frac{\kappa^{2}/\overline{a}}{z^{2}}\,.\label{Potential_in_z_HL_1}
	\end{equation}
	
	Therefore, comparing \cref{Eq_u_of_z,Potential_in_z_HL_1} with the general result of \cref{Biconfluent_Heun_Equation} in the Appendix, we conclude that $u_{\scaleto{HL}{4pt}}^{\scaleto{(1)}{6pt}}(z)$ is determined by
	\begin{equation}
		\begin{aligned}
			u_{\scaleto{HL}{4pt}}^{\scaleto{(1)}{6pt}}(z) 
			&= z^{(1+\alpha)/2}\, e^{-z^{2}/2}
			\left[c_{1}\,\text{HeunB}\left(\alpha,\,\beta,\,\gamma,\,\delta;\,z\right)\right.\\
			&\left.+\, c_{2}\,z^{-\alpha}\,\text{HeunB}\left(-\alpha,\,\beta,\,\gamma,\,\delta;\,z\right)\right]\,,
		\end{aligned}\label{u_HL_1}
	\end{equation}
	where the Heun parameters $\alpha$, $\beta$, $\gamma$, and $\delta$ are:
	\begin{equation}
		\begin{aligned}
			\alpha 
			&= 2\,\sqrt{\frac{1}{4} + \frac{\kappa^{2}}{\overline{a}}}\,,\\
			\beta 
			&= 0\,,\\
			\gamma 
			&= \frac{1}{\sqrt{2}\, \overline{a}}\,\frac{\epsilon^{2} - \overline{a}\, \overline{m}_{\phi}^{2}}{\left(2\epsilon^{2} - \overline{a}\, \overline{m}_{\phi}^{2}\right)^{1/2}}\,,\\
			\delta 
			&= 0\,.
		\end{aligned} \label{Heun_Parameters_HL_1}
	\end{equation}
	By substituting the results of \cref{u_HL_1,HL_Alpha_1} into the relation \eqref{radial-wave-function}, we obtain that the radial solution associated with $\alpha_{\scaleto{Q}{5pt}} = 1$ in the horizonless scenario is given by
	\begin{equation}
		\begin{aligned}
			R_{\scaleto{HL}{4pt}}^{\scaleto{(1)}{6pt}}(z) 
			&= \left(1-\frac{N_{\scaleto{QL}{5pt}}}{2\overline{a}}\frac{z^{2}}{\sqrt{2P}}\right)\,z^{(\alpha-1)/2}\, e^{-z^{2}/2}\,
			\left[c_{1}\,\text{HeunB}\left(\alpha,\,\beta,\,\gamma,\,\delta;\,z\right)\right.\\
			&\left.+\, c_{2}\,z^{-\alpha}\,\text{HeunB}\left(-\alpha,\,\beta,\,\gamma,\,\delta;\,z\right)\right]	
			\,,
		\end{aligned}\label{RWF_HL_1_z}
	\end{equation}
	where $P$ and $z$ are defined by \cref{def-P,z_Def_1}, respectively, while the Heun parameters associated with the BHE are determined by \cref{Heun_Parameters_HL_1}. Moreover, to guarantee a non-singular solution at the origin, compatible with the removal of the black string, we set $c_{2}=0$ to eliminate the contribution of $z^{-\alpha}$ because $\alpha \geq 1$. Also note that the global contribution of $z^{(\alpha-1)/2}$ does not impose problems on the regularity.
	
	For convenience, we can also rewrite \cref{RWF_HL_1_z} (with $c_{2}=0$) in terms of the original variable $\rho$, from which it follows that:
	\begin{equation}
		R_{\scaleto{HL}{4pt}}^{\scaleto{(1)}{6pt}}(\rho) = \overline{R_{0}}\,\left(1-\frac{N_{\scaleto{QL}{5pt}}}{2\overline{a}}\,\rho^{2}\right)\,\rho^{(\alpha-1)/2}\, e^{-\rho^{2}/\lambda_{\scaleto{H}{3pt}}^{2}}\,\text{HeunB}\left(\alpha,\,\beta,\,\gamma,\,\delta;\,(2P)^{-1/4}\rho\right)\,,
		\label{RWF_HL_1_rho}
	\end{equation}
	with
	\begin{equation}
		\lambda_{\scaleto{H}{4pt}}^{-2} = \left(\epsilon^{2} - \frac{1}{2} \overline{a}\, \overline{m}_{\phi}^{2}\right)^{1/2}\, \left(\frac{N_{\scaleto{QL}{5pt}}}{\overline{a}^{3}}\right)^{1/2}\,.\label{def-lambda_H}
	\end{equation}
	Note that $\lambda_{\scaleto{H}{4pt}}^{-2}$ has the dimension of m$^{-2}$ in SI units, making the argument of this exponential dimensionless, as expected. Assuming that $\epsilon^{2} > \overline{a}\,\overline{m}_{\phi}^{2}$ is fixed, we see that this characteristic length $\lambda_{\scaleto{H}{4pt}}$ causes the wave function to decay more rapidly in the case where quintessence is present ($N_{\scaleto{Q}{5pt}} \neq 0$) than in the case where containing only the cosmological constant (when $N_{\scaleto{QL}{5pt}} = 1/l^{2}$). We also emphasize that, after setting $c_{2}=0$ in \cref{RWF_HL_1_z}, we renamed the  arbitrary constant $c_{1}$ to $\overline{R_{0}}$, since the latter had absorbed the multiplicative constants from the change of variables.
	
	Within this class of solutions \footnote{For variable values of $\epsilon$, $\kappa$ and $\overline{m}_{\phi}$.} for $R_{\scaleto{HL}{4pt}}^{\scaleto{(1)}{6pt}}(\rho)$, we can examine the particular set satisfying the $\delta$--condition, defined by \cref{Delta_Condition_EBH}, which also includes the polynomial cases. In this case, it follows that the parameters $\epsilon$, $\kappa$ and $\overline{m}_{\phi}$ are constrained by the equation:
	\begin{equation}
		\begin{aligned}
			\frac{1}{\sqrt{8}\, \overline{a}}\,\frac{\epsilon^{2} - \overline{a}\, \overline{m}_{\phi}^{2}}{\left(2\epsilon^{2} - \overline{a}\, \overline{m}_{\phi}^{2}\right)^{1/2}} \mp \left(\frac{\kappa^{2}}{\overline{a}} + \frac{1}{4}\right)^{1/2} = N + 1\,,
		\end{aligned}\label{Regime_Físico_1_HL}
	\end{equation}
	with $N\in\mathbb{Z}^{+}$. However, this constraint does not encompass all possible eigenstates, only those belonging to $\delta_{N}$. An interesting example of \eqref{Regime_Físico_1_HL} is given by the particular case when $\epsilon^{2} = \overline{a}\,\overline{m}_{\phi}^{2}$, from which it follows that
	\begin{equation*}
		n_{\scaleto{R}{4pt}}^{2} + k_{z}^{2} = \frac{\overline{a}}{4} \left[4\left(N + 1\right)^{2} - 1\right]\,,
	\end{equation*}
	which implies in a quantized linear momentum along the $z$ direction, since $n_{\scaleto{R}{4pt}}$ itself is an integer\footnote{We knew that $n_{\scaleto{R}{4pt}}$ was an integer since the separation of variables in the Klein-Gordon equation. This property follows from the requirement that $\Phi(2\,n_{\scaleto{R}{4pt}}\pi) = \Phi(0)$.}.
	
	In the sequence, we analyze special cases of the effective potential \eqref{Potential_HL_Alpha_1_Shorter} to conclude our analysis of the horizonless scenario (\textsc{hl}) associated with $\alpha_{\scaleto{Q}{5pt}}=1$.
	
	\subsubsection*{Additional Constrained Case:}

	Now that we have the general solution \eqref{RWF_HL_1_rho}, which is valid since\footnote{We recall that, in the case of $\epsilon^{2} = \overline{a}\,\overline{m}_{\phi}^{2}/2$, the change of variables performed to obtain the solution \eqref{RWF_HL_1_rho} is not valid.} $\epsilon^{2}\neq \overline{a}\,\overline{m}_{\phi}^{2}$, the natural sequence for our investigation is to solve the restricted case in which precisely
	\begin{equation}
		\epsilon^{2} = \frac{1}{2}\,\overline{a}\,\overline{m}_{\phi}^{2}\,,\label{Caso_Complementar_HL_1}
	\end{equation}
	to complement the general case. We also emphasize that the constraint \eqref{Caso_Complementar_HL_1} is identical to that of \cref{Vínculo_Energia_Comum}, with the difference that solutions for $\alpha_{\scaleto{Q}{5pt}}=1$ are not bounded above.
	
	Returning to the potential \eqref{Potential_HL_Alpha_1_Shorter}, we see that it is simplified by the constraint \eqref{Caso_Complementar_HL_1}, resulting in:
	\begin{equation*}
		V_{\text{eff}}(\rho) = 	-\frac{\overline{m}_{\phi}^{2}}{2\,\overline{a}} - \frac{\kappa^{2}/\overline{a}}{\rho^{2}}\,.
	\end{equation*}	
	Consequently, we use Bessel's modified equation \cite{Olver2010,Butkov1968} to determine the solution of $u_{\scaleto{HL}{4pt}}^{\scaleto{(1,\,\otimes)}{6pt}}(\rho)$. This procedure results in
	\begin{equation}
		u_{\scaleto{HL}{4pt}}^{\scaleto{(1,\,\otimes)}{6pt}}(\rho) = \sqrt{\rho} \left[c_{1}\,\text{I}_{n}\left(\sqrt{\tfrac{\overline{m}_{\phi}^{2}}{2\, \overline{a}}}\,\rho\right) + c_{2}\,\text{K}_{n}\left(\sqrt{\tfrac{\overline{m}_{\phi}^{2}}{2\, \overline{a}}}\,\rho\right)\right]\,,\label{u_HL_C2}
	\end{equation}
	where, again, $\text{I}_{n}(\rho)$ and $\text{K}_{n}(\rho)$ are the modified Bessel functions, of order $n$, of the first and second kinds, respectively. Furthermore, we emphasize that there is no physical reason why the order $n=\left(\tfrac{\kappa^{2}}{\overline{a}}+1/4\right)^{1/2}$, defined by \cref{order-n}, should be an integer. We also highlight that the symbol \enquote{$\otimes$} was used in $u_{\scaleto{HL}{4pt}}^{\scaleto{(1,\,\otimes)}{6pt}}(\rho)$ to avoid ambiguities regarding the general case.
	
	In addition to the result \eqref{u_HL_C2} for the auxiliary function $u_{\scaleto{HL}{4pt}}^{\scaleto{(1,\,\otimes)}{6pt}}(\rho)$, we also use a first-order approximation in $N_{\scaleto{QL}{5pt}}/\overline{a}$ for $A(\rho)^{-1/2}$, so that the radial solution obtained from \cref{radial-wave-function} is in the form of
	\begin{equation}
		R_{\scaleto{HL}{4pt}}^{\scaleto{(1,\,\otimes)}{6pt}}(\rho) = \rho^{-1/2} \left(1 - \frac{N_{\scaleto{QL}{5pt}}}{2\,\overline{a}}\,\rho^{2}\right)
		\left[c_{1}\,\text{I}_{n}\left(\sqrt{\tfrac{\overline{m}_{\phi}^{2}}{2\, \overline{a}}}\,\rho\right) + c_{2}\,\text{K}_{n}\left(\sqrt{\tfrac{\overline{m}_{\phi}^{2}}{2\, \overline{a}}}\,\rho\right)\right]\,.\label{RWF_HL_Alpha_Q_1_Constraint_2}
	\end{equation}
	To describe particles near the origin, we choose $c_{2} = 0$, for appropriate (non-singular) behavior. For larger values of $\rho$, it is necessary to set $c_{1} = 0$, from which we obtain a monotonically decreasing function.

	With the extended domain solutions established for all scenarios and regimes, we will investigate the emergence of dark phases in some of these solutions. While a comprehensive analysis of both implicit and explicit dark phases across all solutions would be ideal, such an approach would make the text excessively long. Therefore, we will use some examples to draw important conclusions about the occurrence of these phases.

	\section{The emergence of dark phases in extended solutions}\label{Dark_Phases_Investigation}

	Our preceding analyses have explored solutions to the Klein-Gordon equation in various scenarios, including the standard (\textsc{st}), cloudless (\textsc{cl}), and horizonless (\textsc{hl}) cases. When the state parameter $\alpha_{\scaleto{Q}{5pt}}$ takes the values $0$ or $1/2$, these solutions remain valid over a large domain of the radial coordinate, provided that $\rho$ and $\rho_{\scaleto{+}{4pt}}$ are significantly smaller than the AdS radius $l$, that is, $\rho_{\scaleto{+}{4pt}} < \rho \ll l$. Conversely, for $\alpha_{\scaleto{Q}{5pt}} = 1$, our domain is unrestricted, since $N_{\scaleto{QL}{5pt}} = N_{\scaleto{Q}{5pt}} + 1/l^{2}$ incorporates the contributions of both quintessence and the cosmological constant ($\Lambda = -3/l^{2}$). Any further approximations and/or restrictions were chosen to describe the particle's radial solution in specific regimes.
	
	In this section, we examine the implications of our previous results. Specifically, we consider cases where the radial KG solution can be written as a product of two functions: one independent of the quintessence parameter $N_{\scaleto{Q}{5pt}}$, and the other exhibiting dependence (implicit or explicit) on this same parameter, usually in the form of a phase difference.
	
	In a previous work \cite{Deglmann2025}, we had defined this quintessence-induced phase difference as a \enquote{dark phase} and presented a general procedure for determining these possible quantum observables using the Heun parameters $\gamma$ and $\beta$. This procedure was reviewed in Section \ref{RWF_NH_All_Alpha_Q_Near} of the present work, where we discussed the solutions arising from a first-order approximation of the metric function $A(x) = \beta_{\scaleto{+}{4pt}}(x-1)$, which were valid near the event horizon (as $x\to 1^{+}$). Although this was a general result, it could not handle scenarios without an event horizon and could not account for higher-order corrections for farther spin$-0$ particles.
	
	However, our novel extended solutions overcome these limitations. We further demonstrate that, in specific cases, the dependence on $N_{\scaleto{Q}{5pt}}$ can be factored straightforwardly, enabling the determination of dark phases. Moreover, we will explore dark phases in horizonless scenarios and provide examples within standard and cloudless regimes.

    \subsection{Dark phases for the lower bound}

   	To examine the influence of quintessence on the radial solution in the standard scenario of $\alpha_{\scaleto{Q}{5pt}}=0$, we analyze the behavior of \cref{Radial_WF_Alpha_Q_0_ST} as $x\to 1^{+}$. This yields the following approximation:
	\begin{equation}
		R_{\scaleto{ST}{4pt}}^{\scaleto{(0)}{6pt}}(x) = c_{1}\,\exp\left[\frac{\gamma}{2} \ln (x-1) + \frac{\alpha}{2}\right] + c_{2}\,\exp\left[-\frac{\gamma}{2} \ln (x-1) + \frac{\alpha}{2}\right]\,.\label{RWF_0_ST_Approx_DPs}
	\end{equation}
	Higher-order approximations would introduce corrections to the coefficient of $\ln (x-1)$. In contrast, due to the factor of $\exp\left(\alpha x/2\right)$ we would also have corrections to powers of $(x-1)$. However, the logarithmic term dominates the radial wave function as $x$ approaches $1^{+}$ (the event horizon). Consequently, we can use the above approximation as a fundamental term for further analysis of higher-order cases. 
	
	The Heun parameters for the standard scenario with $\alpha_{\scaleto{Q}{5pt}}=0$, given by \cref{Heun_Parameters_Alpha_Q_0_ST}, can be expressed as:
	\begin{equation}
		\begin{aligned}
			\frac{\gamma}{2} 
			&= i\,\frac{\epsilon \rho_{\scaleto{S}{4pt}}}{\left(\overline{a}+N_{\scaleto{Q}{5pt}}\right)^{2}}\,,\\
			\frac{\alpha}{2} 
			&= i \, \frac{\rho_{\scaleto{S}{4pt}}}{\left(\overline{a}+N_{\scaleto{Q}{5pt}}\right)^{2}} \, \sqrt{\epsilon^{2} - \left(\overline{a}+N_{\scaleto{Q}{5pt}}\right) \overline{m}_{\phi}^{2}}\,.
		\end{aligned}\label{Heun_Parameters_AG_0_DPs}
	\end{equation}
	By substituting these expressions for $\gamma$ and $\alpha$ into the radial solution \eqref{RWF_0_ST_Approx_DPs}, we obtain that
	\begin{equation}
		\begin{aligned}
			R_{\scaleto{ST}{4pt}}^{\scaleto{(0)}{6pt}}(x) 
			&= c_{1}\,\exp\left\{i\,\frac{\rho_{\scaleto{S}{4pt}}}{\left(\overline{a}+N_{\scaleto{Q}{5pt}}\right)^{2}} \left[\epsilon \ln(x-1) + \sqrt{\epsilon^{2} - \left(\overline{a}+N_{\scaleto{Q}{5pt}}\right)\overline{m}_{\phi}^{2}}\right]\right\}\\
			&+ c_{2}\,\exp\left\{i\, \frac{\rho_{\scaleto{S}{4pt}}}{\left(\overline{a}+N_{\scaleto{Q}{5pt}}\right)^{2}} \left[-\epsilon \ln(x-1) + \sqrt{\epsilon^{2} - \left(\overline{a}+N_{\scaleto{Q}{5pt}}\right)\overline{m}_{\phi}^{2}}\right]\right\}\,,	
		\end{aligned}\label{ST-0-Dark-phase}
	\end{equation}
    where the expression within the brackets can be interpreted as an \emph{implicit dark phase}. Furthermore, note that the term associated with $\alpha$ is a global phase of \cref{RWF_0_ST_Approx_DPs}, implying that the physical relevance of these \emph{dark phases} arises from the contribution of $\gamma$ due to its occurrence as relative phases.
    
\begin{figure}[ht!]
		\begin{center}
			\includegraphics[width=0.65\textwidth]{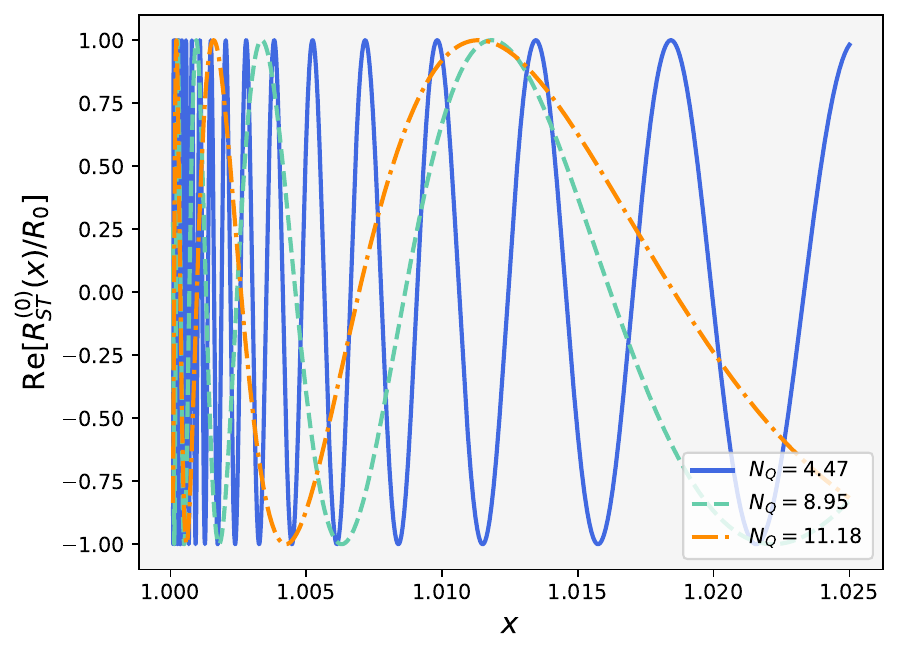}
			\caption{Real part of the non-normalized solution $R_{\scaleto{ST}{4pt}}^{\scaleto{(0)}{6pt}}(x)$ as a function of the dimensionless radial coordinate $x = \rho/\rho_{\scaleto{+}{4pt}}$ for the standard scenario of $\alpha_{\scaleto{Q}{5pt}} = 0$. This result is obtained from \cref{ST-0-Dark-phase}. The parameters are fixed at $\rho_{\scaleto{S}{4pt}} = 10^{2}$ m, $\overline{a} = 10^{-3}$, $\epsilon = n\,\overline{m}_{\phi}$, with $n = 4$, and $\overline{m}_{\phi} = 1$ m$^{-1}$ (chosen for visual clarity). For realistic values of $\overline{m}_{\phi} = m_{\phi} c/\hbar$, the oscillations would be much more rapid, making the distinction between different values of $N_{\scaleto{Q}{5pt}}$ visible only at extremely small distances. Furthermore, note that, besides the compression of these oscillations as $x\to 1^{+}$ (due to $\ln (x-1)$), the wavelength increases with increasing $N_{\scaleto{Q}{5pt}}$.}
			\label{ST_0_NH}
		\end{center}
	\end{figure}

    As an example, the real part of \cref{ST-0-Dark-phase} is represented in Figure \ref{ST_0_NH} for three values of the quintessence parameter: our estimated value of $N_{\scaleto{Q}{5pt}}$, along with lower and upper values. As expected, when $x\to 1^{+}$ the logarithmic term becomes increasingly dominant, compressing the oscillations of the radial wave function and making the different $N_{\scaleto{Q}{5pt}}$ distinguishable only at extremely small distances.
    	
	Still regarding $\alpha_{\scaleto{Q}{5pt}} = 0$, the cloudless scenario (\textsc{cl}) is obtained by setting $\overline{a} = 0$ in the above result. This happens because, when $\alpha_{\scaleto{Q}{5pt}}=0$, both $\overline{a}$ and $N_{\scaleto{Q}{5pt}}$ contribute identically to the metric function $A(x)$, given by \cref{A(x)_Alpha_Q_0}. Thus, we have the freedom to examine the result of \cref{ST-0-Dark-phase} either with $\overline{a}=0$ (cloudless scenario) or with $N_{\scaleto{Q}{5pt}}=0$ (quintessence-free case). In the last case, the \enquote{dark phases} would rather be named \enquote{cloud phases} since the whole solution would be independent of the quintessential fluid.
	
	An interesting particular case is the one where the Heun parameter $\alpha$ is zero, implying that $\epsilon^{2} = \left(\overline{a} + N_{\scaleto{Q}{5pt}}\right) \overline{m}_{\phi}^{2}$, which simplifies \cref{ST-0-Dark-phase} to
	\begin{equation}
		R_{\scaleto{ST}{4pt}}^{\scaleto{(0,\,\circledcirc)}{6pt}}(x) = c_{1}\exp\left[ i\,\frac{\epsilon \rho_{\scaleto{S}{4pt}}}{\left(\overline{a} + N_{\scaleto{Q}{5pt}}\right)^{2}} 
		\ln (x-1)\right] + c_{2}\exp\left[ -i\,\frac{\epsilon \rho_{\scaleto{S}{4pt}}}{\left(\overline{a} + N_{\scaleto{Q}{5pt}}\right)^{2}} 
		\ln (x-1)\right]\,,\label{Vínculo_Energia_Comum_Alpha_Q_0_RWF}
	\end{equation}
    which exhibits the same relative phase as \cref{ST-0-Dark-phase}. The symbol \enquote{$\circledcirc$} was used only to distinguish this particular case, which constrains $\epsilon^{2}$, from the general solution.
	
	In the sequence, we investigate the emergence of dark phases in the regime of $\alpha_{\scaleto{Q}{5pt}} = 1/2$.

	\subsection{Dark phases for intermediate state parameter}
	
	When $\alpha_{\scaleto{Q}{5pt}} = 1/2$, in the standard scenario, the radial wave function $R_{\scaleto{ST}{4pt}}^{\scaleto{(1/2)}{6pt}}(x)$ is determined by \cref{Radial_WF_ST_Middle}, with the Heun parameters specified by \cref{Heun_Parameters_Meio_Intervalo}. Near the event horizon, this radial solution can be approximated as
	\begin{equation}
		R_{\scaleto{ST}{4pt}}^{\scaleto{(1/2)}{6pt}}(x) = \overline{c}_{1}\,\exp\left[\frac{\gamma}{2} \ln (x-1) + \frac{\alpha}{2}\right] + \overline{c}_{2}\,\exp\left[-\frac{\gamma}{2} \ln (x-1) + \frac{\alpha}{2}\right]\,,
	\end{equation}
	considering a zeroth order expansion around $x=1$ (that is, on the event horizon). The constants $\overline{c}_{1}$ and $\overline{c}_{2}$ were renamed to emphasize that they had absorbed the term $(1-N_{\scaleto{Q}{5pt}}\,\rho_{\scaleto{S}{4pt}}/\overline{a}^{2})$. With respect to $\text{HeunC}\left(\alpha,\,\gamma,\,\beta,\,-\delta,\,\delta+\eta;\,1-x\right)$, contained in \cref{Radial_WF_ST_Middle}, we consider only the equivalent\footnote{Although the appendix shows the expression for $\text{HeunC}\left(\alpha,\,\beta,\,\gamma,\,\delta,\,\eta;\,z\right)$, which is centered at $z=0$, the result for the solution centered at $z=1$ is easily obtained by changing $\beta\leftrightarrow\gamma$, $\delta\to-\delta$, $\eta\to\delta+\eta$, and $z\to 1-x$ \cite{Olver2010}.} of \cref{Approx_HeunCl}, which has $c_{0} = 1$. The remaining terms of the Fuchs-Frobenius series, determined by \cref{def-HeunC}, contribute to higher-order corrections of $\left(x-1\right)^{\gamma/2}$. 
	
\begin{figure}[ht!]
	\begin{center}
		\includegraphics[width=0.65\textwidth]{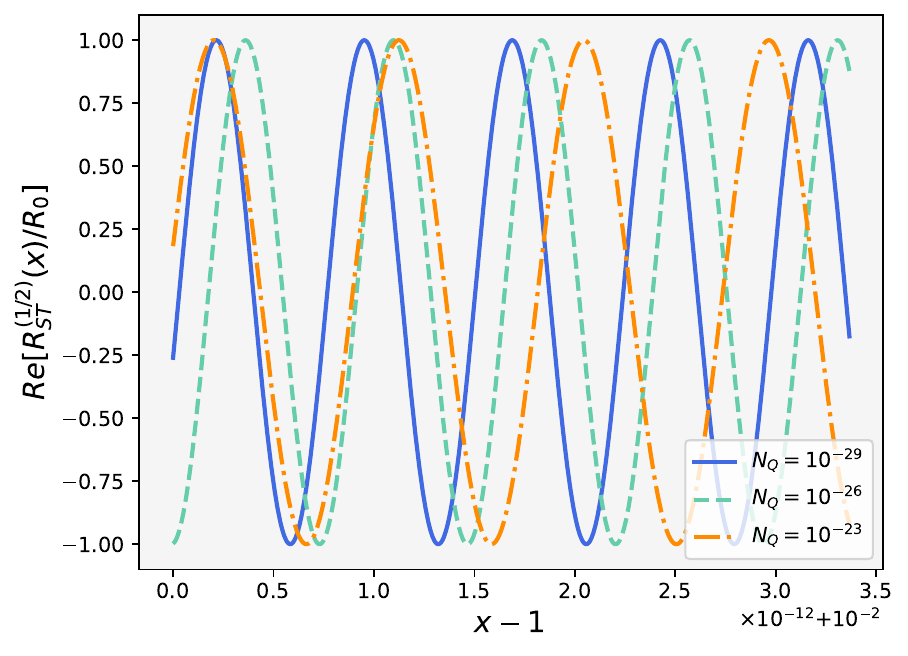}
		\caption{Plot of the real part of $R_{\scaleto{ST}{4pt}}^{\scaleto{(1/2)}{6pt}}(x)/R_{0}$ as a function of the dimensionless coordinate $x-1 = (\rho-\rho_{\scaleto{+}{4pt}})/\rho_{\scaleto{+}{4pt}}$, in the standard scenario of $\alpha_{\scaleto{Q}{5pt}} = 1/2$, for three distinct values of $N_{\scaleto{Q}{5pt}}$. This result is derived from \cref{RWF_ST_Meio_Constrained,Phi_ST_Meio_Extended,DP_Meio_Extended} by setting $c_{+}=R_{0}$ and $c_{-}=0$, while the remaining physical parameters are fixed at $\rho_{\scaleto{S}{4pt}} = 10^{10}$ m, $\overline{a} = 10^{-6}$, $\epsilon = 3\,\overline{m}_{\phi}$, and $\overline{m}_{\phi} = 2.84 \times 10^{-12}$ m$^{-1}$ (implying $m_{\phi} = 10^{-54}$ kg). More realistic values of $\overline{m}_{\phi} = m_{\phi} c/\hbar$ would make the oscillations many orders of magnitude faster, causing the distinction between different values of $N_{\scaleto{Q}{5pt}}$ to be visible only at extremely small distances. In addition, note the compression of oscillations as $x\to 1^{+}$, associated with the $\ln (x-1)$ term, together with the wavelength increase as $N_{\scaleto{Q}{5pt}}$ increases.}
		\label{ST_Meio_Extended}
	\end{center}
\end{figure}

	Furthermore, since the solution of \cref{Radial_WF_ST_Middle} is valid only in the \emph{physical} regime, where $N_{\scaleto{Q}{5pt}}\,\rho_{\scaleto{S}{4pt}}/\overline{a}^{2} \ll 1$, the Heun parameters $\gamma$ and $\alpha$, defined in \cref{Heun_Parameters_Meio_Intervalo}, can be written as:
	\begin{equation}
		\begin{aligned}
			\frac{\gamma}{2} &= i\, \frac{\epsilon \rho_{\scaleto{S}{4pt}}}{\overline{a}^{2}} \left(1 - \frac{2 N_{\scaleto{Q}{5pt}}\,\rho_{\scaleto{S}{4pt}}}{\overline{a}^{2}}\right)\,,\\
			\frac{\alpha}{2} &= \frac{\epsilon \rho_{\scaleto{S}{4pt}}}{\overline{a}^{2}} \left(1 + \frac{N_{\scaleto{Q}{5pt}}\,\rho_{\scaleto{S}{4pt}}}{\overline{a}^{2}}\right)\,.
		\end{aligned}
	\end{equation}
	Taking into account that the term $\exp(\alpha/2)$ is a global phase independent of $x$, we move its contribution to the constants $\overline{c}_{1}$ and $\overline{c}_{2}$. In this way, the radial wave function $R_{\scaleto{ST}{4pt}}^{\scaleto{(1/2)}{6pt}}(x)$ behaves as
	\begin{equation}
		\begin{aligned}
			R_{\scaleto{ST}{4pt}}^{\scaleto{(1/2)}{6pt}}(x) = \Phi_{+}(x)
			\exp\left[-i\,\delta_{\scaleto{ST}{4pt}}^{\scaleto{\,(1/2)}{6pt}}(x)\right] + \Phi_{-}(x)
			\exp\left[+i\,\delta_{\scaleto{ST}{4pt}}^{\scaleto{\,(1/2)}{6pt}}(x)\right]\,,
		\end{aligned} \label{RWF_ST_Meio_Constrained}
	\end{equation}
	when close to the event horizon. The functions $\Phi_{\pm}(x)$ and $\delta_{\scaleto{ST}{4pt}}^{\scaleto{\,(1/2)}{6pt}}(x)$ are, respectively,
	\begin{equation}
		\Phi_{\pm}(x) 
		= c_{\pm}\,
		\exp\left[\pm i\, \frac{\epsilon \rho_{\scaleto{S}{4pt}}}{\overline{a}^{2}} \ln (x-1) \right]\label{Phi_ST_Meio_Extended}
	\end{equation}
	and
	\begin{equation}
		\delta_{\scaleto{ST}{4pt}}^{\scaleto{\,(1/2)}{6pt}}
		= \frac{2\epsilon\,N_{\scaleto{Q}{5pt}}\,\rho_{\scaleto{S}{4pt}}^{2}}{\overline{a}^{4}}\,
		\ln\left(x-1\right)\,.\label{DP_Meio_Extended}
	\end{equation}
	Moreover, it should be noted that $\overline{c_{1,\,2}}$ were renamed to $c_{\pm}$ after absorbing the constant term of $\exp\left(\alpha/2\right)$. In addition, it is relevant to stress that the functions $\Phi_{\pm}(x)$ are independent of the quintessence parameter $N_{\scaleto{Q}{5pt}}$, while $\delta_{\scaleto{ST}{4pt}}^{\scaleto{\,(1/2)}{6pt}}(x)$ is a dark phase, induced by the presence of quintessence. To illustrate this case, we provide Figure \ref{ST_Meio_Extended}, which is a detailed numerical example.

	In addition to this standard case, we show an example of a dark phase occurring in the cloudless scenario for a massless spin$-0$ boson ($\overline{m}_{\phi} = 0$), with $\alpha_{\scaleto{Q}{5pt}}=1/2$. In this situation, we consider the radial wave function $R_{\scaleto{CL}{4pt}}^{\scaleto{(1/2,\,\triangle)}{6pt}}(x)$, given by \cref{Radial_WF_CL_Middle_Extras,U_WF_CL_Middle_Extra_1}, as
	\begin{equation}
		R_{\scaleto{CL}{4pt}}^{\scaleto{(1/2,\,\triangle)}{6pt}}(x) = x^{-1/2}\left[c_{1}\, x^{s/2} + c_{2}\, x^{-s/2}\right]\,.
	\end{equation}
	The symbol \enquote{$\scaleto{\triangle}{8pt}$} denotes that this radial solution is a restricted case, valid only for $\overline{m}_{\phi} = 0$. Additionally, we recall that this solution was obtained for values of $x$ where $x\gg 1/x$, that is, for particles which are reasonably distant from the event horizon. 
	
	Furthermore, since $N_{\scaleto{Q}{5pt}}$ is on the order of $10^{-26}$ m$^{-1}$ when $\alpha_{\scaleto{Q}{5pt}}=1/2$, we see that the parameter $s$, defined by \cref{def-s}, can be written as
	\begin{equation*}
		\frac{s}{2} = i\,\frac{|\epsilon|}{N_{\scaleto{Q}{5pt}}} \left(1 - \frac{N_{\scaleto{Q}{5pt}}^{2}}{\epsilon^{2}}\right) + \mathcal{O}\left(N_{\scaleto{Q}{5pt}}^{4}\right)\,,\label{def-s-puiseux}
	\end{equation*}
	implying that
	\begin{equation}
		R_{\scaleto{CL}{4pt}}^{\scaleto{(1/2,\,\triangle)}{6pt}}(x) = \Phi_{\scaleto{+}{4pt}}(x)\, \exp\left[i\,\delta_{\,\scaleto{CL}{4pt}}^{\scaleto{(1/2,\,\triangle)}{6pt}}\right] + \Phi_{\scaleto{-}{4pt}}(x)\, \exp\left[-i\,\delta_{\scaleto{CL}{4pt}}^{\,\scaleto{(1/2,\,\triangle)}{6pt}}\right]\,.
	\end{equation}
	Hence, the functions $\Phi_{\scaleto{\pm}{5pt}}(x)$ and $\delta_{\scaleto{CL}{4pt}}^{\scaleto{\,(1/2,\,\triangle)}{6pt}}(x)$ are determined by
	\begin{equation*}
		\Phi_{\scaleto{\pm}{5pt}}(x) = c_{\scaleto{\pm}{5pt}} \, x^{-1/2}\,,
	\end{equation*}
	while
	\begin{equation}
		\delta_{\scaleto{CL}{4pt}}^{\scaleto{\,(1/2,\,\triangle)}{6pt}}(x) = \frac{|\epsilon|}{N_{\scaleto{Q}{5pt}}} \left(1 - \frac{N_{\scaleto{Q}{5pt}}^{2}}{\epsilon^{2}}\right) \ln x \,, \label{delta_CL}
	\end{equation}
	respectively.
	
	The relevance of this result lies in its validity for a large domain of the variable $x$, that is, for all $x$ satisfying the following conditions: $x \ll 1/x$ and $x \, \rho_{\scaleto{+}{4pt}} \ll |l|$. In the case where $N_{\scaleto{Q}{5pt}} \ll \epsilon$ (which is strongly plausible from a physical point of view), the phase difference $\delta_{\scaleto{CL}{4pt}}^{\scaleto{\,(1/2,\,\triangle)}{6pt}}(x)$ simply becomes $\delta_{\scaleto{CL}{4pt}}^{\scaleto{\,(1/2,\,\triangle)}{6pt}}(x) =(|\epsilon|/N_{\scaleto{Q}{5pt}}) \ln x$. In this particular example, we see the occurrence of a dark phase without the need to move the scalar particle close to the event horizon. We also highlight that the occurrence of a $\ln\left(x\right)$ instead of $\ln\left(x-1\right)$ indicates that this massless scalar particle ignores the existence of the event horizon, which is consistent with the regime of $x\gg 1/x$. Certainly, these examples do not exhaust all the possibilities associated with $\alpha_{\scaleto{Q}{5pt}} = 1/2$, but they illustrate the occurrence of dark phases in this regime. 
	
	In the next subsection, we will investigate the limiting case where $\alpha_{\scaleto{Q}{5pt}}=1$, both in the standard (\textsc{st}) and cloudless (\textsc{cl}) scenarios.

	\subsection{Dark phases at the upper bound}

	After exploring the emergence of dark phases in the extended solutions for $\alpha_{\scaleto{Q}{5pt}} = 0$ and $1/2$ (with examples from both standard and cloudless scenarios), the analysis now turns to the limiting case of $\alpha_{\scaleto{Q}{5pt}}=1$ (i.e., $\omega_{\scaleto{Q}{5pt}} = -1$), which represents the most physically relevant regime. This focus is motivated by observational constraints indicating that the dark energy state parameter ($\omega_{\scaleto{DE}{4pt}}$) is very close to $-1$ \cite{DESI62024,DESCollaboration2025}. And, although our solutions are derived for a cylindrically symmetric theoretical universe, we are interested in analyzing the behavior of these dark phases near $\alpha_{\scaleto{Q}{5pt}} = 1$ to enable a future comparison between cylindrical and spherically symmetric solutions. This approach facilitates an assessment of the relative influence of spacetime geometry and the quintessence fluid on the emergence of these phases.
	
	As an example of the standard scenario, we consider the radial solution $R_{\scaleto{ST}{4pt}}^{\scaleto{(1,\, \oslash)}{6pt}}(x)$, given by \cref{RWF_1_Constrained_Bessel}:
	\begin{equation*}
		R_{\scaleto{ST}{4pt}}^{\scaleto{(1,\, \oslash)}{6pt}}(x) = \frac{R_{0}}{\sqrt{x}}\left(1 - \frac{N_{\scaleto{QL}{5pt}}\rho_{\scaleto{S}{4pt}}^{2}}{2\,\overline{a}^{2}}\,x^{2}\right) \text{K}_{n}\left(\sqrt{V_{0}} \,x\right)\,,
	\end{equation*}
	which is a valid description provided $x^{3} \gg \left(N_{\scaleto{QL}{5pt}}\,\rho_{\scaleto{S}{4pt}}^{2}/\overline{a}^{3}\right)^{-1}$. This solution describes spin$-0$ particles which are distant from the event horizon and  satisfy the constraint $\epsilon^{2} = \overline{a}\,\overline{m}_{\phi}^{2}/2$. 
	
	For instance, let us suppose that $N_{\scaleto{QL}{5pt}} = 6.9 \times 10^{-53}$ m$^{-2}$ (in accordance with \cref{Estimate-N_Q,N_QL_Definition}), $\rho_{\scaleto{S}{4pt}} = 10^{15}$ m, $\overline{a} = 10^{-6}$, $\kappa = 1$ (particle moving in a fixed-$z$ plane), and $\overline{m}_{\phi} = 6.37 \times 10^{17}$ m$^{-1}$ (using the Higgs boson as a reference). Under these conditions, the dimensionless scale factor $\sqrt{V_{0}}$, as defined by \cref{def-V_0}, is on the order of $5 \times 10^{41}$, which represents an extremely large value. Consequently, the radial \enquote{wave function} decays rapidly, making it nearly impossible to distinguish between scenarios with or without the quintessence fluid. However, in a hypothetical case regarding a massless particle, it may be possible to obtain an improved result by considering \cref{RWF_1_Far_EH_x}. In the present case, the contribution from quintessence remains quite small, as demonstrated below.
	
	Given the condition $x^{3} \gg \left(N_{\scaleto{QL}{5pt}}\,\rho_{\scaleto{S}{4pt}}^{2}/\overline{a}^{3}\right)^{-1}$, together with the estimates for $V_{0}$, we can investigate the behavior of the radial solution $R_{\scaleto{ST}{4pt}}^{\scaleto{(1,\, \oslash)}{6pt}}(x)$ by means of an asymptotic expansion for large values of the argument $\sqrt{V_{0}} \,x$. For better understanding, we first recall the expansion of $\text{K}_{n}\left(\sqrt{V_{0}} \,x\right)$ for $\sqrt{V_{0}} \,x\to\infty$, with the order $n$ fixed \cite{Olver2010}:
	\begin{equation}
		\text{K}_{n}\left(\sqrt{V_{0}} \,x\right) = 
		\left(\frac{\pi}{2\sqrt{V_{0}}\,x}\right)^{1/2} e^{-\sqrt{V_{0}}\,x} \sum_{k=0}^{\infty} \frac{a_{k}(n)}{\sqrt{V_{0}^{k}}\,x^{k}}\,.
	\end{equation}
	Having this point established, we see that the asymptotic expansion of $R_{\scaleto{ST}{4pt}}^{\scaleto{(1,\, \oslash)}{6pt}}(x)$ must contain a dominant term of the form $e^{-\sqrt{V_{0}}\, x}$, derived from the modified Bessel function $\text{K}_{n}\left(\sqrt{V_{0}}\, x\right)$. From this result, we see that, as the argument  increases, it becomes harder to perceive the alteration induced in the wave function by the quintessence fluid.
	
	In an attempt to determine a dark phase for this regime, we calculate the expansion of $R_{\scaleto{ST}{4pt}}^{\scaleto{(1,\, \oslash)}{6pt}}(x)$ for $\sqrt{V_{0}} \,x \to \infty$, with fixed $n$, using a computer algebra system (CAS), from which we obtain that:
	\begin{equation}
		R_{\scaleto{ST}{4pt}}^{\scaleto{(1,\, \oslash)}{6pt}}(x) = \overline{R_{0}}\,x\,e^{-\sqrt{V_{0}}\, x} + \mathcal{O}\left(x^{-k}\right)\,,\label{RWF_1_Far_Expand}
	\end{equation}
	with $k \in \mathbb{Z}^{+}$ and $\sqrt{V_{0}} \in \mathbb{R}$. Observe that $\overline{R_{0}}$ had absorbed, besides $R_{0}$, the remaining $x-$independent multiplicative factors from the expansion. 
	
	Considering that $N_{\scaleto{QL}{5pt}}(\kappa^{2}/\overline{a} - 3) \ll \overline{m}_{\phi}^{2}$, which is valid for all $\overline{m}_{\phi} \neq 0$, we see that\footnote{The scale factor $\sqrt{V_{0}}$ will always be a real number because, if $\overline{m}_{\phi}=0$ occurs, the solution of \eqref{RWF_1_Constrained_Bessel} is no longer valid, and it is necessary to resort to the general solution \eqref{RWF_1_Far_EH_x}.}:
	\begin{equation*}
		\sqrt{V_{0}} = 	\frac{\rho_{\scaleto{S}{4pt}}\,\overline{m}_{\phi}}{\left(2\,\overline{a}^{3}\right)^{1/2}}\,
		\left[1 - \frac{N_{\scaleto{QL}{5pt}}}{\overline{m}_{\phi}^{2}}\left(\frac{\kappa^{2}}{\overline{a}} - 3\right)\right]\,.
	\end{equation*} 
	Thus, it follows that the asymptotic expansion of $R_{\scaleto{ST}{4pt}}^{\scaleto{(1,\, \oslash)}{6pt}}(x)$, given by the result \eqref{RWF_1_Far_Expand}, can be rewritten as:
	\begin{equation}
		R_{\scaleto{ST}{4pt}}^{\scaleto{(1,\, \oslash)}{6pt}}(x) = \Phi^{\scaleto{(1,\, \oslash)}{6pt}}_{\scaleto{ST}{4pt}}(x)\,
		\exp\left[ \delta_{\scaleto{ST}{4pt}}^{\scaleto{(1,\, \oslash)}{6pt}}(x)\right]\,,\label{RWF_1_Far_DP}
	\end{equation}
	where $\Phi^{\scaleto{(1,\, \oslash)}{6pt}}_{\scaleto{ST}{4pt}}(x)$ is given by
	\begin{equation*}
		\Phi^{\scaleto{(1,\, \oslash)}{6pt}}_{\scaleto{ST}{4pt}}(x) = \overline{R_{0}}\,x\,\exp\left[-\frac{\rho_{\scaleto{S}{4pt}}\,\overline{m}_{\phi}}{\left(2\,\overline{a}^{3}\right)^{1/2}}\, x\right]\,,\label{Phi_1_Far_Expansion}
	\end{equation*}
	while $\delta^{\scaleto{(1,\, \oslash)}{6pt}}_{\scaleto{ST}{4pt}}(x)$ takes the form of
	\begin{equation}
		\delta^{\scaleto{(1,\, \oslash)}{6pt}}_{\scaleto{ST}{4pt}}(x) = \frac{N_{\scaleto{QL}{5pt}}}{\overline{m}_{\phi}}\,\frac{\rho_{\scaleto{S}{4pt}}}{\left(2\,\overline{a}^{3}\right)^{1/2}}  \left(\frac{\kappa^{2}}{\overline{a}} - 3\right) x\,.\label{DP_1_Far}
	\end{equation}
	Note that the function $\Phi^{\scaleto{(1,\, \oslash)}{6pt}}_{\scaleto{ST}{4pt}}(x)$ is independent of $N_{\scaleto{QL}{5pt}}$, while $\delta^{\scaleto{(1,\, \oslash)}{6pt}}_{\scaleto{ST}{4pt}}(x)$ depends explicitly on this parameter. In addition, we see that $\delta_{\scaleto{ST}{4pt}}^{\scaleto{(1,\oslash)}{6pt}}(x)$ is linear on $N_{\scaleto{QL}{5pt}}$, $\rho_{\scaleto{S}{4pt}}$, and $x$. 
	
	With the result of eqs. \eqref{RWF_1_Far_DP}--\eqref{DP_1_Far}, we show how the dark energy parameter $N_{\scaleto{QL}{5pt}}$ interferes with the dominant term of $R_{\scaleto{ST}{4pt}}^{\scaleto{(1,\, \oslash)}{6pt}}(x)$. \textit{A priori}, we can denote the function $\delta^{\scaleto{(1,\, \oslash)}{6pt}}_{\scaleto{ST}{4pt}}(x)$ as a \enquote{dark phase} for particles which are far from the event horizon. However, we emphasize that this \enquote{phase} is, in fact, a real function that has a quite small contribution to the amplitude change in the wave function. 
    
\begin{figure}[ht!]
	\begin{center}
		\includegraphics[width=0.65\textwidth]{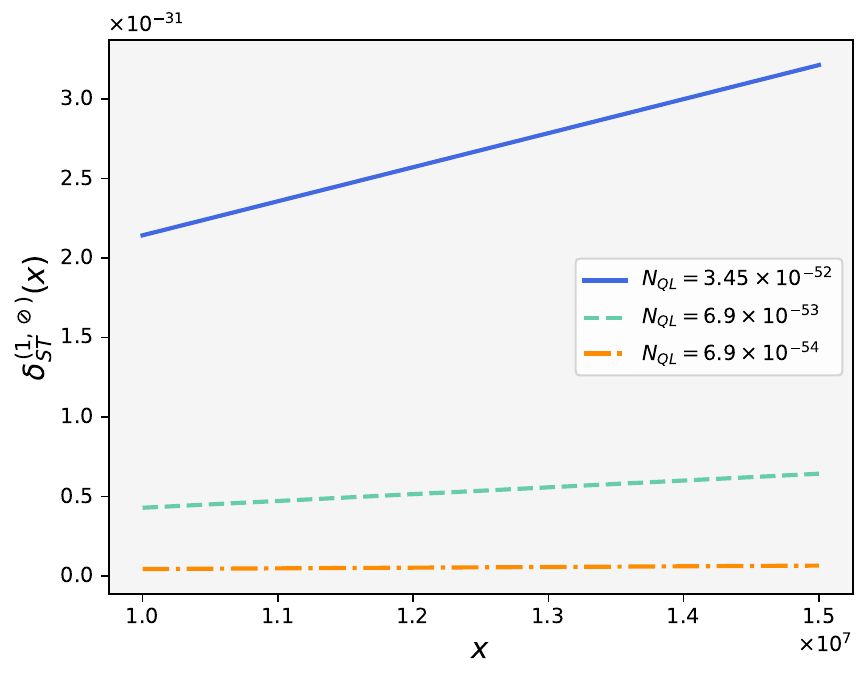}
		\caption{Plot of $\delta_{\scaleto{ST}{4pt}}^{\scaleto{(1,\oslash)}{6pt}}(x)$, defined by \cref{DP_1_Far}, as a function of the dimensionless coordinate $x = \rho/\rho_{\scaleto{+}{4pt}}$, in the standard scenario of $\alpha_{\scaleto{Q}{5pt}} = 1$. The physical parameters are fixed at $\rho_{\scaleto{S}{4pt}} = 10^{15}$ m, $\overline{a} = 10^{-6}$, $\kappa^{2} = 25$, and $\overline{m}_{\phi} = 2.84 \times 10^{17}$ m$^{-1}$ (implying $m_{\phi} = 10^{-25}$ kg). The three distinct values for  $N_{\scaleto{QL}{5pt}}$ were chosen to represent the following cases: the orange \enquote{dashdotted} line refers to a case where $N_{\scaleto{QL}{5pt}} = 1/l^{2}$, i.e., when $N_{\scaleto{Q}{5pt}}=0$ and the quintessential fluid is absent; the green \enquote{dashed} line is associated with the occurrence of both quintessence and the cosmological constant; at last, the solid blue line represents a case where quintessence is more abundant. Note that these lines differ only by their slope since $\delta_{\scaleto{ST}{4pt}}^{\scaleto{(1,\oslash)}{6pt}}(x)$ is linear in $N_{\scaleto{QL}{5pt}}$. In addition, we reinforce that the condition $x\gg \left(N_{\scaleto{QL}{5pt}}\rho_{\scaleto{S}{4pt}}^{2}/\overline{a}^{3}\right)^{-1}$ was satisfied for all chosen values of $N_{\scaleto{QL}{5pt}}$.}
		\label{ST_1_Extended}
	\end{center}
\end{figure}

    From a numerical point of view, it is not trivial to show the difference induced in $R_{\scaleto{ST}{4pt}}^{\scaleto{(1,\, \oslash)}{6pt}}(x)$ for the cases when $N_{\scaleto{Q}{5pt}}\neq 0$ and those where $N_{\scaleto{Q}{5pt}}=0$ (making $N_{\scaleto{QL}{5pt}} = 1/l^{2}$). For this reason, we present a plot of \cref{DP_1_Far} in Figure \ref{ST_1_Extended}, where the linearity on both $N_{\scaleto{QL}{5pt}}$ and $x$ is evident. We reinforce that, in this specific case, $N_{\scaleto{QL}{5pt}}$ determines the slope of each line.
	
	In our next example, we consider the cloudless scenario, with $\alpha_{\scaleto{Q}{5pt}}=1$, described by \cref{RWF_CL_1_Constrained}. This solution is valid as long as $\epsilon^{2} = \kappa^{2}\,N_{\scaleto{QL}{5pt}}$, where the parameter $\ell$, present in the radial solution, was defined by \cref{ell-m}. We choose this particular case to demonstrate that $R_{\scaleto{CL}{4pt}}^{\scaleto{(1,\,\boxtimes)}{6pt}}(x)$ can be rewritten in the form of
	\begin{equation}
		R_{\scaleto{CL}{4pt}}^{\scaleto{(1,\,\boxtimes)}{6pt}}(x) = \Phi^{\scaleto{(1,\,\boxtimes)}{6pt}}_{\scaleto{CL}{4pt}}(x)\,\exp\left[-\delta_{\scaleto{CL}{4pt}}^{\scaleto{(1,\,\boxtimes)}{6pt}}(x)\right]\,, \label{RWF_1_CL_Far}
	\end{equation}
	where
	\begin{equation}
		\Phi^{\scaleto{(1,\,\boxtimes)}{6pt}}_{\scaleto{CL}{4pt}}(x) = R_{0}\,x^{-3/2}\,,\label{Phi_1_CL_Far}
	\end{equation}
	and, by using the definitions in \cref{ell-m}, we conclude that
	\begin{equation}
		\delta_{\scaleto{CL}{4pt}}^{\scaleto{(1,\,\boxtimes)}{6pt}}(x) = \left(\frac{9}{4} + \frac{\overline{m}_{\phi}^{2}}{N_{\scaleto{QL}{5pt}}}\right)^{1/2} \ln \left(x\right)\,.\label{DP_1_CL_Far}
	\end{equation}
	Therefore, eqs. \eqref{RWF_1_CL_Far}--\eqref{DP_1_CL_Far} show that, in the absence of the cloud of strings, and under the constraint $\epsilon^{2} = \kappa^{2}\,N_{\scaleto{QL}{5pt}}$, the \enquote{dark phase} 	
	$\delta_{\scaleto{CL}{4pt}}^{\scaleto{(1,\,\boxtimes)}{6pt}}(x)$ is indeed a real function, responsible for changing the amplitude of $R_{\scaleto{CL}{4pt}}^{\scaleto{(1,\,\boxtimes)}{6pt}}(x)$. 
	
	If we consider the orders of magnitude of $\overline{m}_{\phi}$ and $N_{\scaleto{QL}{5pt}}$, we can rewrite \cref{DP_1_CL_Far} as
	\begin{equation*}
		\delta_{\scaleto{CL}{4pt}}^{\scaleto{(1,\,\boxtimes)}{6pt}}(x) = \frac{\overline{m}_{\phi}}{N_{\scaleto{QL}{5pt}}^{1/2}}\,\ln \left(x\right)\,.
	\end{equation*}
	Again, this wave function decays so rapidly that numerically distinguishing $N_{\scaleto{Q}{5pt}} = 0$ from $N_{\scaleto{Q}{5pt}} \neq 0$ is a challenge.
	
	In conclusion, in the limiting case of $\alpha_{\scaleto{Q}{5pt}}=1$, we examined examples in which there is no occurrence of dark phases in the original sense of the term. These \enquote{phases} are associated with real functions and are responsible for modifying the amplitude of radial solutions. In any case, we presented examples of the procedure for determining these functions. 
	
	However, given the infinitely small intensity of the quintessence parameter $N_{\scaleto{Q}{5pt}} = N_{\scaleto{QL}{5pt}} - 1/l^{2}$, it becomes impractical to approach the problem numerically. For further clarification, we only need to recall that the difference (in orders of magnitude) between $\overline{m}_{\phi}^{2}$ and $N_{\scaleto{QL}{5pt}}$ would be on the order of $10^{87}$, which points to the ineffectiveness of numerical methods for this physical problem.
	
	Our next, and final, analysis will be on the possible emergence of \emph{dark phases} in the horizonless scenario, with $\alpha_{\scaleto{Q}{5pt}}$ assuming the values $0,\,1/2,\,1$. Unlike the results of Section \ref{RWF_NH_All_Alpha_Q_Near}, our extended solutions can handle these cases.

	\subsection{Horizonless scenarios}
	
	The extended solutions we have presented through this section have an important characteristic: the possibility of exploring the scenario where $\rho_{\scaleto{S}{4pt}} = 0$, that is, in the absence of an event horizon. This horizonless scenario (which has no black string) could not be investigated from the general result of \eqref{RWF_NH_DPs}, which was only valid for scalar particles near the event horizon. However, considering the solutions of the \textsc{hl} scenario, given by eqs. \eqref{RWF_HL_0_Full}, \eqref{Radial_WF_Intermediate_HL} and \eqref{RWF_HL_1_z}, we can now investigate whether dark phases also emerge in the absence of the event horizon.
	
\begin{figure}[ht!]
		\begin{center} 
			\includegraphics[width=0.65\textwidth]{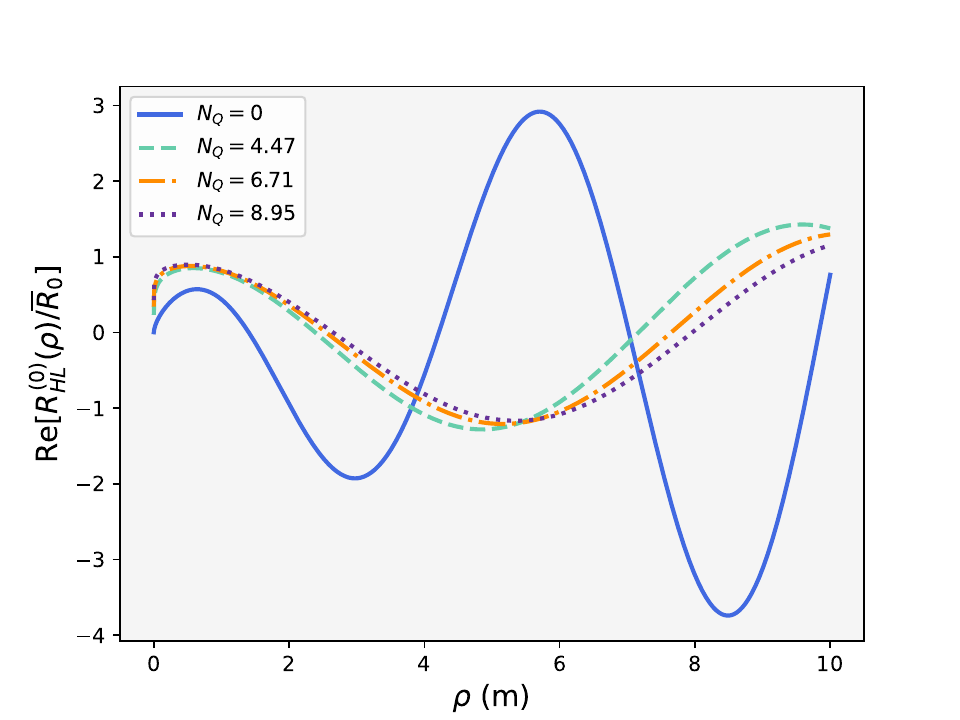}
			\caption{The real part of the radial solution $R_{\scaleto{HL}{4pt}}^{\scaleto{(0)}{6pt}}(\rho)$, given by eqs. \eqref{R_WF_HL_0_NO} and \eqref{Dark-phase-HL-0}, up to normalization, in the horizonless scenario with $\alpha_{\scaleto{Q}{5pt}} = 0$, for various values of the quintessence parameter $N_{\scaleto{Q}{5pt}}$. In the absence of an event horizon (and of the \emph{black string} itself), this behavior is valid near the origin (when $\rho = 0$). For simplicity, we choose $\kappa = 1$, which describes a spin$-0$ particle moving in a plane with fixed $z = z_{0}$. The remaining parameters are set at: $\rho_{\scaleto{S}{4pt}} = 10^{5}$ m, $\overline{a} = 1$, $\overline{m}_{\phi} = 1$ m$^{-1}$, $\epsilon = 4\,\overline{m}_{\phi}$. Observe that we choose an unrealistic value for $\overline{m}_{\phi}$ in order to facilitate visualization.}
			\label{HL_0_Near_Origin}
		\end{center}
	\end{figure}

\begin{figure}[ht!]
		\begin{center} 
			\includegraphics[width=0.65\textwidth]{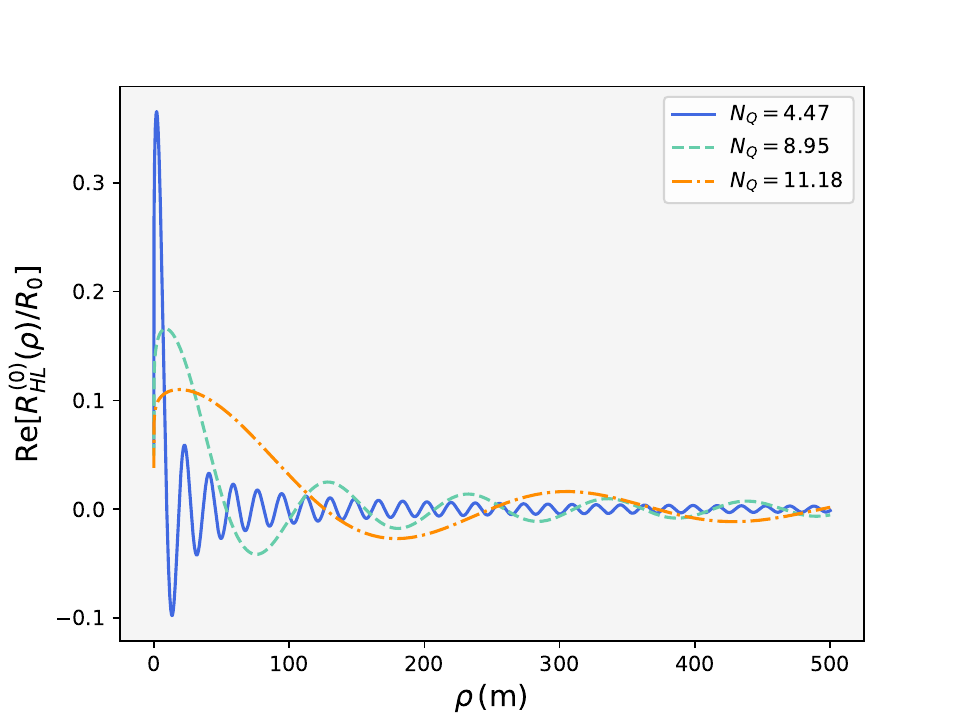}
			\caption{The real component of $R_{\scaleto{HL}{4pt}}^{\scaleto{(0)}{6pt}}(\rho)$, given by \cref{RWF_HL_Alpha_Q_0_Bessel} (with $c_{1} = R_{0}$ and $c_{2} = 0$). This result is valid provided that $\rho>0$ is significantly smaller than the AdS radius $l$ and it complements the plot from Figure \ref{HL_0_Near_Origin}, which describes a horizonless scenario with $\alpha_{\scaleto{Q}{5pt}} = 0$ near the origin. For consistency, we use again that: $\kappa = 1$, $\rho_{\scaleto{S}{4pt}} = 10^{5}$ m, $\overline{a} = 1$, $\overline{m}_{\phi} = 1$ m$^{-1}$, $\epsilon = 4\,\overline{m}_{\phi}$. Finally, note that the oscillation frequency decreases with increasing $N_{\scaleto{Q}{5pt}}$, as expected.}
			\label{HL_0_FD}
		\end{center}
	\end{figure}

	Let us first consider the regime of $\alpha_{\scaleto{Q}{5pt}} = 0$, where the solution can be equally well described by either \cref{RWF_HL_0_Full} or \cref{RWF_HL_Alpha_Q_0_Bessel}. For convenience, we choose the approach with Bessel functions and expand $R_{\scaleto{HL}{4pt}}^{\scaleto{(0)}{6pt}}(\rho)$, with $c_{2}=0$, around $\rho = 0$ to examine the influence of quintessence on the scalar particle's wave function near the origin. This results in:
	\begin{equation}
		R_{\scaleto{HL}{4pt}}^{\scaleto{(0)}{6pt}}(\rho) = \Phi_{\scaleto{HL}{4pt}}^{\scaleto{(0)}{6pt}}(\rho)\, \exp\left[i\,\delta_{\scaleto{HL}{4pt}}^{\scaleto{\,(0)}{6pt}}(\rho)\right]\,,\label{R_WF_HL_0_NO}
	\end{equation}
	with
	\begin{equation}
		\Phi_{\scaleto{HL}{4pt}}^{\scaleto{(0)}{6pt}}(\rho) = \overline{R_{0}}\,\rho^{-1/2}\,,\label{Phi_HL_0_Expand}
	\end{equation}
	where the normalization constant $\overline{R_{0}}$ had absorbed the other multiplicative factors (independent of $\rho$), such that:
	\begin{equation*}
		\overline{R_{0}} = c_{1}\,\frac{2^{-m} k^{m}}{\Gamma(m+1)}\,\exp\left[-i m (k+\pi)\right]\,.
	\end{equation*}
	Here, $c_{1}$ is a normalization constant, $\Gamma(m+1)$ is a gamma function \cite{Olver2010,Butkov1968}, while $k$ and $m$ are defined by eqs. \eqref{def-k} and \eqref{def-m}, respectively. Additionally, note that the dark phase $\delta_{\scaleto{HL}{4pt}}^{\scaleto{(0)}{6pt}}(\rho)$ has the form of
	\begin{equation}
		\delta_{\scaleto{HL}{4pt}}^{\scaleto{\,(0)}{6pt}} = \sqrt{\frac{\kappa^{2}}{\overline{a}+N_{\scaleto{Q}{5pt}}} + \frac{1}{4}}\,\left(\rho - i\,\ln \rho\right)\,,\label{Dark-phase-HL-0}
	\end{equation}
	which is responsible for changes in both the phase and amplitude of $R_{\scaleto{HL}{4pt}}^{\scaleto{(0)}{6pt}}(\rho)$. 
	Furthermore, we choose not to expand the square root in \cref{Dark-phase-HL-0} around $N_{\scaleto{Q}{5pt}} = 0$, since $N_{\scaleto{Q}{5pt}}$ is not small within this regime; after all, we expect $N_{\scaleto{Q}{5pt}}$ to be close to $8.95$, according to Table \ref{table_N_Q_values}, when $\alpha_{\scaleto{Q}{5pt}}=0$. 
	In Figures \ref{HL_0_Near_Origin} and \ref{HL_0_FD}, we show the behavior of the real part of $R_{\scaleto{HL}{4pt}}^{\scaleto{(0)}{6pt}}(\rho)$, except for normalization, for different values of the quintessence parameter $N_{\scaleto{Q}{5pt}}$. It is relevant to note that Figure \ref{HL_0_Near_Origin} refers to the approximation near the origin, given by \cref{Dark-phase-HL-0}, while Figure \ref{HL_0_FD} is valid throughout the radial domain, as determined by \cref{RWF_HL_Alpha_Q_0_Bessel}, in which the amplitude of $R_{\scaleto{HL}{4pt}}^{\scaleto{(0)}{6pt}}(\rho)$ correctly decreases with increasing radial distance $\rho$.
	
    After examining the case where $\alpha_{\scaleto{Q}{5pt}} = 0$, we turn our attention to the horizonless scenario with $\alpha_{\scaleto{Q}{5pt}} = 1/2$. The radial solution in the extended domain is given by \cref{Radial_WF_Intermediate_HL}, subject to the constraint $\epsilon^{2} = \overline{a}\,\overline{m}_{\phi}^{2}/2$. However, to analyze its behavior near the origin, we expand $R_{\scaleto{HL}{4pt}}^{\scaleto{(1/2)}{6pt}}(\rho)$ in series up to the first order, which gives
	\begin{equation}
		R_{\scaleto{HL}{4pt}}^{\scaleto{(1/2)}{6pt}}(\rho) = \overline{R_{0}}\,\rho^{\beta/2}\,\rho^{-1/2}\left[-1 + \left(1 + \frac{N_{\scaleto{Q}{4pt}}}{2\overline{a}} - \frac{\alpha}{2} - \nu\right)\rho + \mathcal{O}(\rho^{2})\right]\,,\label{hl_meio_series}
	\end{equation}
    where the coefficient $c_{2}$ is determined by \cref{recorrência-z-0,As}:
    \begin{equation}
        c_{2} = -\frac{1}{2}\,\frac{\left(\alpha-1+\alpha\beta-2\beta - 2\eta\right)}{\left(1+\beta\right)}\,,
    \end{equation}
    while the Heun parameters were defined by \cref{Heun_Parameters_Middle_Alpha_Q}. 
	Note that the Heun parameters $\alpha$ and $\beta$ are independent of $N_{\scaleto{Q}{5pt}}$, but $\eta$ is not.

    In this case, the factorization of \cref{hl_meio_series} is quite direct, implying that
    \begin{equation}
        R_{\scaleto{HL}{4pt}}^{\scaleto{(1/2)}{6pt}}(\rho) = \Phi_{\scaleto{HL}{4pt}}^{\scaleto{(1/2)}{6pt}}(\rho)\,\delta_{\scaleto{HL}{4pt}}^{\scaleto{(1/2)}{6pt}}(\rho)\,,\label{RWF_Meio_HL_Expand_Product}
    \end{equation}
    with
    \begin{equation*}
        \Phi_{\scaleto{HL}{4pt}}^{\scaleto{(1/2)}{6pt}}(\rho) = \overline{R_{0}}\,\rho^{\beta/2}\,\rho^{-1/2}\,,
    \end{equation*}
    while
    \begin{equation}
        \delta_{\scaleto{HL}{4pt}}^{\scaleto{(1/2)}{6pt}}(\rho) = -1 + \left(1 + \frac{N_{\scaleto{Q}{4pt}}}{2\overline{a}} - \frac{\alpha}{2} - \nu\right)\rho + \mathcal{O}(\rho^{2})\,.
    \end{equation}
    The expression for $\delta_{\scaleto{HL}{4pt}}^{\scaleto{(1/2)}{6pt}}(\rho)$ show that it is not a standard \enquote{dark phase} as it changes the amplitude of $R_{\scaleto{HL}{4pt}}^{\scaleto{(1/2)}{6pt}}(\rho)$. It would be appropriate to name it as a \enquote{dark amplitude} instead.
    
	To conclude our analysis, we investigate the horizonless scenario associated with $\alpha_{\scaleto{Q}{5pt}}=1$. The radial solution in the extended domain, as determined by \cref{RWF_HL_1_rho}, is given by
	\begin{equation*}
		R_{\scaleto{HL}{4pt}}^{\scaleto{(1)}{6pt}}(\rho) = \frac{\overline{R_{0}}}{\rho}\,\left(1-\frac{N_{\scaleto{QL}{5pt}}}{2\overline{a}}\,\rho^{2}\right)\,\rho^{(1+\alpha)/2}\, e^{-\rho^{2}/\lambda_{\scaleto{H}{3pt}}^{2}}\,\text{HeunB}\left(\alpha,\,\beta,\,\gamma,\,\delta;\,(2P)^{-1/4}\rho\right)\,,
	\end{equation*}
	with the characteristic length $\lambda_{\scaleto{H}{4pt}}$ determined in \cref{def-lambda_H}, while $P$ was defined by \cref{def-P}. We recall that this solution is valid for all $\rho \in \mathbb{R}^{+}$, provided that $2\epsilon^{2} \neq \overline{a}\, \overline{m}_{\phi}^{2}$.
	
	The Heun parameters, $\alpha,\,\beta,\,\gamma$ and $\delta$, are determined by \cref{Heun_Parameters_HL_1} and are all independent of $N_{\scaleto{QL}{5pt}}$. However, the mandatory change of variables from $z$ back to $\rho$ introduced the direct dependence on $N_{\scaleto{QL}{5pt}}^{1/4}$.
	
	Then, in accordance with the previous cases which depended on solutions of the CHE, our best strategy is to check for the occurrence of dark phases in a particular limit, such as the case where $\rho\to 0$. In this limit, a second-order series expansion\footnote{We reinforce that a first-order expansion would not be useful, because the coefficient accompanying $\rho^{1}$ is identically null.} of $R_{\scaleto{HL}{4pt}}^{\scaleto{(1)}{6pt}}(\rho)$ reads
	\begin{equation}
		R_{\scaleto{HL}{4pt}}^{\scaleto{(1)}{6pt}}(\rho) = \overline{R_{0}}\,\rho^{(\alpha-1)/2}\left[1 - \left(\frac{N_{\scaleto{QL}{4pt}}}{2\overline{a}} + \frac{1}{\lambda_{\scaleto{H}{4pt}}} - \frac{v_{2}}{\sqrt{2P}}\right) \rho^{2} + \mathcal{O}\left(\rho^{3}\right)\right]\,,\label{RWF_1_HL_Expanded}
	\end{equation}
	provided that $2 \epsilon^{2} \neq \overline{a}\,\overline{m}_{\phi}^{2}$. Recalling $\alpha \geq 1$, in accordance with \cref{Heun_Parameters_HL_1}, it follows that $R_{\scaleto{HL}{4pt}}^{\scaleto{(1)}{6pt}}(\rho)$ is regular at $\rho=0$, as desired. The remaining parameters $\lambda_{\scaleto{H}{4pt}}$ and $P$ are defined by \cref{def-lambda_H,def-P}, while $v_{2}$ is the third term of the series \eqref{def-HeunB}, which determines HeunB$\left(\alpha,\beta,\,\gamma,\delta;\,(2P)^{-1/4}\rho\right)$.
    
    Considering the specific Heun parameters of \cref{Heun_Parameters_HL_1} and the recursion relation \eqref{recorrência-EBH}, it is simple to show that
    \begin{equation}
        v_{2} = \frac{1}{2}\,\frac{2-\gamma+\alpha}{2+\alpha}\,,
    \end{equation}
    where $\alpha$ and $\gamma$ are independent of
    $N_{\scaleto{QL}{4pt}} = N_{\scaleto{Q}{4pt}} + 1/l^{2}$.
	
	Therefore, we can factor \eqref{RWF_1_HL_Expanded} to rewrite it as
	\begin{equation}
		R_{\scaleto{HL}{4pt}}^{\scaleto{(1)}{6pt}}(\rho) = \Phi_{\scaleto{HL}{4pt}}^{\scaleto{(1)}{6pt}}(\rho)\,\delta_{\scaleto{HL}{4pt}}^{\scaleto{(1)}{6pt}}(\rho)\,,\label{RWF_1_HL_Expand_Product}
	\end{equation}
	where 
	\begin{equation*}
		\Phi_{\scaleto{HL}{4pt}}^{\scaleto{(1)}{6pt}}(\rho) = \overline{R_{0}}\,\rho^{(\alpha-1)/2}\,,
	\end{equation*}
	so that $\delta_{\scaleto{HL}{4pt}}^{\scaleto{(1)}{6pt}}(\rho)$ is given by
	\begin{equation}
		\delta_{\scaleto{HL}{4pt}}^{\scaleto{(1)}{6pt}}(\rho) = 1 - \left(\frac{N_{\scaleto{QL}{4pt}}}{2\overline{a}} + \frac{1}{\lambda_{\scaleto{H}{4pt}}} - \frac{v_{2}}{\sqrt{2P}}\right) \rho^{2} + \mathcal{O}\left(\rho^{3}\right)\,.
	\end{equation}
	Similarly to the previous case of $R_{\scaleto{HL}{4pt}}^{\scaleto{(1/2)}{6pt}}(\rho)$, $\delta_{\scaleto{HL}{4pt}}^{\scaleto{(1)}{6pt}}(\rho)$ should not be interpreted merely as a phase, because it does change the amplitude of $R_{\scaleto{HL}{4pt}}^{\scaleto{(1)}{6pt}}(\rho)$. Actually, since it is a dark energy-induced effect, we call it a \enquote{dark amplitude}.
	
	Our final task is to investigate the remaining particular case when $\epsilon^{2} = \overline{a}\,\overline{m}_{\phi}^{2}/2$. The radial solution associated with both $\alpha_{\scaleto{Q}{5pt}}=1$ and this spectral constraint is given by \cref{RWF_HL_Alpha_Q_1_Constraint_2}, where the order $n$ of the modified Bessel functions is defined by \cref{order-n}. From this result for $R_{\scaleto{HL}{4pt}}^{\scaleto{(1,\,\otimes)}{6pt}}(\rho)$, we observe the absence of a dark phase, while we see that $R_{\scaleto{HL}{4pt}}^{\scaleto{(1,\,\otimes)}{6pt}}(\rho)$ also exhibits a dark energy-induced change in the factor $\left(1- N_{\scaleto{QL}{5pt}}/2\overline{a} \, \rho^{2}\right)$, which is, again, a \enquote{dark amplitude}. This factor is valid for the entire radial domain, since the Bessel functions occurring in \cref{RWF_HL_Alpha_Q_1_Constraint_2} are independent of $N_{\scaleto{QL}{5pt}}$. In other words, as $\rho\to 0$ the effect becomes null and increases for larger values of $\rho$.
	
	Finally, we recall the main results of this section, where we determined analytical solutions for the radial Klein-Gordon equation \eqref{general-equation-for-R}, which may be interpreted as a wave function, in an extended domain. We explored solutions for the standard (\textsc{st}), cloudless (\textsc{cl}), horizonless (\textsc{hl}), and quintessence-free (\textsc{ql}) scenarios and we were able to verify the emergence of dark phases in the standard and cloudless cases provided $\alpha_{\scaleto{Q}{5pt}} = 0,\, 1/2,$ and $1$. Furthermore, we demonstrated that in horizonless (\textsc{hl}) scenarios, the induction of dark phases is more subtle, generally manifesting as an attenuation of the wave function amplitude, and not primarily as a relative phase. We emphasize that these results are important because they describe the physics of scalar particles subject to the influence of a dark energy candidate. The purely analytical nature of these solutions is essential, since we expect the induced quantum effects to be extremely small. In other words, analytical precision is vital for identifying such phenomena, something that can hardly be captured by numerical solutions in this context. Thus, the contributions arising from our extended solutions not only represent an advance over the class of our previous cases (near the event horizon), but also lay a foundation for future developments.

    \section{Conclusion}\label{Conclusion}	

    Understanding the intricate interplay between quantum phenomena and the nature of dark energy is of the highest importance for a comprehensive understanding of our universe. In this study, we have extended our previous analysis on the Klein-Gordon equation for a scalar particle within the BCK spacetime, which features the core components of our investigation: mass distribution, an event horizon, and a Kiselev-like fluid. Our analysis focused on exploring the influence of quintessence, represented by the parameters $\alpha_{\scaleto{Q}{5pt}}$ and $N_{\scaleto{Q}{5pt}}$, on the radial KG solutions with broader radial domains, defined separately for each case. By doing so, we were able to investigate the emergence of \enquote{dark phases} in these extended solutions.
    
    For $\alpha_{\scaleto{Q}{5pt}} = 0$ and $\alpha_{\scaleto{Q}{5pt}} = 1/2$, the solutions revealed that the quintessence contribution significantly affects the wave function, especially at larger distances. Likewise, the regime of $\alpha_{\scaleto{Q}{5pt}} = 1$ combined the influence of the quintessential fluid and the cosmological constant by means of the parameter $N_{\scaleto{QL}{5pt}}=N_{\scaleto{Q}{5pt}}+ 1/l^{2}$, showing how the quintessence contribution would dominate over the cosmological constant. Moreover, the regime of $\alpha_{\scaleto{Q}{5pt}} = 1$ is of great importance, even within our hypothetical cylindrical universe, since these solutions allow for valuable future comparisons with the realistic case of spherical symmetry, for which we have recent observational constraints \cite{DESI2025,DESCollaboration2025}.
    
    By employing these solutions, we have demonstrated the existence of quintessence-induced phase shifts in the radial wave functions for the black string scenarios (both standard and cloudless). We termed them \enquote{dark phases} since they are directly associated with the quintessence parameter $N_{\scaleto{Q}{5pt}}$. These phases are particularly evident near the event horizon.
    
    Additionally, our extension to horizonless scenarios revealed that, while dark phases are usually absent near the origin (except for \cref{Dark-phase-HL-0}), quintessence still influences the overall radial solution through a \enquote{dark amplitude}. The effect increases for larger distances, and it is consistent with our knowledge that dark energy predominantly affects physics at large scales. But at the same time, it is intriguing how this extension to horizonless cases differed from our expectations for dark phases. These results suggest that the event horizon plays a critical role in the quintessence-induced phase shifts.

    We further note that our investigation provides generally applicable results, offering opportunities for experimental detection. In this regard, techniques based on quantum particle interferometry are well-suited for testing our findings. For example, the Hanbury Brown and Twiss (HBT) effect \cite{HBT1956,baym1998,Hama1988}, already widely used in stellar astronomy and heavy-ion collisions, relies on quantum particles emitted with random phases, making it an excellent candidate for experimental applicability. Therefore, the primary goal would be to identify specific physical systems where the parameter configurations yield a statistically significant signal.
    
    In addition, the emergence of dark phases in scalar particle solutions may modify established results in quantum field theory regarding particle production. Specifically, phenomena such as the Unruh effect, Hawking radiation, and energy levels are directly affected and are expressed in terms of the Heun parameters. For instance, the exact dependence of the Hawking temperature on these dark parameters could provide interesting thermodynamic features; however, a detailed analysis of these aspects is deferred to future work.
	
    In summary, our findings contribute to a deeper understanding of the potential influence of dark energy on quantum phenomena. Future numerical studies are necessary to explore our current results and their theoretical predictions.

	\section*{Acknowledgments}
	
	The authors appreciate the valuable questions raised by both reviewers, which contributed to the improvement of this work.
	B.V.S. acknowledges financial support from the National Council for Scientific and Technological Development (CNPq, Brazil), under grant number 141549/2023-8.
	
	The authors used Grammarly\textsuperscript{\textsc{tm}} exclusively to enhance the language in selected sections of this manuscript. All data, results, and analyses were produced entirely by the authors.

	\appendix

	\section{The Confluent and Biconfluent Heun Equations}\label{appendix}
	
	To characterize the physical systems in our study, this appendix presents important results on the confluent and biconfluent Heun equations and their associated local solutions, which are extensively used in Sections \ref{Sec-2-Results-Paper-One} and \ref{Sec-3-Extended-Solutions}. These results can be derived from references \cite{Olver2010,Ronveaux1995} and are also discussed in the appendix of \cite{Simao2025}.
	
	This section is organized into three subsections that introduce the confluent and biconfluent Heun equations and their Liouville normal forms.

	\subsection{Confluent Heun Equation}\label{appendix_CHE}
	
	We begin by examining the canonical form of the confluent Heun equation (CHE), given by \cite{Ronveaux1995,Olver2010}:
	\begin{equation}
		\frac{d^{2} y}{d z^{2}} + \left(\alpha + \frac{\beta + 1}{z} + \frac{\gamma + 1}{z - 1}\right)\frac{d y}{d z} + \left(\frac{\mu}{z} + \frac{\nu}{z - 1}\right)y = 0\,.\label{HeunConfluentCanonical}
	\end{equation}
	This equation exhibits regular singularities at $z=0$ and $z=1$, as well as an irregular singularity as $z$ approaches infinity. The parameters $\mu$ and $\nu$ are expressed in terms of the Heun parameters $\alpha$, $\beta$, $\gamma$, $\delta$, and $\eta$ as follows:
	\begin{align}
		\mu &= \frac{1}{2}\left(\alpha - \beta - \gamma + \alpha\beta - \beta\gamma\right) - \eta\,,\label{Mu_Condition}\\
		\nu &= \frac{1}{2}\left(\alpha + \beta + \gamma + \alpha\gamma + \beta\gamma\right) + \delta + \eta\,.\label{Nu_Condition}
	\end{align}
	Mathieu functions, spheroidal wave functions, and Coulomb spheroidal functions are special cases of solutions of the confluent Heun equation \cite{Olver2010}. The solutions to the CHE belong to following classes: the local solutions (obtained with the Fuchs-Frobenius method), the Heun functions (where the local solutions are extended to encompass two regular singularities), the path-multiplicative solutions, and the Heun polynomials (which belong to a subclass of solutions known as \enquote{$\delta_{N}$} \cite{Fiziev2010}). In this work, we are interested in the local solutions (with their analytic continuation) and the $\delta_{N}$ subclass.
	
	Local solutions to the CHE can be determined about a regular singular point, in this case $z_{0} = 1$, using the Fuchs-Frobenius method \cite{Olver2010}. These solutions are valid within the radius of convergence, which for the CHE is $|z-z_{0}|<1$. Then, considering the analytic continuation of these local solutions, it can be demonstrated that the formal solution to \cref{HeunConfluentCanonical} is given by \cite{Olver2010}:
	\begin{equation}
		y(z) = \mathcal{C}_{1}\,\text{HeunC}\left(\alpha,\,\beta,\,\gamma,\,\delta,\,\eta;\,z\right) + \mathcal{C}_{2}\,z^{-\beta}\,\text{HeunC}\left(\alpha,\,-\beta,\,\gamma,\,\delta,\,\eta;\,z\right)\,,\label{Sol_Geral_ECH_0}
	\end{equation}
	where $\beta \notin \mathbb{Z}$, which includes the case $\beta \neq 0$, and
	\begin{equation}
		\begin{aligned}
			y(z) 
			&= \mathcal{C}_{1}\,\text{HeunC}\left(-\alpha,\,\gamma,\,\beta,\,-\delta,\,\delta+\eta;\,1-z\right)\\
			&+ \mathcal{C}_{2} \left(z-1\right)^{-\gamma}\text{HeunC}\left(-\alpha,\,-\gamma,\,\beta,\,-\delta,\,\delta+\eta;\,1-z\right)\,,
		\end{aligned}\label{Sol_Geral_ECH_1}
	\end{equation}
	if $\gamma \notin \mathbb{Z}$, which similarly includes $\gamma \neq 0$. The constants $\mathcal{C}_{1}$ and $\mathcal{C}_{2}$ are arbitrary and must be specified by boundary conditions or initial values. It is important to stress that \cref{Sol_Geral_ECH_0,Sol_Geral_ECH_1} represent physical solutions centered at $z_{0} = 0$ and $z_{0} = 1$, respectively.
	
	Furthermore, $\text{HeunC}\left(\alpha,\,\beta,\,\gamma,\,\delta,\,\eta;\,z\right)$ should be interpreted as the analytic continuation of the local solution $\text{HeunC}\ell\left(\alpha,\,\beta,\,\gamma,\,\delta,\,\eta;\,z\right)$, as defined by \cite{Simao2025}:
	\begin{equation}
		\text{HeunC}\ell\left(\alpha,\,\beta,\,\gamma,\,\delta,\,\eta;\,z\right) = \sum_{m=0}^{+\infty} c_{m}\,z^{m}\,,\label{def-HeunC}
	\end{equation}
	where $c_{0} = 1$, $c_{1} = -\mu/\left(1+\beta\right)$, and the three-term recurrence relation is given by:
	\begin{equation}
		A_{2}\,c_{m+2} = A_{0}\,c_{m} + A_{1}\,c_{m+1}\,,\quad m\geq 0\,.\label{recorrência-z-0}
	\end{equation}
	The explicit forms of $A_{i}$, for $i=1,\,2,\,3$, are as follows:
	\begin{equation}
		\begin{aligned}
			A_{0} &= m\,\alpha + \frac{\alpha}{2}\left(\beta + \gamma + 2\right) + \delta\,,\\[4pt]
			A_{1} &= \frac{\beta}{2} + \eta - \frac{1}{2}\left(\alpha - \gamma\right)\left(1 + \beta\right) + \left(m + 1\right)\left(m + \Delta_{c}\right)\,,\\[4pt]
			A_{2} &= \left(m + 2\right)\left(m + 2 + \beta\right)\,,
		\end{aligned}\label{As}
	\end{equation}
	where $\Delta_{c} = \beta + \gamma - \alpha + 2$. Additionally, as $z \to 0$, the following holds:
	\begin{equation}
		\text{HeunC}\ell\left(\alpha,\,\pm\beta,\,\gamma,\,\delta,\,\eta;\,z\to 0\right) \approx 1\,.\label{Approx_HeunCl}
	\end{equation}
	
	After establishing the main results regarding the CHE, we subsequently present the $\delta_{N}-$condition, used in the series break-off.

	\subsubsection{Series break-off and $\delta_{N}-$subclass}\label{sec-vínculos-espectrais}

	In physics, obtaining solutions with spectral constraints is of great importance, as these often provide details about energy levels and other relevant physical parameters. This is also true for Heun-type equations, including the CHE, leading to frequent use of constraints in the literature.
	
	These constrained cases originate from the Fuchs-Frobenius solutions\footnote{For more details, please see our appendix in \cite{Simao2025}.}, determined by \cref{def-HeunC,recorrência-z-0,As}, when we impose that, for a given $m = N \in \mathbb{Z}^{+}$, $A_{0}=0$ and $c_{m+1}\left(\alpha,\,\beta, \,\gamma, \,\delta, \,\eta\right) = 0$. Consequently, it follows from the general recurrence relation of \cref{recorrência-z-0} that the coefficients $c_{N+1}$ and $c_{N+2}$ will be all zero, terminating the series \eqref{def-HeunC}.
	
	The first requirement, $A_{0}=0$, leads to the following constraint:
	\begin{align}
		\alpha\left(N + 1 + \frac{\beta + \gamma}{2}\right) + \delta = 0\,,\quad N\in\mathbb{Z}^{+}\,.\label{Delta_Condition}
	\end{align}
	This result, referred to as the $\delta$-condition, defines a subclass of solutions, termed the $\delta_{N}-$subclass \cite{Fiziev2010}. When this condition is combined with the requirement that $c_{N+1}^{(1)}= 0$, the confluent Heun polynomials are obtained. Conversely, if only the condition \eqref{Delta_Condition} (a spectral constraint) is imposed, the resulting solutions belong to the $\delta_{N}$-subclass, but are not necessarily the Heun polynomial.
	
	In the restricted case where the Heun parameter $\alpha$ is zero ($\alpha = 0$), the condition \eqref{Delta_Condition} requires that $\delta$ is also zero. Under these circumstances, Heun polynomials can still be obtained; however, non-polynomial solutions within the $\delta_{N}-$ subclass do not exhibit spectrum quantization.
	
	Having reviewed the main properties of the CHE, the next subsection addresses the biconfluent Heun equation.

	\subsection{Biconfluent Heun equation}\label{appendix_BHE}
	
	The canonical form of the Biconfluent Heun equation (BHE) is defined as \cite{Ronveaux1995,Olver2010}:
	\begin{equation}
		\frac{d^{2}y}{d z^{2}} + \left(\frac{1+\alpha}{z} - \beta - 2 z\right)\frac{d y}{d z} + \left\{\left(\gamma - \alpha - 2\right) - \frac{\left[\delta + \left(1 + \alpha\right)\beta\right]/2}{z}\right\}y(z) = 0\,.\label{Biconfluent_Heun_Equation}
	\end{equation}
	This equation exhibits a regular singularity at $z = 0$ and an irregular singularity as $z$ approaches infinity. Its solutions are classified analogously to those of the Confluent Heun Equation (CHE). Since the biconfluent form lacks a singularity at $z = 1$, unlike the CHE, the Fuchs-Frobenius (local) solution is centered solely at $z_{0} = 0$. The radius of convergence is determined only by $|z| < \infty$, as the only remaining singularity within the domain is at infinity. The absence of the regular singularity at $z = 1$ also means the BHE has one fewer Heun parameter than the confluent case.
	
	The formal solution to \cref{Biconfluent_Heun_Equation} can be expressed as follows:
	\begin{equation}
		y(z) = \mathcal{C}_{1}\,\text{HeunB}\left(\alpha,\,\beta,\,\gamma,\,\delta;\,z\right) + \mathcal{C}_{2}\,z^{-\alpha}\,\text{HeunB}\left(-\alpha,\,\beta,\,\gamma,\,\delta;\,z\right)\,,\label{Sol_Geral_EBH}
	\end{equation}
	where $\mathcal{C}_{1}$ and $\mathcal{C}_{2}$ are arbitrary constants to be determined by the boundary conditions or initial values. Moreover, the function $\text{HeunB}\left(\alpha,\,\beta,\,\gamma,\,\delta;\,z\right)$ is defined as
	\begin{equation}
		\text{HeunB}\left(\alpha,\,\beta,\,\gamma,\,\delta;\,z\right) = \sum_{m=0}^{+\infty} v_{m}\,z^{m}\,.\label{def-HeunB}
	\end{equation}
	The coefficients $v_{m}$ satisfy $v_{0} = 1$, $v_{1} = \left(\beta + \alpha\beta + \delta\right)/\left(2+2\alpha\right)$, and the general recurrence relation
	\begin{equation}
		L_{2}\,v_{m+2}  = L_{1}\,v_{m+1} + L_{0}\,v_{m}\,,\quad m\geq 0\,.\label{recorrência-EBH}
	\end{equation}
	Note that, similarly to the CHE, the solutions to the biconfluent Heun equation also satisfy a three-term recursion relation. The expressions for $L_{i}$, where $i=0,\,1,\,2$, are given by
	\begin{equation}
		\begin{aligned}
			L_{0} &= 2\left(m + 1\right) - \gamma + \alpha\,,\\[5pt]
			L_{1} &= \left(m + 1\right)\beta + \frac{\delta}{2} + \frac{\beta}{2}\left(1 + \alpha\right)\,,\\[5pt]
			L_{2} &= \left(m + 2\right)\left(m + 2 + \alpha\right)\,.\label{Ls}
		\end{aligned}
	\end{equation}
	\vspace{1mm}
	
	Furthermore, we can define an equivalent $\delta-$condition for the solutions of the BHE. To do so, we impose that, for $m=N \in \mathbb{Z}^{+}$:
	\begin{equation}
		\gamma - \alpha = 2N + 2\,,\label{Delta_Condition_EBH}
	\end{equation}
	implying that $L_{0}=0$. This condition defines the $\delta_{N}-$subclass of solutions regarding the BHE and, if combined with the constraint $v_{N+1}(\alpha,\,\beta,\,\gamma,\,\delta)=0$, originates the biconfluent Heun polynomials.	
	
	This completes the presentation of the fundamental aspects of the solution to the BHE using the Frobenius method. We recall that the two solutions comprising $y(z)$ are linearly independent provided that $\alpha$ is not an integer ($\alpha \notin \mathbb{Z}$), which includes the case of $\alpha \neq 0$. Conversely, if $\alpha$ is an integer, an alternative linearly independent solution must be constructed in accordance with Fuchs' Theorem.

	In the final subsection, we present the general Liouville normal form as well as the corresponding normal forms for the confluent and biconfluent Heun equations, which are essential for our extended solutions.

	\subsection{Liouville normal form} \label{appendix_Liouville_Normal_Form}
	
	Transforming second-order linear ordinary differential equations (ODEs) into Liouville normal form is highly beneficial for numerous physics applications. This transformation is particularly significant for two reasons. First, the literature on spectral analysis frequently employs the Liouville normal form as a basis for determining whether the spectrum is continuous or quantized. Second, in this form, the resulting ODE is equivalent to a one-dimensional time-independent Schrödinger equation, which yields more detailed insights into the system's dynamics from a physics perspective.
	
	The present work focuses specifically on the normal forms of the confluent and biconfluent Heun equations. The results presented in this section are essential for obtaining the radial solution of the Klein-Gordon equation in a BCK spacetime. To this end, the procedure is revisited, beginning from first principles. 
	
	Consider the following second-order linear ordinary differential equation:
	\begin{equation}
		y''(z) + p(z)\, y'(z) + q(z)\, y(z) = 0\,.\label{General_2nd_Order_ODE}
	\end{equation}
	If we consider that
	\begin{equation}
		y(z) = I(z)\,u(z)\,,\label{y-I-u}
	\end{equation}
	then, \cref{General_2nd_Order_ODE} can be rewritten as follows:
	\begin{equation}
		\begin{aligned}
			0 
			&= I(z)\, u''(z) + \left[2 I'(z) + p(z)\,I(z)\right]u'(z)\\[5pt]
			&+ \left[I''(z) + p(z)\,I'(z) + q(z)\,I(z)\right] u(z)\,.
		\end{aligned}\label{Expand-2nd-order-ODE}
	\end{equation}
	To eliminate the first derivative term of $u(z)$, the coefficient $2 I'(z) + p(z)\,I(z)$ must be set to zero. This condition yields the following differential equation for $I(z)$:
	\begin{equation}
		I'(z) = -\frac{1}{2}\, p(z)\, I(z)\,,\label{Eq_I_Deriv}
	\end{equation} 
	whose solution is given by
	\begin{equation}
		I(z) = I_{0}\exp\left(-\frac{1}{2} \int_{z_0}^{z} p(z')\,dz'\right)\,.\label{Recipe_for_I}
	\end{equation}
	Therefore, since $2 I'(z) + p(z)\,I(z)$ is zero, with $I(z) \neq 0$, the differential equation for $u(z)$ takes the form:
	\begin{equation}
		u''(z) + \left[\frac{I''(z)}{I(z)} + p(z)\,\frac{I'(z)}{I(z)} + q(z)\right]u(z) = 0\,.\label{Eq_U_Normal}
	\end{equation}
	By substituting the result from \cref{Eq_I_Deriv} into \cref{Eq_U_Normal}, we obtain the Liouville normal form of \cref{General_2nd_Order_ODE}:
	\begin{equation}
		u''(z) + \left[-\frac{1}{2}\, p'(z) - \frac{1}{4}\, p(z)^{2} + q(z)\right] u(z) = 0\,.\label{General_Equation_Liouville_Normal_Form}
	\end{equation}
	As previously noted, this transformation is relevant in various contexts, including the analysis of the eigenvalue spectrum of a differential equation \cite{Titchmarsh1962}. From a physical perspective, the normal form is particularly advantageous because it is equivalent to a one-dimensional time-independent Schroedinger equation, with an effective potential $V_{\text{eff}}(z)$ defined as follows:
	\begin{equation}
		V_{\text{eff}}(z) = -\frac{1}{2}\, p'(z) - \frac{1}{4}\, p(z)^{2} + q(z)\,,\label{def-potencial-efetivo-appendix}
	\end{equation}
	in accordance with our previous arguments.
	
	In the following subsection, we apply Liouville's normal form to provide a more convenient description of Heun's confluent equation. Subsequently, the same method will be used for the biconfluent equation.

	\subsubsection{The normal form of the confluent Heun equation}

	Given that the Liouville normal form is described in \cref{General_Equation_Liouville_Normal_Form}, it can be directly applied to the Heun confluent equation, as given by \eqref{HeunConfluentCanonical}. In this context, the coefficients $p(z)$ and $q(z)$, defined in \cref{General_2nd_Order_ODE}, are identified as follows:
	\begin{equation}
		\begin{aligned}
			p(z) &= \alpha + \frac{\beta + 1}{z} + \frac{\gamma + 1}{z - 1}\,,\\
			q(z) &= \frac{\mu}{z} + \frac{\nu}{z - 1}\,.
		\end{aligned}\label{p-e-z}
	\end{equation}
	By employing the substitution $y(z)=I(z)\,u(z)$, and utilizing the expression for $p(z)$ along with the results from \eqref{y-I-u} and \eqref{Recipe_for_I}, the solution for $y(z)$ assumes the following form:
	\begin{equation*}
		y(z) = y_{0}\,e^{-\alpha z/2}\,z^{-(1+\beta)/2}\,\left(z - 1\right)^{-(1 + \gamma)/2}\,u(z)\,.
	\end{equation*}
	Here, $y_{0}$ denotes an integration constant and $u(z)$ represents an auxiliary function. The explicit form of $u(z)$ is therefore given by
	\begin{equation}
		u(z) = u_{0}\,e^{\alpha z/2}\,z^{(1+\beta)/2}\,\left(z - 1\right)^{(1 + \gamma)/2}\,y(z)\,,\label{u-from-y}
	\end{equation}
	where $u_{0}$ is a constant. This representation is particularly relevant for the analysis of Klein-Gordon solutions.
	
	To derive the normal form of the CHE in terms of $u(z)$, the expressions for $p(z)$ and $q(z)$ from \cref{p-e-z} are substituted into \cref{General_Equation_Liouville_Normal_Form}, yielding
	\begin{equation}
		u''(z) + \left[\frac{A}{z} + \frac{B}{(z - 1)} + \frac{C}{z^{2}} + \frac{D}{(z - 1)^{2}} + E\right] u(z) = 0\,,\label{Eq_u_Normal_Form}
	\end{equation}
	with
	\begin{equation}
		\begin{aligned}
			A &= \tfrac{1}{2} - \eta\,,\\
			B &= -\tfrac{1}{2} + \delta + \eta\,,\\
			C &= \left(1 - \beta^{2}\right)/4\,,\\
			D &= \left(1 - \gamma^{2}\right)/4\,,\\
			E &= -\alpha^{2}/4\,.
		\end{aligned} \label{Parameters_A_to_E_CHE}
	\end{equation}
	
	By using the solutions of the CHE provided in \eqref{Sol_Geral_ECH_0} and \eqref{Sol_Geral_ECH_1}, along with the relation \eqref{u-from-y}, we obtain that the solution for an ordinary differential equation in the form \cref{Eq_u_Normal_Form} is
	\begin{equation}
		\begin{aligned}
			u(z) &= e^{\alpha z/2}\,z^{(1+\beta)/2}\,\left(z - 1\right)^{(1 + \gamma)/2}
			\left[c_{1}
			\text{HeunC}\left(\alpha,\,\beta,\,\gamma,\,\delta,\,\eta;\,z\right) \right.\\
			&\left.+\, c_{2} \, z^{-\beta}\,
			\text{HeunC}\left(\alpha,\,-\beta,\,\gamma,\,\delta,\,\eta;\,z\right)\right] \,,\label{Sol-Geral-Normal-ECH_z-0}
		\end{aligned}
	\end{equation}
	for the case centered at $z_{0}=0$. It is important to note that $\text{HeunC}\left(\alpha,\,\beta,\,\gamma,\,\delta,\,\eta;\,z\right)$ denotes the analytical continuation of the local solution of the series \eqref{def-HeunC}. For the case centered at $z_{0}=1$, the solution is given by
	\begin{equation}
		\begin{aligned}
			u(z) &= e^{\alpha z/2}\,z^{(1+\beta)/2}\,\left(z - 1\right)^{(1 + \gamma)/2}
			\left[c_{1}
			\text{HeunC}\left(-\alpha,\,\gamma,\,\beta,\,-\delta,\,\delta+\eta;\,1-z\right) \right.\\
			&\left.+\, c_{2}\,\left(1-z\right)^{-\gamma}
			\text{HeunC}\left(-\alpha,\,-\gamma,\,\beta,\,-\delta,\,\delta+\eta;\,1-z\right)\right] \,.\label{Sol-Geral-Normal-ECH_z-1}
		\end{aligned}
	\end{equation}
	In both cases, $c_{1}$ and $c_{2}$ are arbitrary constants determined by the boundary conditions or initial values.
	
	With the formal solution for \cref{Eq_u_Normal_Form} established, the analysis now proceeds to the normal form of the biconfluent Heun equation.

	\subsubsection{The normal form of the biconfluent Heun equation}

	In this final subsection, Liouville's method is applied to derive the normal form of the biconfluent Heun equation. The procedure follows the same procedure as the confluent case.
	
	First, we identify the coefficients $p(z)$ and $q(z)$, defined by the \cref{General_2nd_Order_ODE}, referring to the canonical form of the BHE, given by \cref{Biconfluent_Heun_Equation}, from which it follows that:
	\begin{equation}
		\begin{aligned}
			p(z) &= \frac{1+\alpha}{z} - \beta - 2z\,,\\
			q(z) &= \gamma-\alpha-2 -\frac{\delta+\left(1+\alpha\right)\beta}{2z}\,.
		\end{aligned}\label{p-e-z-EBH}
	\end{equation}
	
	Next, we perform the substitution $y(z)=I(z)\,u(z)$, where $I(z)$ satisfies \cref{Recipe_for_I} and $u(z)$ is an auxiliary function. According to the result above for $p(z)$, $I(z)$ becomes
	\begin{equation*}
		I(z) = \exp\left[-\frac{1}{2} \int_{z_{0}}^{z} \left(\frac{1+\alpha}{z'} - \beta - 2 z'\right) dz'\right]\,,
	\end{equation*}
	so that
	\begin{equation*}
		I(z) = I_{0}\, z^{-(1+\alpha)/2}\, e^{\beta z/2}\, e^{z^{2}/2}\,,
	\end{equation*}
	where $I_{0}$ is a constant. Therefore, the solution to the biconfluent equation, $y(z)$, takes the form
	\begin{equation*}
		y(z) = y_{0}\,z^{-(1+\alpha)/2}\, e^{\beta z/2}\, e^{z^{2}/2}\,u(z)\,.
	\end{equation*}
	Consequently, the auxiliary function $u(z)$ can be expressed as
	\begin{equation}
		u(z) = u_{0}\,z^{(1+\alpha)/2}\, e^{-\beta z/2}\, e^{-z^{2}/2}\,y(z)\,.\label{forma-u-EBH}
	\end{equation}
	The equation governing $u(z)$ is obtained from the prescription \eqref{General_Equation_Liouville_Normal_Form}, leading to
	\begin{equation}
		u''(z) + \left[A\, z^{2} + B\, z + C + \frac{D}{z} + \frac{E}{z^{2}}\right] u(z) = 0\,,\label{Biconfluent_Normal_Form}
	\end{equation}
	where now the parameters $A$, $B$, $C$, $D$, and $E$ are given by
	\begin{equation}
		\begin{aligned}
			A &= - 1/2\,,\\
			B &= -\beta\,,\\
			C &= \gamma - \beta^{2}/4\,,\\
			D &= -\delta/2\,,\\
			E &= \left(1 - \alpha^{2}\right)/4\,. 
		\end{aligned}\label{Biconfluent_Parameters}
	\end{equation}
	It is important to note that these results were obtained using \cref{p-e-z-EBH}.
	
	Finally, by applying \eqref{Sol_Geral_EBH}, it follows that every differential equation of the form \eqref{Biconfluent_Normal_Form} admits the solution
	\begin{equation}
		\begin{aligned}
			u(z) 
			&= u_{0}\,z^{(1+\alpha)/2}\, e^{-\beta z/2}\, e^{-z^{2}/2}
			\left[c_{1}\,\text{HeunB}\left(\alpha,\,\beta,\,\gamma,\,\delta;\,z\right)\right.\\ 
			&\left. + c_{2}\,z^{-\alpha}\,\text{HeunB}\left(-\alpha,\,\beta,\,\gamma,\,\delta;\,z\right)\right]\,, \label{EBH_Normal_Form_Solution}
		\end{aligned}
	\end{equation}
	where $c_{1}$ and $c_{2}$ are constants determined by the boundary conditions of the physical problem. This result enables a more direct and convenient analysis of solutions to the Klein-Gordon equation in specific physical regimes.
	
	The exploration of the properties and solutions of the confluent and biconfluent Heun equations completes the necessary mathematical foundation for obtaining the radial Klein-Gordon extended solutions in the BCK spacetime.

\bibliography{References_A2TM.bib}
\bibliographystyle{unsrt}

\end{document}